\documentclass[11pt, showpacs]{article}
\pdfoutput=1

\usepackage{amsmath,amssymb,amsfonts,bm,marginnote,cite,graphicx,slashed,bbm}
\usepackage[colorlinks=true, pdfstartview=FitV, linkcolor=black, citecolor=black, urlcolor=black]{hyperref}
\usepackage[usenames,dvipsnames]{xcolor}

\usepackage{epstopdf}
\usepackage[justification=centering]{caption}
\usepackage{subcaption}
\usepackage{dcolumn}
\usepackage{bbm}
\usepackage{amscd}
\usepackage{mathrsfs}
\usepackage{dsfont}
\usepackage{comment}
\usepackage{setspace}
\usepackage[margin=0.75in]{geometry}
\usepackage{tensor}
\usepackage{mathrsfs}
\usepackage{cancel}
\usepackage{hyperref}
\usepackage{empheq}
\usepackage{youngtab}
\usepackage{empheq}
\usepackage{MnSymbol}
\usepackage{float}
\usepackage[all]{xy}
\usepackage{graphbox}

\DeclareMathAlphabet{\mathbbold}{U}{bbold}{m}{n}

\edef\marginnotetextwidth{\the\textwidth}

\parskip=\baselineskip

\newcommand{\thistitle}{
Interfaces and the extended Hilbert space of Chern-Simons theory}

\newcommand{\addressuiuc}{
	Department of Physics, University of Illinois,
 	1110 West Green St., Urbana IL 61801, U.S.A.
	}
	
\newcommand{\addressam}{
	Institute of Physics, University of Amsterdam,
	904 Science Park, 1098 XH Amsterdam, The Netherlands
	}

\newcommand{\be}{\begin{equation}}
\newcommand{\ee}{\end{equation}}
\newcommand{\beq}{\begin{eqnarray}}
\newcommand{\eeq}{\end{eqnarray}}
\newcommand{\bea}{\begin{eqnarray}}
\newcommand{\eea}{\end{eqnarray}}
\newcommand{\beqn}{\begin{eqnarray}}
\newcommand{\eeqn}{\end{eqnarray}}

\newcommand{\bs}{\boldsymbol}

\def\pa{\partial}

\newcommand{\mTr}{\mathrm{Tr}}

\newcommand{\mf}[1]{\mathfrak{#1}}
\newcommand{\mc}[1]{\mathcal{#1}}

\newcommand{\omf}[1]{\overline{\mathfrak{#1}}}
\newcommand{\tmc}[1]{\tilde{\mathcal{#1}}}

\newcommand{\bmc}[1]{\boldsymbol{\mathcal{#1}}}
\newcommand{\tbmc}[1]{\boldsymbol{\widetilde{\mathcal{#1}}}}

\newcommand{\obmf}[1]{\overline{\boldsymbol{\mathfrak{#1}}}}
\newcommand{\varpepsilon}{\varepsilon}

\newcommand{\myfig}[3]{
	\begin{figure}[ht]
	\centering
	\includegraphics[width=#2cm]{#1}\caption{\small{\textsf{#3}}}\label{fig:#1}
	\end{figure}
	}

\begin{document}

\title{\thistitle}
\author{
	{Jackson R. Fliss$^{1,2}$ and Robert G. Leigh$^{1}$}\\
	\\
	{$^1$\small \emph{\addressuiuc}}
	\\
	{$^2$\small \emph{\addressam}}
\\}
\date{\today}
\maketitle\thispagestyle{empty}

\begin{abstract}
The low energy effective field theories of $(2+1)$ dimensional topological phases of matter provide powerful avenues for investigating entanglement in their ground states.  In \cite{Fliss:2017wop} the entanglement between distinct Abelian topological phases was investigated through Abelian Chern-Simons theories equipped with a set of topological boundary conditions (TBCs).  In the present paper we extend the notion of a TBC to non-Abelian Chern-Simons theories, providing an effective description for a class of gapped interfaces across non-Abelian topological phases.  
These boundary conditions furnish a defining relation for the extended Hilbert space of the quantum theory and allow the calculation of entanglement directly in the gauge theory.  Because we allow for trivial interfaces, this includes a generic construction of the extended Hilbert space in any (compact) Chern-Simons theory quantized on a Riemann surface.  Additionally, this provides a constructive and principled definition for the Hilbert space of effective ground states of gapped phases of matter glued along gapped interfaces.  Lastly, we describe a generalized notion of surgery, adding a powerful tool from topological field theory to the gapped interface toolbox.

\end{abstract}

\newpage

\section{Introduction}\label{sect:Intro}

The entanglement of spatial subregions of ground-state wavefunctions of $(2+1)$ dimensional gapped phases of matter provides a clear signature of non-trivial topological order.  One aspect of this signature is reflected in the entanglement entropy, which leads to a universal correction to the area law: the so-called \emph{topological entanglement entropy} (TEE) \cite{Kitaev:2005dm,Levin:2006zz}.  
Another aspect is the role of the entanglement spectrum in the bulk-edge correspondence.  Indeed, the boundaries of topological phases of matter generically host gapless degrees of freedom often referred to as ``edge modes."  The dynamics of these edge modes is intimately tied to bulk entanglement through a variety of mechanisms (e.g., spectrum matching \cite{PhysRevLett.108.196402} and  entanglement inflow \cite{Hughes:2015ora}).  The edge modes paint an appealing heuristic, and by now, well-understood, picture of how bulk entanglement can arise even if all (bulk) dynamics is gapped: arising from an imaginary boundary, the modes on the edge of the spatial tensor factors (the {\it entangling surface}) must be invisible to the bulk state.  The short-range correlations that erase or ``gap out" these imaginary edge modes is counted by the area law of the bulk entanglement entropy; however it is the global constraints on the state that lead to the subleading correction and provides the signal of topological order. 

Reversing this logic, the above picture also allows for a construction of a bulk Hilbert space assembled from ``gluing" spatial subregions through gapping out their respective edge degrees of freedom.  Interestingly, this construction works for gluing together systems in possibly different topological phases.  This has led to an increasing interest in a classification of which systems can allow such mutual gapped boundaries, (which are often called \emph{gapped interfaces}) and in the presence of which interactions the edge modes of these systems are unstable to mass generation \cite{Levin:2013gaa,Wang:2013vna}.  Because the edge dynamics are related to bulk entanglement, it is perhaps unsurprising that this classification is reflected in the bulk entanglement entropy.  This idea was first posited in \cite{Cano:2014pya} through an explicit microscopic construction of $(2+1)$ Abelian topological phases using ``coupled wires."  In that paper, the authors found that the choice of gapping interactions for wires straddling the entangling cut modified the universal subleading correction leading to a new, effective, TEE.  This effect was explored further in \cite{Fliss:2017wop} from the point of view of the low-energy effective field theory which is governed by Abelian $K$-matrix Chern-Simons theory.  The gapped interactions were mirrored by a set of \emph{topological boundary conditions} (TBCs)\cite{Kapustin:2010hk} that label Lagrangian subspaces of $\mathbb K=K_L\oplus (-K_R)$ (where $K_L$ and $K_R$ are the $K$-matrices of the theories to the left and the right of the interface).  This story ties in well with the known connections of gapped interfaces, and the description of anyon condensation by Lagrangian subsets \cite{Kapustin:2010hk,Bais:2008ni,Bais:2008xf,Levin:2013gaa,Barkeshli:2013yta}.  In that paper, it was also explained, directly from consideration of the bulk Hilbert space, the special role Ishibashi states \cite{Ishibashi:1988kg} play in reproducing bulk entanglement.  We will return to this point shortly.

This story has been extended, notably in \cite{Lou:2019heg,Shen:2019rck}, to non-Abelian topological phases described by the quantum doubles of finite groups.  There it was argued that the ground state entanglement displays a modified TEE.  In this paper, we aim to supplement this story with the point of view of the low-energy topological field theory, this time a non-Abelian Chern-Simons theory.  As a primary goal, we will understand the results of \cite{Lou:2019heg} directly in terms of the bulk gapped Hilbert space, in a similar vein as \cite{Fliss:2017wop}.  As we shall see below, the role of the $K$-matrix is replaced by the level, $k$, and the Killing form of a Lie algebra $\mf g$. Correspondingly, we will classify TBCs through an analogous (though relaxed) notion of Lagrangian subspace.  Although a generic gapped interface does not have to arise through such a construction\footnote{In particular, the methods of this paper are most naturally stated in terms of {\it symmetry breaking}.  There are also classes of interfaces based upon {\it symmetry extension}, \cite{Wang:2017loc}, whose interpretation in Chern-Simons theory is not clear.  We thank Juven Wang for pointing this out.} we will explain how an interface constructed this way displays all the hallmarks of a gapped interface.  Although it is our belief that any gapped interface between theories with non-Abelian Chern-Simons descriptions can be obtained through our construction, we do not prove this claim\footnote{Additionally we do not claim to provide a generic description of anyon condensation.}.  To be careful about this distinction, we refer to our construction as an {\it isotropic interface.} 

The ability to explore isotropic interfaces in the low-energy field theory comes along with many of the powerful tools of topological field theory.  One such tool is \emph{surgery}, which allows for the evaluation of the Chern-Simons path integral on an arbitrarily complicated compact three manifold in terms of a few simple ``ingredient path integrals," (for example the path integrals on the three sphere $S^3$ or on $S^2\times S^1$).  Using these tools, the R\'enyi path integrals (and henceforth the entanglement entropy) of homogenous Chern-Simons theory corresponding to various entanglement cuts on various spatial manifolds were evaluated in \cite{Dong:2008ft}, extending the results of \cite{Kitaev:2005dm,Levin:2006zz}.  One result of this paper is to show that the entropy across subregions separated by isotropic interfaces can be evaluated in a similar manner.  In doing so, we must evaluate a new novel set of ``ingredient" path integrals.

In addition to surgery methods, the field theory approach to these interfaces allows for bulk manifestation of the Ishibashi states of \cite{Lou:2019heg}, in what we regard as the first main result of this paper.  
All gauge theories face a fundamental obstruction to writing the Hilbert space as a local tensor factor \cite{Buividovich:2008gq,Donnelly:2011hn,Donnelly:2014gva,Donnelly:2016auv,Soni:2015yga,Casini:2013rba}.  Before we can discuss entanglement entropy, we must address this issue and define what it is that we are computing.  In \cite{Fliss:2017wop} we showed that this could be done explicitly using the \emph{extended Hilbert space} prescription \cite{Donnelly:2014fua,Donnelly:2015hxa} and that the bulk state is realized in the extended tensor product precisely as an Ishibashi state.  In this paper, we show that this construction carries over naturally for bulk theories populated by isotropic interfaces.  Because this includes trivial interfaces, this gives a systematic extension of the extended Hilbert space construction to all (compact) Chern-Simons theories on closed manifolds.  This is a significant addition to the small (but growing) handful of existing extended Hilbert space constructions in continuum field theory \cite{Donnelly:2016jet,Fliss:2017wop,Donnelly:2018ppr,Jafferis:2019wkd,Belin:2019mlt,Geiller:2019bti,Hung:2019bnq}.  

Lastly, we remark that although there is much literature about the classification, the anyon excitations, and the ground state degeneracies of the theories populated by interfaces \cite{Wang:2012am,Levin:2013gaa,Wang:2013vna,Lan:2014uaa,Hung:2014tba,Kapustin:2013nva}, we are not aware of any systematic constructions of the Hilbert spaces for these theories.  We show that the extended Hilbert space provides a natural construction of the Hilbert spaces of these theories (and the ground states that furnish them) that, without another independent construction, can be taken as their \emph{definition}.  We regard this as the second main result of the paper.  This is elaborated on in Section \ref{sect:QGint} with further details in Appendix \ref{app:B}.

The structure of the paper is as follows.  In Section \ref{sect:BCs} we begin with a discussion of classical boundary conditions in non-Abelian Chern-Simons theory and introduce the notion of an isotropic subalgebra which we use to construct interfaces between separate Chern-Simons theories.  We elaborate on the physical intuition of why these correspond to gapped interfaces from the point of view of the WZW description of the wavefunctions.  
This story includes a description of what anyon excitations do as they approach the interface which we phrase in terms of the branching of representations upon restriction to a subalgebra.  In Section \ref{sect:QG} we review the aforementioned obstruction to Hilbert space factorization and the extended Hilbert space resolution.  In doing so, we promote the classical boundary conditions of Section \ref{sect:BCs} to quantum operators whose kernel defines the embedding of the bulk Hilbert space into the tensor product.  We use this construction to compute the entanglement entropy across isotropic interfaces for several examples in Section \ref{sect:ExamplesQG}.  Although these results are both regulated and exact, we present a geometric surgery perspective on the  same examples in Section \ref{sect:surgery}.  We finish the main body of the paper with a discussion of these results and their implications for condensed matter theory and for AdS/CFT duality.  Lastly, in 
Appendix \ref{app:B}, we present details on the construction of bulk Hilbert spaces for Riemann surfaces supporting isotropic interfaces.

\section{Classical boundary conditions and gapped interfaces}\label{sect:BCs}

We begin with the action of Chern-Simons theory with a simple gauge group\footnote{For the purposes of this paper, we will always take the gauge group to be compact, but we will offer comments on non-compact groups in the discussion. } $G$ on a compact manifold $M$:
\begin{equation}
S_{CS}=\frac{k}{4\pi}\int_M\mTr\left(A\wedge dA+\frac{2}{3}A\wedge A\wedge A\right)
\end{equation}
where $\mTr$ is taken over a fixed representation of $G$.  The discussions of this paper will be phrased in terms of the equivalent Killing form on $\mf g$, the Lie algebra of $G$:
\begin{equation}
K^{ab}\equiv \mTr\left(t^at^b\right)\qquad\qquad \mf g=\text{span}_{\mathbb R}\{t^a\}.
\end{equation}
Of course, when $\mathfrak g$ is simple we can always find a basis in which $K^{ab}=\delta^{ab}$, but this is not necessary (for instance, we might be interested in expressing $\mf g$ in a Cartan-Weyl basis).  We will also, for convenience, introduce what we will call the \emph{level-Killing form}:
\beq\label{eq:LKFdef}
\kappa^{ab}\equiv k\,K^{ab},
\eeq
which in many ways in the following discussion plays an analogous role to the ``$K$-matrix" familiar in Abelian Chern-Simons theories.  Equation \eqref{eq:LKFdef} extends naturally to {\it semi}-simple Lie algebras $\mathfrak g=\mathfrak g_1\oplus\mathfrak g_2\oplus\ldots$ by regarding it as tensor sum $\kappa=k_1K_1\oplus k_2K_2\oplus\ldots$; this is the generic framework that we will keep in mind.

When $M$ possesses a boundary,  variations of $S_{CS}$ produce a boundary term on-shell\footnote{Note that $\delta a_a$ is both a form on $\pa M$ and on the phase space. The wedge product denotes antisymmetrization of both; for example an expression like $\delta a_a\wedge \delta a_b$ is {\it symmetric} in $a,b$.}
\begin{equation}\label{eq:CSsymponeform}
\delta S_{CS}=\frac{\kappa^{ab}}{4\pi}\int_{\pa M}a_a\wedge \delta a_b.
\end{equation}
where $a_a$ is the pullback of $A_a$ to $\pa M$.  As usual, we interpret this variation as a one-form on the space of field configurations and the resulting boundary term  then gives rise to the \emph{pre-symplectic one-form}, ${\boldsymbol \theta}$. 
Equation \eqref{eq:CSsymponeform} can be modified by the addition of boundary terms that contribute exact variations.  Independent of these additions, the second variation defines the \emph{pre-symplectic two-form}\footnote{In gauge theories, this form possesses null directions (along variations corresponding to gauge transformations).  Only after modding out by these null directions does the pre-symplectic two-form yield a symplectic form on the phase space.  We will not be concerned with this distinction now because we will be interested in identifying a subgroup of these null directions to preserve on the boundary.} on the space of field configurations
\begin{equation}
{\boldsymbol \Omega}=\delta {\boldsymbol \theta}=\frac{\kappa_{ab}}{4\pi}\int_{\pa M}\delta a_a\wedge \delta a_b.
\end{equation}
whose inverse determines the Poisson brackets on phase space.   Boundary conditions for the fields $a_a$ can be classified by \emph{Lagrangian subspaces of ${\boldsymbol \Omega}$}.  A subspace of variational vectors, $\mathcal L_{\bs \Omega}$, is Lagrangian iff it is {\it isotropic} and {\it co-isotropic} with respect to ${\boldsymbol \Omega}$:
\begin{equation}
{\bs \Omega}(v,w)=0\qquad \forall\;\;w\in\mc L_{\bs\Omega}\qquad\Leftrightarrow \qquad v\in\mc L_{\bs\Omega}.
\end{equation}
The vanishing of ${\bs\theta}$ when restricted to $\mc L_{\bs\Omega}$ identifies the ``canonical coordinates" of this subspace.  

\begin{figure}
\centering
 \begin{tabular}{ c c c}
    \includegraphics[align=c,height=5cm]{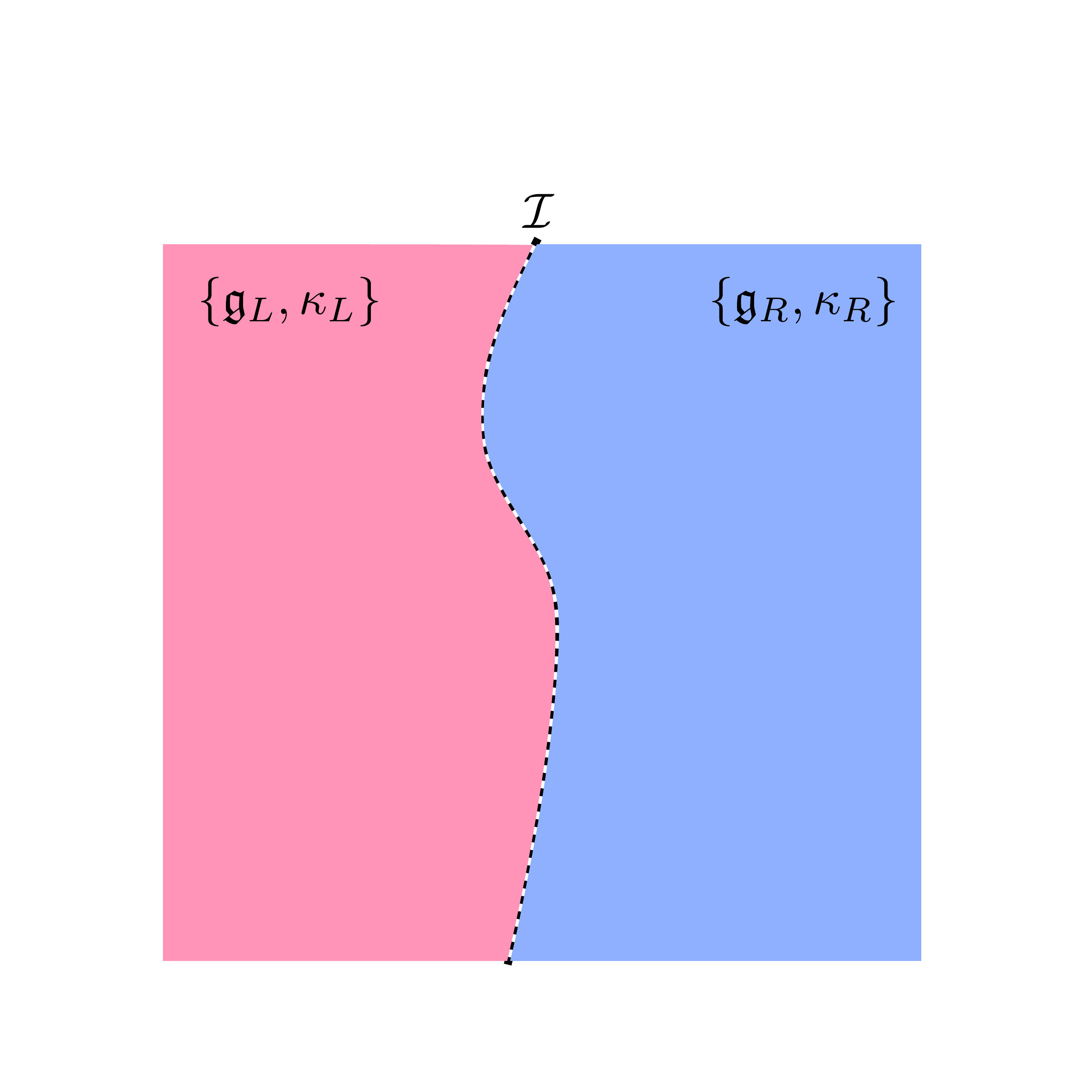} & \hspace{3 cm} & \includegraphics[align=c,height=5cm]{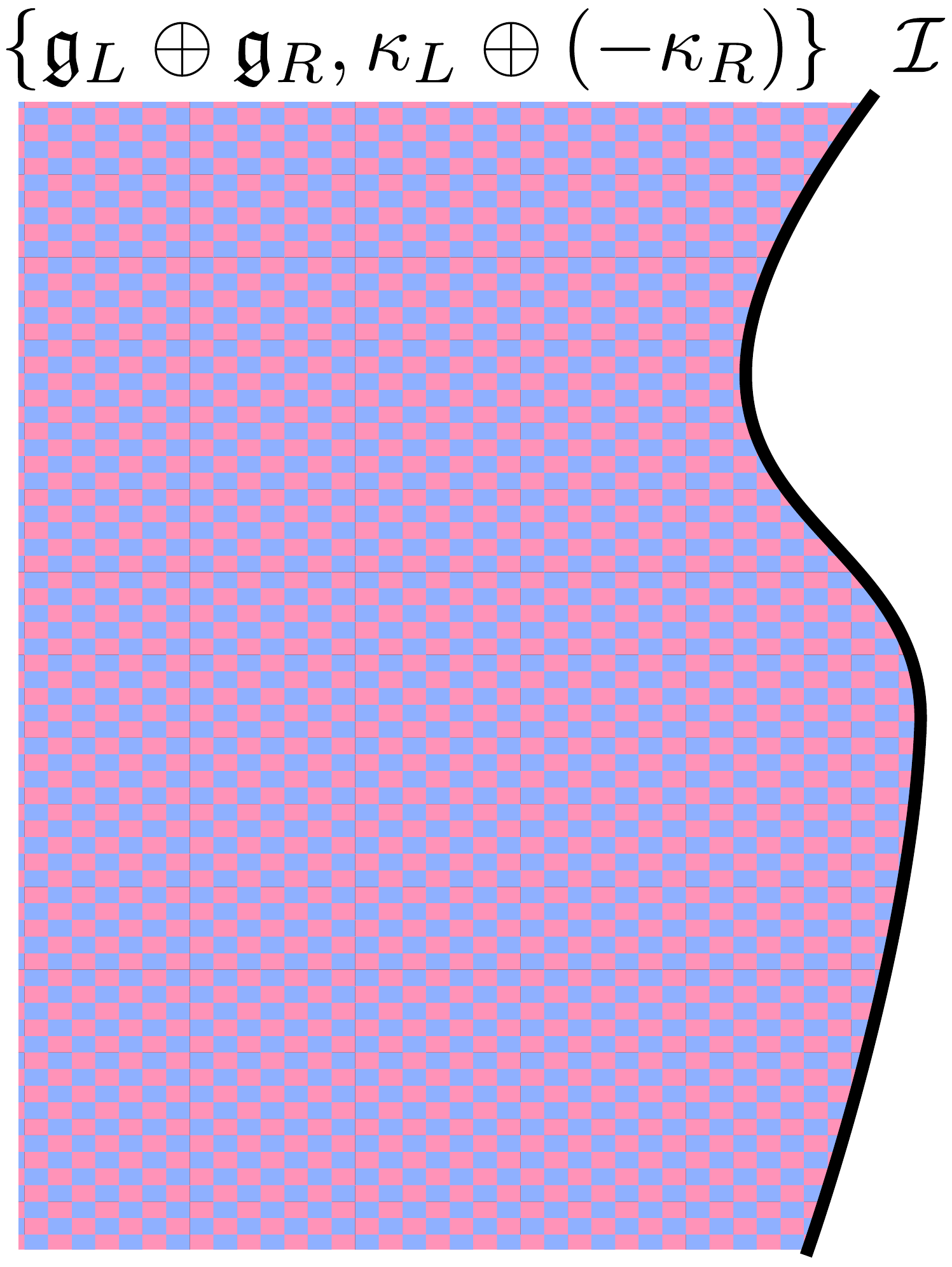}
    \end{tabular}\caption{\small{\textsf{On the left two Chern-Simons theories joined on an interface.  Alternatively, once folded, this is a theory valued in the tensor sum with a hard boundary.}}}\label{LRsetup}
\end{figure}

Generic boundary conditions require additional structure to be introduced.  This is typically a complex structure (or alternatively a metric structure) on $\pa M$.  Indeed, the common procedure when $\pa M$ is a Riemann surface is to introduce complex coordinates $\{z,\bar z\}$ and fix either $a_{z}$ or $a_{\bar z}$ to zero, implemented at the level of the symplectic one-form by the addition\footnote{We write $\pm$ here, but for a given signature of $\kappa$, only one sign choice leads to a unitary boundary theory.\cite{Andrade:2011sx}} of $S_{bndy}=\pm\frac{1}{4\pi}\kappa^{ab}\int_{\pa M}a_a\wedge\star a_b$.  Here $\star$ is the Hodge star with respect to the boundary volume form,  $i\,dz\wedge d\bar z$.  After gauge fixing, this yields the boundary variables corresponding to a chiral Wess-Zumino-Witten (WZW) theory.  This is a standard procedure for defining holomorphic wave-functionals of Chern-Simons theory when we interpret $\pa M$ as a Cauchy slice. For the present case, however, this is not what we are interested in: we are looking for interfaces upon which such degrees of freedom pair up and become massive as a result of interactions between them. Such a situation is sketched in Fig. \ref{LRsetup}; associated with the Chern-Simons on the left is the Lie algebra $\mf g_L$ and level-Killing form $\kappa_L$ and that on the right with $\mf g_R$ and $\kappa_R$. At least locally, we can alternatively view this as folded over to a Chern-Simons theory with algebra $\mf g=\mf g_L\oplus \mf g_R$ and level-Killing form ${\boldsymbol\kappa}=\kappa_L\oplus(-\kappa_R)$, with a hard boundary, as shown in fig.\ref{LRsetup} (we will thus interchangeably refer to interface and boundary in what follows).  Let us now describe a class of {\it topological} boundary conditions that can arise in such a case and do not require the introduction of an auxiliary metric structure on the interface.



Returning to the symplectic two-form, let us look for a subalgebra ${\obmf g}\subset\mf g$ that is Lagrangian with respect to the level-Killing form, $\bs\kappa$.  
A necessary condition for the existence of such a subalgebra is that the number of positive eigenvalues of $\bs\kappa$ is equal to the number of negative eigenvalues.  Thus, for a simple Lie algebra this is a futile effort: the Killing form is positive definite and so cannot admit any Lagrangian subspaces.  However, in our context,  $\mf g=\mf g_L\oplus\mf g_R$ being a direct sum of two Lie algebras with ${\boldsymbol\kappa}=(\kappa_L)\oplus(-\kappa_R)$, it is possible that such a subalgebra $\obmf g$ exists. A simple example would occur if $\mf g_L$ and $\mf g_R$ were isomorphic, in which case $\obmf g$ could be the diagonal subalgebra $\obmf g=\mf g_{L,diag}=\text{span}\left\{t^a\oplus t^a\right\}\subset \mf g_L\oplus \mf g_R$.

At first glance, if $\kappa_L$ and $\kappa_R$ are both positive, then the isotropic and co-isotropic conditions impose $\dim\obmf g=\dim\mf g_L=\dim\mf g_R$.  Lagrangian subspaces of this type are known as \emph{Lagrangian Lie subalgebras} (with respect to $\bs\kappa$)\cite{MR1764438,MR2223160,Evens:aa,Evens:ab,Severa:2016prq}.  As restrictive as this condition is, there can still exist non-trivial interfaces even when $\mf g_L\simeq\mf g_R$ (as we find in \textbf{example 1} of Section \ref{sect:examples}).  However, in the interest of constructing a more general class of gapped interfaces, {\it we will relax the co-isotropic condition} and allow for subspaces that are not half-dimensional.  To this end we will search for a subalgebra, $\omf g\subseteq \mf g_L$ and $\omf g\subseteq\mf g_R$ such that $\left.\kappa_L\right|_{\omf g}=\left.\kappa_R\right|_{\omf g}$.  If such a subalgebra exists then the diagonal $\obmf g:=\omf g_{diag}\subset\omf g\oplus\omf g\subset \mf g_L\oplus \mf g_R$ is an {\it isotropic subalgebra} with respect to ${\bs\kappa}$.  

If there exists an isotropic subalgebra of $\mf g_L\oplus\mf g_R$ with respect to $\bs\kappa$ then we define a corresponding \emph{isotropic interface}, $\mc I$, via the following boundary conditions.  The fields from the left and the right of $\mc I$, once pulled back to the interface (denoted $a^L$ and $a^R$, respectively) will be fixed to lie in $\omf g$ and will be continuous within this subalgebra.  To be more specific, let us denote the embedding of $\omf g$ into $\mf g_{L,R}$ as $\iota_{L,R}$ (respectively):
\beq
\iota_{L,R}:\omf g\hookrightarrow \mf g_{L,R}.
\eeq
Then $a^{L,R}$ must be expressed in terms of a continuous field $\omf a\in\omf g$ as $a^{L,R}=\iota_{L,R}\circ \omf a$.  In a particular basis $\{\bar t^{\bar a}\}$ of $\omf g$, and $\{t_{L,R}^a\}$ of $\mf g_{L,R}$, we can describe this embedding as
\beq
\iota_{L,R}\circ \bar t^{\bar a}={\left(v_{L,R}\right)_a}^{\bar a}\;t^a_{L,R}.
\eeq
such that the components of $a^{L,R}$ in these bases satisfy
\begin{equation}
a^L_a={(v_L)_a}^{\bar b}\omf a_{\bar b}\qquad\qquad a^R_a={(v_R)_a}^{\bar b}\omf a_{\bar b}\qquad\qquad \omf a\in\omf g
\end{equation}
These conditions hold {\it locally} on the interface\footnote{However, for Abelian theories $v_{L,R}$ must, in fact, be constant.  See footnote \ref{footnote:abelianlocality}.}.  In the previously mentioned simple case where $\mf g_L$ is isomorphic to $\mf g_R$ and $\obmf g=\mf g_{L,diag}$, we can take ${(v_{L,R})_a}^{\bar b}=\delta_a{}^{\bar b}$, which we call a \emph{trivial interface.}


The continuity of the symplectic form across $\mc I$ is then 
\begin{equation}\label{eq:keff}
\int_{\mc I}\delta \omf a\cdot(v_L)^t\cdot\kappa_L\cdot v_L\cdot \delta\omf a=\int_{\mc I}\delta\omf a\cdot(v_R)^t\cdot \kappa_R\cdot v_R\cdot \delta\omf a\equiv \int_{\mc I}\delta\omf a\cdot\kappa_{eff}\cdot\delta\omf a
\end{equation}
These conditions are the same whether the gauge groups are Abelian or non-Abelian. However, in the non-Abelian case we have more structure and thus potentially extra conditions. That is, not only do we have matching at the level of vector spaces, but also at the level of the algebras:
\beq
[\iota_{L,R}\circ \bar t^{\bar a},\iota_{L,R}\circ\bar t^{\bar b}]=\iota_{L,R}\circ [\bar t^{\bar a},\bar t^{\bar b}]
\eeq 
Given structure constants $f_{L,R}{}^{ab}{}_c$ and ${\omf f^{\bar a\bar b}}_{\bar c}$ for $\mf g_{L,R}$ and $\omf g$, respectively, then
\beq\label{eq:structurecons}
{\left(v_{L,R}\right)_a}^{\bar a}{\left(v_{L,R}\right)_b}^{\bar b}f_{L,R}{}^{ab}{}_c={\omf f^{\bar a\bar b}}_{\bar c}{\left(v_L\right)_c}^{\bar c}
\eeq
If these conditions can be satisfied, then the effective algebra at the interface is $\omf g$ with structure constants $\omf f$ and level-Killing form $\kappa_{eff}$. 

Let us pause to address the following concern.  From the viewpoint of the classical symplectic form, relaxing co-isotropy seems to be a perverse direction to take: the choice of a half-dimensional subspace is the guide for choosing a polarization for wavefunctions in the quantum theory.  For non-Abelian Chern-Simons theory however, the dimension of its Lie algebra is a poor measure of its quantum degrees of freedom; the proper measure is the Sugawara central charge associated to its affine algebra.  Indeed one would imagine that in order for all of the light degrees of freedom at the interface to be gapped then the matching of the chiral central charges is a necessary condition.  This is guiding principle we will take in this paper. To this end we will require that our subalgebra, $\omf g\subseteq \mf g_{L,R}$ admits an affine subalgebra extension\footnote{Note that this extension is typically easy to find:  the matching of  central terms in the corresponding affine algebras are automatically satisfied via \eqref{eq:keff}.}  $\hat{\omf g}\subseteq\hat{\mf g}_{L,R}$ such that $c_{\omf g}=c_L=c_R$.  
We note if $\hat{\omf g}$ is a proper subalgebra then it must be {\it conformally embedded} into both $\hat{\mf g}_L$ and $\hat{\mf g}_R$.  Conformal embeddings have been well studied and are extremely constrained (see \cite{DiFrancesco:1997nk} for a nice overview).  Here we see that they appear as a natural class of topological boundary conditions.



\begin{figure}
\centering
 \begin{tabular}{ c c c}
    \includegraphics[align=t,width=8.2cm]{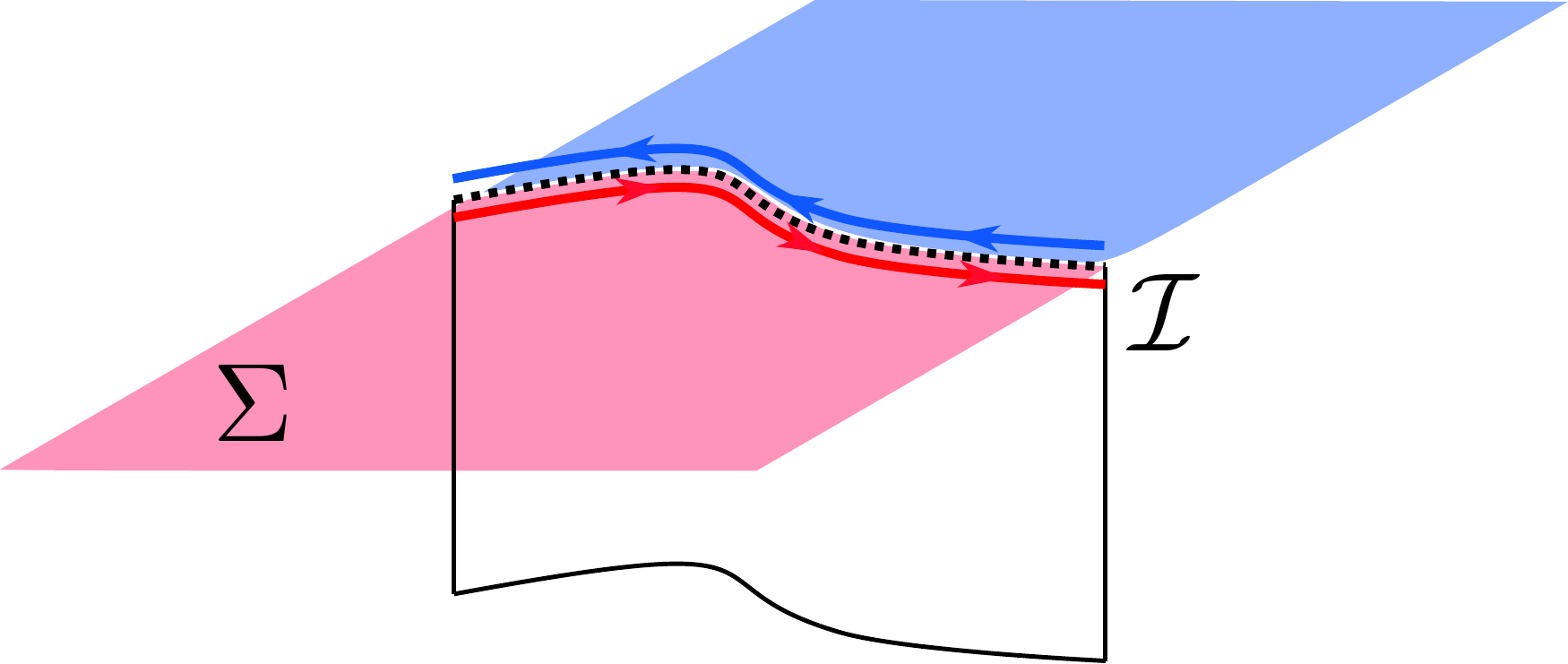} & \hspace{1 cm} & \includegraphics[align=t,width=7cm]{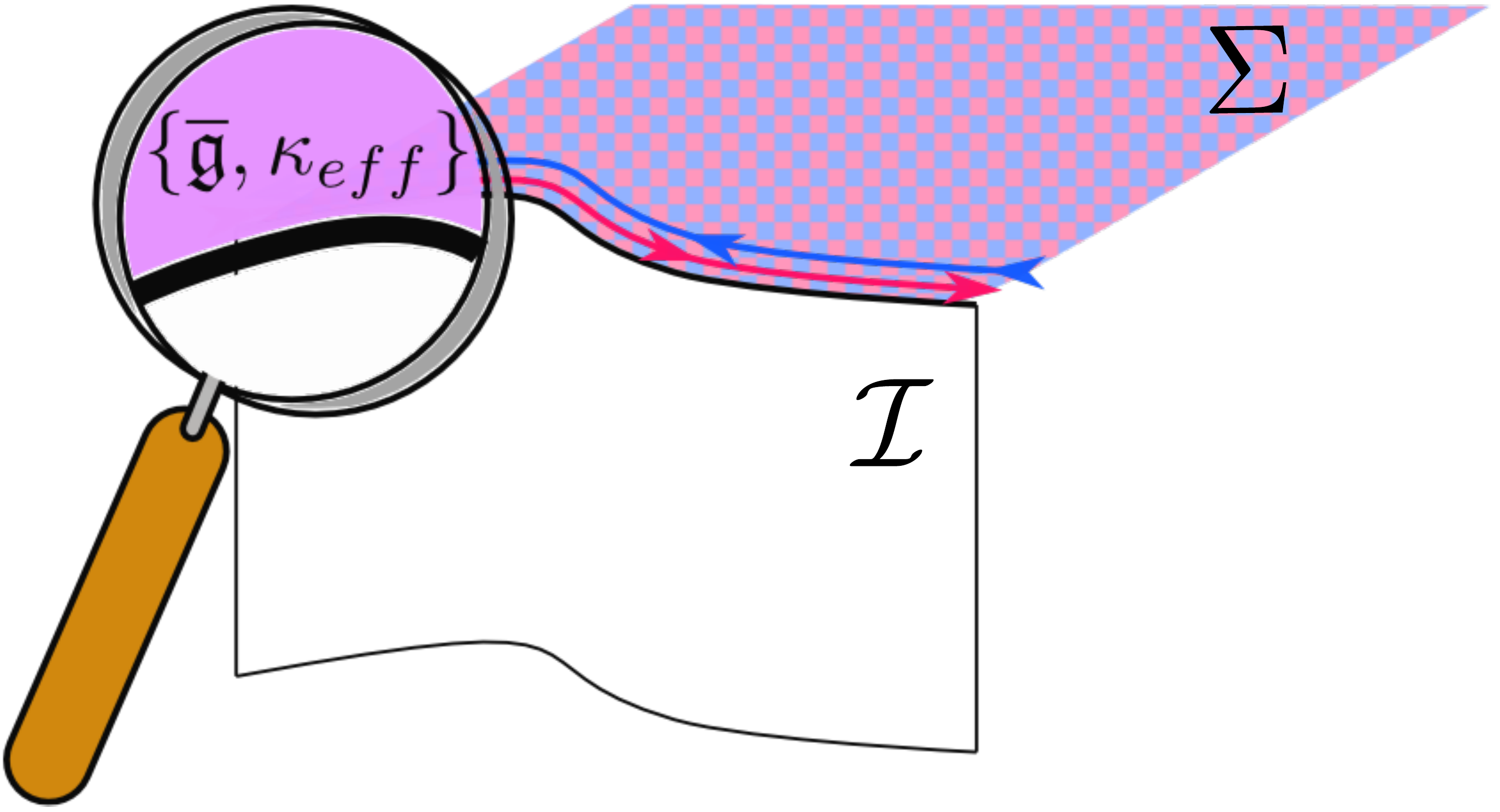}
    \end{tabular}\caption{\small{\textsf{On a Cauchy slice $\Sigma$ the bulk theory can be rewritten as a chiral WZW theory on either side of the interface. In the folded theory, we have a chiral and anti-chiral WZW whose restriction to $\omf g$ close to the interface coalesces into a single non-chiral WZW consistent with the interface (now treated as a boundary) being gapped.}}}\label{fig:cauchyWZW}
\end{figure}

So far the discussion of these boundary conditions has been on the level of consistency of embedded subalgebras.  It is perhaps instructive to illustrate why these correspond to ``gapped interfaces" in the usual sense.  In that vein, consider the path integral on a manifold with boundary $\Sigma$ that intersects transversely with $\mathcal I$, as in Figure \ref{fig:cauchyWZW}.  As is standard, the theory can be rewritten (with holomorphic boundary conditions) as a theory of chiral WZW fields living on $\Sigma$ and in principle we have two separate WZW theories that meet at $\Sigma\cap\mc I$.  In the folded theory, the interface is treated as a hard boundary and the WZW$_L$ and WZW$_R$ theories would 
give rise to chiral and anti-chiral massless modes at $\Sigma\cap\mc I$.  The topological boundary conditions, however, ensure that these modes are gapped out in the following way.  The currents for the two theories, as they approach the interface, are restricted to span the subalgebra $\hat{\omf g}$; the conditions \eqref{eq:keff} and \eqref{eq:structurecons} ensure that this can be done consistently.  In particular, since the chiral central charges of the two theories are equal, 
$c_L=c_R$, the two WZW's can be joined into a single non-chiral WZW via the standard Polyakov-Wiegmann trick \cite{Polyakov:1983tt,Polyakov:1984et} which is unstable to gap formation.  For the Abelian theory this is explicit: the identification of the currents from WZW$_L$ (once appropriately mapped to $\hat{\omf g}$) with those from WZW$_R$ can be implemented at the level of the elementary compact bosons via a massive deformation whose IR limit becomes a hard delta-functional \cite{Fliss:2017wop}.

Now let us discuss the inclusion of anyon punctures.  We recall that the states on surfaces pierced by anyon excitations are related to integrable highest weight representations of the affine Lie algebra $\hat{\mf g}_k$.  We will label generic anyons by their corresponding highest weights $\alpha,\beta,\ldots$, and their corresponding integrable representations by $\mc R_\alpha$; for Abelian anyons specifically we will denote both the anyon and its representation by a lowercase Latin letter (e.g., $q$).  In both the Abelian and non-Abelian theories we will always denote the identity anyon by ``$0$". When we have interfaces between distinct phases, we can have anyon punctures in each phase, and thus we can associate an integrable representation to each side of an interface. After imposing topological boundary conditions, we must restrict fields at the interface to lie in the isotropic subalgebra $\obmf g$, with its own irreducible highest weight representations (irreps).  The restriction of a representation, $\mc R_\alpha$, of $\hat{\mf g}_k$ to a subalgebra then will be a linear combination of the irreps of $\hat{\obmf g}$, schematically:
\begin{equation}\label{eq:represtrict}
\left.\mc R^{(\hat{\mf g})}_\alpha\right|_{\hat{\obmf g}}=\bigoplus_{\bar \alpha}m_\alpha^{\bar \alpha}\mc R^{(\hat{\obmf g})}_{\bar \alpha}
\end{equation}
If the identity representation of $\hat{\obmf g}$ appears on the right-hand side of this decomposition, then we say that the anyon, $\alpha,$ condenses at the boundary.  The multiplicity with which the identity representation appears in \eqref{eq:represtrict} tells us the number of channels in which the anyon can condense:
\begin{equation}
\mathcal W_{\alpha}\equiv m_\alpha^{0}.
\end{equation}
There is an intuitive picture for this in the folded setup of an isotropic interface between two topological phases.  Since $\hat{\omf g}$ is a subalgebra of both $\hat{\mf g}_L$ and $\hat{\mf g}_R$, each have their own decomposition upon restriction to $\hat{\omf g}$.  Such decompositions are called \emph{conformal branchings}\footnote{We note that because representations of affine algebras are infinite dimensional, for a restriction to a generic subalgebra there is no guarantee that the branching coefficients are finite.  However, happily for conformal embeddings (and only conformal embeddings) this happens to be the case.  This is known as the {\it finite reducibility theorem}\cite{Bouwknegt:1987mr,Goddard:1984hg,Goddard:1985jp}.} in the literature \cite{DiFrancesco:1997nk,Bais:2008ni,Bais:2008xf,Lou:2019heg}.  The identity irrep of $\hat{\obmf g}=\left(\hat{\omf g}\oplus\hat{\omf g}\right)_{diag}$ can then appear at the interface only if $\hat{\mf g}_L$ and $\hat{\mf g}_R$ branch into a representation and its conjugate (respectively):
\begin{equation}\label{eq:TBCtunnelmat}
{\mc W_{\alpha_L}}^{\alpha_R}=\sum_{\bar \beta}m_{\alpha_L}^{\bar \beta}m_{\alpha^\ast_R}^{\bar \beta^\ast}
\end{equation}
These {\it tunneling matrices} intertwine the modular data of the $L$ and the $R$ topological phases and so provide a realization of the tunneling matrices established in \cite{Lan:2014uaa}:
\begin{align}\label{eq:modintertwine}
\sum_{\beta_L}{\mc S^{(L)}_{\alpha_L}}^{\beta_L}{\mc W_{\beta_L}}^{\alpha_R}=&\sum_{\beta_R}{\mc W_{\alpha_L}}^{\beta_R}{\mc S^{(R)}_{\beta_R}}^{\alpha_R}\nonumber\\
\sum_{\beta_L}{\mc T^{(L)}_{\alpha_L}}^{\beta_L}{\mc W_{\beta_L}}^{\alpha_R}=&\sum_{\beta_R}{\mc W_{\alpha_L}}^{\beta_R}{\mc T^{(R)}_{\beta_R}}^{\alpha_R}
\end{align}
where $\mc S^{(L,R)}$ and $\mc T^{(L,R)}$ are the modular $S$ and $T$ matrices of the left and right theories.  This is easy to show from \eqref{eq:TBCtunnelmat} and we do so briefly.  For a {\it generic} embedding $\hat{\omf g}\hookrightarrow \hat{\mf g}_L$, the restriction of the affine character of an integrable highest weight representation of $\hat{\mf g}_L$ admits the following decomposition \cite{DiFrancesco:1997nk}:
\beq
\chi^{(\hat{\mf g}_L)}_{\alpha_L}(\tau)=\sum_{\gamma}\chi_{\{\alpha_L;\gamma\}}(\tau)\chi^{(\hat{\omf g})}_\gamma(\tau)
\eeq
where the sum is over integrable highest weights $\gamma$ of $\hat{\omf g}$.  The functions $\chi_{\{\alpha_L;\gamma\}}(\tau)$ behave nicely under modular transformations
\beq
\chi_{\{\alpha_L,\gamma\}}\left(-1/\tau\right)=\sum_{\beta_L}\sum_{\delta}{\mc S^{(\hat{\mf g}_L)}_{\alpha_L}}^{\beta_L}\chi_{\{\beta_L,\delta\}}(\tau){\mc S^{(\hat{\omf g})\dagger}_\delta}^\gamma\qquad\qquad \chi_{\{\alpha_L,\gamma\}}(\tau+1)=\sum_{\beta_L}\sum_{\delta}{\mc T^{(\hat{\mf g}_L)}_{\alpha_L}}^{\beta_L}\chi_{\{\beta_L;\delta\}}(\tau){\mc T^{(\hat{\omf g})\dagger}_{\delta}}^{\gamma}
\eeq
For a {\it conformal embedding} however, these functions are {\it constant}\footnote{In fact, for a generic embedding, $\chi_{\{\alpha_L,\gamma\}}$ are characters of the coset theory $\hat{\mf g}_L/\hat{\omf g}$.  The fact that these are constant for conformal embeddings indicates that this would-be coset is trivial: this is consistent with the statement that the TBCs are gapping out the degrees of freedom at the interface.} and equal to our branching coefficients \cite{Bouwknegt:1987mr}:
\beq
\chi_{\{\alpha_L;\gamma\}}(\tau)=m_{\alpha_L}^\gamma.
\eeq
This implies
\beq
m_{\alpha_L}^\gamma={\left(\mc S^{(\hat{\mf g}_L)}\cdot m\cdot \mc S^{(\hat{\omf g})\dagger}\right)_{\alpha_L}}^{\gamma}\qquad\qquad m_{\alpha_L}^\gamma={\left(\mc T^{(\hat{\mf g}_L)}\cdot m\cdot \mc T^{(\hat{\omf g})\dagger}\right)_{\alpha_L}}^{\gamma}
\eeq
Similar statements apply for the $R$ phase branching coefficients.  Equation \eqref{eq:modintertwine} then follows directly from the definition, \eqref{eq:TBCtunnelmat}, and the unitarity of $\mc S^{(\hat{\omf g})}$ and $\mc T^{(\hat{\omf g})}$.

The coefficients ${\mc W_{\alpha_L}}^{\alpha_R}$ provide a map from the fusion spaces across the interface; Ref. \cite{Lan:2014uaa} introduced them as the dimension of the Hilbert space on the two-sphere with anyon punctures $\alpha_L$ and $\alpha_R$ on either side of an equatorial interface, as pictured in Figure \ref{fig:LR3sphereanyons}.  While this is {\it a priori} a different definition than equation \eqref{eq:TBCtunnelmat}, it is easy to convince oneself that physically these two concepts are the same.  We will see that this is true in Section \ref{sect:S2singleint} by explicitly constructing the Hilbert space $\mc H_{S^2_{\alpha_L,\alpha_R}}$.  
\myfig{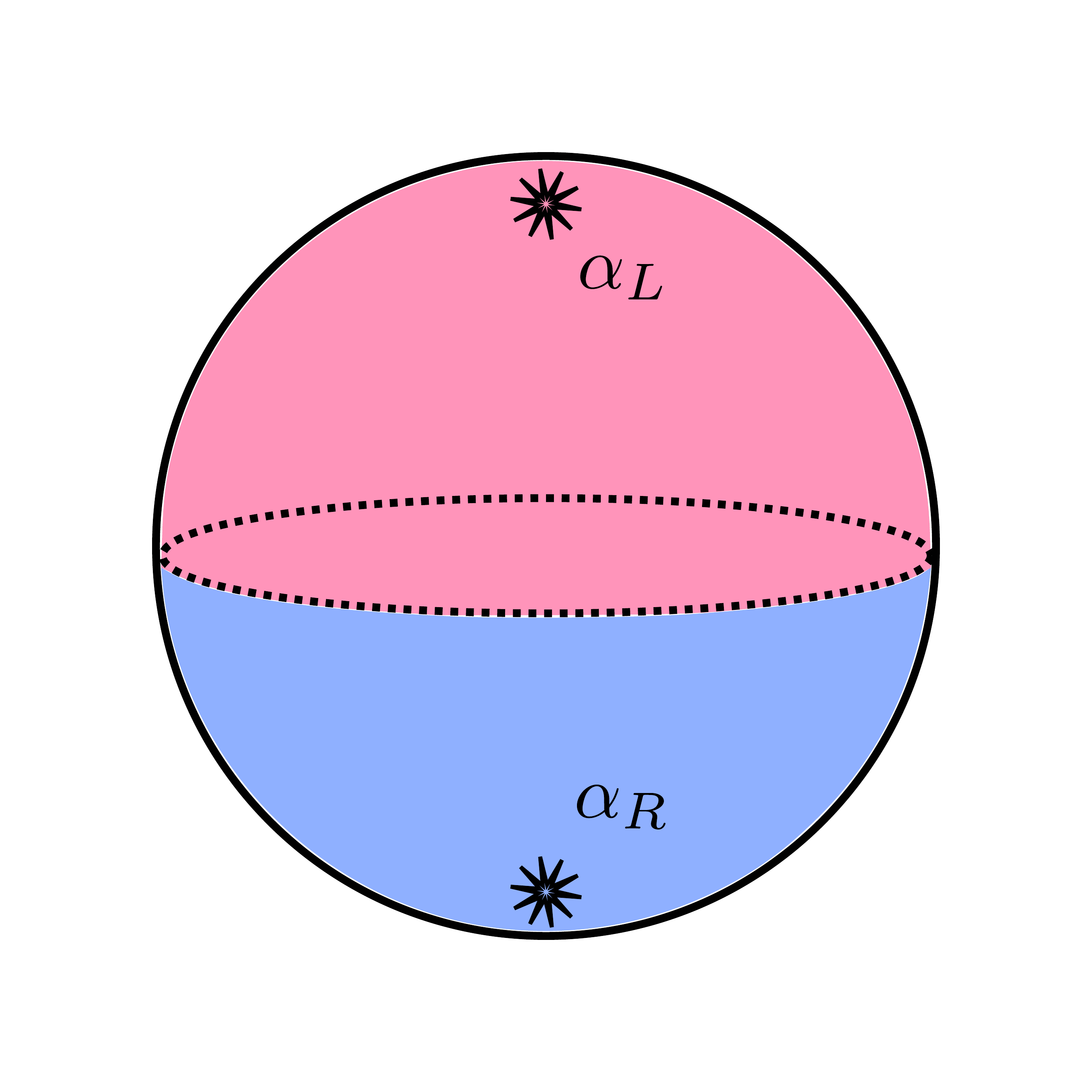}{3.5}{The dimension of the Hilbert space for the above configuration gives the tunneling matrix coefficients: ${\mathcal W_{\alpha_L}}^{\alpha_R}=\dim\mc H_{S^2_{\alpha_L,\alpha_R}}$.}

\subsection{Examples}\label{sect:examples}

Although it is not our intention to classify all possible isotropic interfaces, in this section we give a number of examples to orient the reader. For the first set of examples, let us see what interfaces we can construct between the simplest non-Abelian Chern-Simons theories.

{\bf Example 1: $\hat{\mf g}_L=\mf{su}(2)_{k_L}$, $\hat{\mf g}_R=\mf{su}(2)_{k_R}$}\\
The matching of the chiral central charges tells us immediately that the two levels must match, $k_L=k_R\equiv k$.  Looking at conformally embedded subalgebras of $\mf{su}(2)_k$ we find two possibilities: $\mf{u}(1)_2$ (when $k=1$) and $\mf{su}(2)_k$ itself.

The former is the (trivial) identification of the Cartan algebras of each phase.  This admits a nice physical picture: the vertex operator algebra of the $\mf{u}(1)_2$ theory is in fact extended to $\mf{su}(2)_1$ and so this gapped interface is really the extension of a gapped interface formed from identifying the currents of the Abelian theory (along the lines of \cite{Fliss:2017wop}) at its critical radius.  We will revisit this more generically in example 3.

For the latter, we can construct (at least) two independent interfaces.  Firstly, there is the trivial interface.  The anyons that can permeate the interface are of the form $\alpha_L=\alpha_R$ (in the folded picture the set of condensed anyons are of the form $\{\alpha,\alpha^\ast\}=\{\alpha,\alpha\}$).  

Secondly, at the level of the algebra $\mf{su}(2)$, we might consider the subalgebra spanned by
\begin{equation}\label{eq:finitesu2ex}
\text{span}\left\{J^L_3\oplus(-J^R_3),J^L_{\pm}\oplus J^R_{\mp}\right\}.
\end{equation}
One immediate extension of this subalgebra to the affine $\mf{su}(2)_k$ is the direct matching of the modes:
\beq
\hat{\obmf g}=\text{span}\left\{J^L_{3,m}\oplus\left(-J^R_{3,m}\right),J^L_{\pm,m}\oplus J^R_{\mp, m}\right\}
\eeq
Note, however, that the $L$ and $R$ highest weight representations are related by Weyl reflection and so correspond to the same integrable representation.  Thus we see that the set of permeable anyons are again of the form $\{\alpha,\alpha\}$ and so from the point of view of anyon condensation, this interface is indistinguishable from the trivial interface.  In fact this subalgebra is related to the trivial subalgebra by a global $SU(2)$ conjugation.

A non-trivial affine extension of \eqref{eq:finitesu2ex} can be constructed as 
\beq
\hat{\obmf g}=\text{span}\left\{J^L_{3,m}\oplus\left(\frac{k}{2}\delta_{m,0}-J^R_{3,m}\right),J^L_{\pm, m}\oplus J^R_{\mp,m\mp 1}\right\}
\eeq
Here $\hat{\mf g}_R$ weights are related to $\hat{\mf g}_L$ weights via an outer automorphism \cite{DiFrancesco:1997nk}.  When folded, this interface is equivalent to the anyon condensation characterized by the $\mathbb Z_2$ (or permutation) modular invariant of $\mf{su}(2)_k$\cite{Hung:2015hfa}.  This is a peculiar example in which the mode expansions of the currents are mixed.  {\it This can only happen in the non-Abelian Chern-Simons theory}: the topological boundary conditions relating the Cartan subalgebras are generically described by the Abelian story in \cite{Fliss:2017wop}.\footnote{There is a supposition of {\it locality} built into this.  If an Abelian gauge transformation of the $L$ phase having support on a circular interface is related to that of the $R$ phase via $\lambda_{L,a}(\theta)\sim {f_{a}}^b(\theta)\lambda_{R,b}(2\pi-\theta)$ then the only solution consistent with unitarity and the commutation relations of the Cartan sublagebras is ${f_{a}}^b=\text{constant}$ and from there the story follows \cite{Fliss:2017wop}.\label{footnote:abelianlocality}}
%
%

Although we will not go into detail here, the use of outer automorphisms can be extended to larger algebras with $\mf g_L\simeq \mf g_R$. 
Instead, in the following examples we focus on interfaces formed from the conformal embedding $\omf g\simeq \mf g_L\subset \mf g_R$.  Generically, it is a necessary condition that the level of $\mf g_R$ is one \cite{DiFrancesco:1997nk}.  

{\bf Example 2: $\hat{\mf g}_L=\mf{su}(2)_{k_L}$, $\hat{\mf g}_R=\mf{su}(N)_{1}$}\\
In this example we see that the possible gapped domain walls are sparse: matching of the central charge\footnote{Recall that $c=k(N^2-1)/(k+N)$ for $\mf{su}(N)_k$. Thus here we have $c_L=3k_L/(k_L+2)$ and $c_R=N-1$.} with $k\in\mathbb Z_+$ only allows $N=3$ and thus $k_L=4$.  $\mf{su}(3)$ has an $\mf{su}(2)$ subalgebra associated to each simple root.  It is a simple exercise to see that we cannot match the central terms of the corresponding affine algebras when $k_L=4$.  However there is another embedding of $\mf{su}(2)$ given by (up to a global $SU(3)$ conjugation) by
\beq
\hat{\obmf g}=\text{span}\left\{J^3_{m}\oplus H^{(\alpha_1+\alpha_2)}_m, J^+_m\oplus2(E^{\alpha_1}_m+E^{\alpha_2}_m), J^-_m\oplus(E^{-\alpha_1}_m+E^{-\alpha_2}_m)\right\}.
\eeq
where for a given root $\alpha$, $H^\alpha=\sum_{i=1}^r\alpha_iH^i$.  This is a proper conformal embedding\footnote{This conformally embedded $\mf{su}(2)$ differs from those associated to simple roots by its \emph{embedding index} \cite{DiFrancesco:1997nk}.  
For the purposes of our paper, we will always implicitly set the norm of all highest roots to 2; this index is then equivalently encoded in $v_{L,R}$.} and thus we can define a corresponding isotropic interface after imposing boundary conditions.

The branchings of the integral representations of $\mf{su}(3)_1$ upon restriction to $\mf{su}(2)_4$ are given by the following:
\begin{equation}
(\lambda_1,\lambda_2)=(0,0)\rightarrow (2j=0)\oplus (2j=4)\qquad\qquad (0,1)\rightarrow 2\qquad\qquad (1,0)\rightarrow 2
\end{equation}
where $(\lambda_1,\lambda_2)$ are the Dynkin labels for the simple roots $\alpha_{1,2}$ of $\mf{su}(3)$.  Thus we have the following permeable anyons:
\begin{equation}
\{(0,0); 0\}\qquad\{(0,0); 4\}\qquad \{(1,0);2\}\qquad \{(0,1);2\}
\end{equation}
This is equivalently stated via the tunneling matrix
\begin{equation}
{\mc W_{\alpha_{L}}}^{\alpha_R}=\left(\begin{array}{ccc}1&0&0\\0&0&0\\0&1&1\\0&0&0\\1&0&0\end{array}\right)
\end{equation}
where the ordering of the columns is $(0,0),(1,0),(0,1)$.  It is easy to verify that 
\beq
{\mc W}\cdot {\mc S^{\mf{su}(3)_1}}={\mc S}^{\mf{su}(2)_4}\cdot \mc W\qquad\qquad {\mc W}\cdot{\mc T}^{\mf{su}(3)_1}={\mc T}^{\mf{su}(2)_4}\cdot\mc W
\eeq
where
\beq
{\mc S^{\mf{su}(3)_1}}=\frac{1}{\sqrt 3}\left(\begin{array}{ccc}1&1&1\\1&q&q^2\\1&q^2&q\end{array}\right)\qquad q=e^{2\pi i/3};\qquad\qquad {{\mc S^{\mf{su}(2)_4}}_{j_1}}^{j_2}=\frac{1}{\sqrt{3}}\sin\left(\frac{\pi(2j_1+1)(2j_2+1)}{6}\right)
\eeq
and
\beq
{\mc T}^{\mf{su}(3)_1}=e^{\frac{-2\pi i c_R}{24}}\text{diag}\left(1,e^{\frac{2\pi i}{3}},e^{\frac{2\pi i}{3}}\right)\qquad {\mc T}^{\mf{su}(2)_4}=e^{-\frac{2\pi i c_L}{24}}\text{diag}\left(1,e^{\frac{\pi i}{4}},e^{\frac{2\pi i}{3}},e^{\frac{5\pi i}{4}},1\right)\qquad c_L=c_R=2.
\eeq
{\bf Example 3: $\hat{\mf g}_L=\mf{u}(1)^N_{K_L}$, $\hat{\mf g}_R=\mf{su}(N+1)_{1}$}\\
In this last example we consider an interface between the Abelian theory with $K$-matrix
\beq
K_L=\left(\begin{array}{cccc}2&-1&0&\ldots\\-1&2&-1&\ldots\\ 0&-1&2&\ldots\\\vdots&\ddots&\ddots&\ddots\end{array}\right)
\eeq
and the non-Abelian $\mf{su}(N+1)_1$.  That these two theories admit a gapped interface is perhaps not surprising: for $N=2$, the $L$ phase describes\footnote{up to $GL(2,\mathbb Z)$ conjugation} the Halperin (221) state which has a hidden extended $SU(3)_1$ symmetry \cite{ardonne1999new}; analogous $\mf u(1)^N$ states with extended $SU(N+1)_1$ symmetry have since been constructed using coupled wires \cite{Fuji:2017elm}.  Since the $K$-matrix of the $\mf{u}(1)^N$ is precisely the Cartan matrix of the $\mf{su}(N+1)$ algebra, we find an isotropic algebra with respect to $K_L\oplus(-\kappa_R)$ by simply identifying the $\mf{u}(1)^N$ generators, $\{\phi^i\}$, with the Cartan subalgebra of $\mf{su}(N+1)$ in the Chevalley basis:
\beq
\hat{\obmf g}=\text{span}\left\{\phi^i_m\oplus H^{\alpha_i}_m\right\}_{i=1,\ldots, N}
\eeq
The $N+1$ anyons of the $\mf{u}(1)^N_{K_L}$ theory map  straightforwardly into the $N+1$ anyons of the $\mf{su}(N+1)_1$ by identifying the Abelian charge vector with the corresponding  Dynkin label of an $\mf{su}(N+1)_1$ integrable representation; that is, the tunneling matrix, $\mc W$, is the identity.

Given the general setup for Abelian theories \cite{Fliss:2017wop} we can look for interesting interfaces between these two theories by searching for integer matrices $\{v_L\}$ and $\{v_R\}$ obeying
\beq
(v_L)^t\cdot K_L\cdot v_L=(v_R)^t\cdot \kappa_R\cdot v_R.
\eeq
One such example in the $N=2$ case is given by 
\beq
v_L=\left(\begin{array}{cc}1&3\\3&2\end{array}\right)\qquad\qquad v_R=\left(\begin{array}{cc}2&3\\3&1\end{array}\right)
\eeq
and the effective theory at the interface is an Abelian theory with 
\beq
\kappa_{eff}=7\left(\begin{array}{cc}2&1\\1&2\end{array}\right)
\eeq
As we will see in the following section, this topological boundary condition is distinguished from the above by the topological entanglement entropy across the interface.

\section{Quantum gluing operators}\label{sect:QG}
The above discussion has been focused on classical boundary conditions.  As we move to discussions of entanglement, we will implement these boundary conditions with the action of quantum operators.  Before doing so, let us briefly remark about the state of affairs for entanglement in gauge theories.

At a formal level, a basic ingredient in discussing the entanglement of a subsystem $A$ is the factorization of the Hilbert space:
\begin{equation}\label{eq:Hilbertfactorize}
\mc H=\mc H_A\otimes \mc H_{A^c}.
\end{equation}
We are interested in the case in which  $A$ is a spatial subregion on the Cauchy surface on which we are defining the state, $\rho$.  The subregion $A$ is then separated from $A^c$ by an auxiliary co-dimension 2 surface that we refer to as the \emph{entangling surface}.  In a typical local quantum field theory this factorization is ill-defined in the following sense.  Due to short range correlations at the entangling surface, the division of $\mc H$ into $\mc H_A$ and $\mc H_{A^c}$ is extremely sensitive to the UV cutoff of the theory.  However, we regard this obstruction to be minimal and there are suitable prescriptions for overcoming it.  For instance, we can put the theory on a lattice, with spacing $\varepsilon$ or thicken the entangling surface to width $\varepsilon$ and impose boundary conditions.  With such a regularization in hand, we can take \eqref{eq:Hilbertfactorize} literally and compute the reduced density matrix and its subsequent entanglement entropy:
\begin{equation}
S_{ent}=-\mTr_{\mc H_A}\left(\rho_A\log\rho_A\right)\qquad\qquad \rho_A=\mTr_{\mc H_{A^c}}\rho.
\end{equation}
The above mentioned sensitivity to the UV cutoff is signaled by divergences in $S_{ent}$.  For instance, in the vacuum state
\begin{equation}
S_{ent}\sim \frac{L^{d-2}}{\varepsilon^{d-2}}+\ldots
\end{equation}
In quantum gauge theories however, the situation is even more complicated.  Even after regulating the local pile-up of modes at the entangling surface, the Hilbert space of gauge invariant states refuses a local factorization\footnote{
While this is the generic story for local spatial subsystems, there are other possible non-local partitions of the Hilbert space that appear to be perfectly well-defined even in gauge theories (for example the multi-boundary setups in \cite{Balasubramanian:2016sro,Balasubramanian:2018por}.)} of the form \eqref{eq:Hilbertfactorize}.  One manifestation of this is that the constraints of gauge invariance are non-locally realized and prevent a gauge invariant state existing on a sole tensor factor \cite{Buividovich:2008gq,Donnelly:2011hn}.  A separate but equivalent manifestation of this fact is that any association of an algebra of gauge invariant operators to a subregion $A$ (call it $\mc A_A$) possesses a common center with its complement algebra $\mc A_{A^c}$ \cite{Casini:2013rba,Casini:2014aia}; a simple result for von Neumann algebras then necessarily precludes a factorization \eqref{eq:Hilbertfactorize}.

This complication is present in Chern-Simons theory in a very stark way.  Let us first ignore the introduction of interfaces and consider the Hilbert space of gauge invariant states on a Riemann surface, $\Sigma_g$, of genus $g$.  For compact groups, Chern-Simons theory is special in that the dimension of this space is finite dimensional \cite{Verlinde:1988sn}:
\begin{equation}
\dim\mc H_{\Sigma_g}=\sum_{\alpha}\frac{1}{\left|{\mc S_0}^\alpha\right|^{2g-2}}
\end{equation}
Now we bisect $\Sigma_g$ into two subregions with the entangling surface consisting of several disconnected circles: $\Sigma_g=A\#_{\{S_i^1\}}A^c$ and ask if
\begin{equation}\label{eq:Hilbertfactorize?}
\mc H_{\Sigma_g}\stackrel{?}{=}\mc H_A\otimes \mc H_{A^c}.
\end{equation}
The answer is defiantly {\bf no}.  In fact both $\mc H_A$ and $\mc H_{A^c}$ have to be \emph{infinite dimensional}.  Indeed, considering the generator of a gauge transformation $A\rightarrow A+d\lambda +[A,\lambda]$ on a Cauchy surface with some set of circular boundaries we have (after imposing Gauss' law)
\begin{equation}\label{eq:circlegaugegen}
\hat Q[\lambda]=\sum_i\frac{\kappa^{ab}}{2\pi}\int_{S^1_i}\lambda^{(i)}_{a}a_{b}
\end{equation}
where $\lambda^{(i)}$ is the pullback of $\lambda$ to the $i^{th}$ circle.  The presence of these boundary terms (which are responsible for the central extension of the current algebra) changes $\hat Q[\lambda]$ from a first class to a second class constraint.  Simply put, $\hat Q$ now acts as a global symmetry at the circular boundaries.  As such, states carry a representation of this symmetry, which is a collection of integrable representations of extended K\v ac-Moody algebras.  Thus we have, schematically\footnote{The factor $V^{(A)}_{\alpha_1,\alpha_2,\ldots, \alpha_n}$ is the fusion space for the handle-body with punctures that $A$ forms when its boundaries are shrunk down to anyon punctures.  It is a finite non-zero factor that counts the conformal blocks on this space.}, 
\begin{equation}\label{eq:HSsurfacewithholes}
\mc H_A=V^{(A)}_{\alpha_1,\alpha_2,\ldots}\bigotimes_{i}\mc H_{S^1_i}[\alpha_i]\qquad\qquad \mc H_{S^1_i}[\alpha_i]=\text{span}\left\{|\alpha_i\rangle, J^a_{-n}|\alpha_i\rangle, J^a_{-n}J^b_{-m}|\alpha_i\rangle,\ldots\right\}
\end{equation}
which is infinite dimensional.

A natural resolution to this problem and the approach that we will adopt in this paper is called the \emph{extended Hilbert space approach}\footnote{This is complemented by an alternative approach that might be called the \emph{algebraic approach} (see for instance \cite{Casini:2013rba,Casini:2014aia}).  The algebraic entropy seems to be extremely constrained in Chern-Simons theory.  See Section \ref{sect:Discussion} for comments on this.}\cite{Buividovich:2008gq,Donnelly:2011hn,Donnelly:2014fua,Donnelly:2014gva,Donnelly:2015hxa,Ghosh:2015iwa,Soni:2015yga}.  Although \eqref{eq:Hilbertfactorize?} cannot exist as an equality, we can embed $\mc H_\Sigma$ as a subspace in $\mc H_A\otimes\mc H_{A^c}$.  $\mc H_A\otimes \mc H_{A^c}$, as an extended Hilbert space, contains states that are not gauge invariant. However, it is a simple matter to find the subspace of gauge invariant states by looking at the kernel of the operators \eqref{eq:circlegaugegen}:
\begin{equation}
|\psi\rangle_{(\in\mc H)}\hookrightarrow|\tilde\psi\rangle\in\mc H_A\otimes\mc H_{A^c}\qquad\qquad \left(\hat Q[\lambda_A]\otimes \hat{\mathbbm 1}_{A^c}+\hat{\mathbbm 1}_A\otimes\hat Q[\lambda_{A^c}]\right)|\tilde\psi\rangle=0.
\end{equation}
Of course we have to say how we identify the gauge parameters across the entangling surface.  For each circular component parameterized by $\theta$, we can express $\lambda_A=\sum_{m\in\mathbb Z}(\lambda_{A}^{(i)})_me^{im\theta}$ (and similarly $\lambda_{A^c}^{(i)}$).  The natural identification, once accounting for a flip in orientation, is
\begin{equation}
(\lambda^{(i)}_A)_{m}=(\lambda^{(i)}_{A^c})_{-m}
\end{equation} 
(we are suppressing Lie algebra indices).  Denoting $\frac{\kappa^{ab}}{2\pi}\int_{S^1_i}e^{-im\theta}a_{b}\equiv J^{(i),a}_m$, this translates to the preservation of the chiral symmetry algebra across that component of the entangling surface:
\beq
\left(\hat J^{(i),a}_{A,m}\otimes\hat{\mathbbm 1}^{(i)}_{A^c}+\hat{\mathbbm 1}^{(i)}_A\otimes\hat J^{(i),a}_{A^c,-m}\right)|\tilde\psi\rangle=0.
\eeq
The solution to this equation is the non-Abelian generalization of the Ishibashi state, and the full state is the tensor product of such Ishibashi states for each circular component of the entangling surface:
\beq
|\tilde\psi\rangle=\bigotimes_i|\alpha_i\rrangle\qquad\qquad |\alpha_i\rrangle=\sum_{M}|\alpha_i, M\rangle\otimes\overline{|\alpha_i, M\rangle}
\eeq
where $M$ labels an orthonormal basis of descendants of the conformal module with primary $\alpha_i$.  In this embedding, the notion of tracing out $A^c$ is now clear: we simply trace $|\alpha_i\rrangle\llangle\alpha_i|$ over basis states $\overline{|\alpha_i, M\rangle}$. In \cite{Fliss:2017wop}, this result for Abelian Chern-Simons provided a universal explanation for the equivalence between bulk entanglement spectrum and the spectrum of chiral edge modes and for the efficacy of the calculation of spatial entanglement using left-right entanglement of Ishibashi states \cite{Wen:2016snr}.  Happily, we find that this explanation persists into the non-Abelian theories.

\subsection{Including interfaces}\label{sect:QGint}
Now let us consider the same line of inquiry when states contain interfaces of the type above, taking the entangling surface along some set of interfaces.  As one might guess the above paradox appears in this context as well: the Hilbert space of gauge invariant states of a Riemann surface with interfaces is finite dimensional, while the ``would-be" tensor factors are infinite dimensional and correspond to integrable representations on which operators of the form
\begin{equation}\label{eq:LRglobaltxops}
\hat Q_L[\lambda_L]\otimes \hat{\mathbbm 1}+\hat{\mathbbm 1}\otimes\hat Q_R[\lambda_R]
\end{equation}
 act as a global symmetry (for each circular component of the entangling surface).  
Although this is morally true, we do not have a generic construction of the Hilbert spaces in question.  What we show in Appendix \ref{app:B} is that the extended Hilbert space \emph{provides a construction} of the Hilbert space of gauge invariant states when interfaces are involved, and this Hilbert space is finite dimensional (although both of its factors are infinite dimensional).  The identification of the embedded Hilbert space follows from our discussion of TBCs.  Focusing on a single component of the entangling surface coinciding with an interface $\mc I_i$, the unbroken gauge group is generated by an isotropic subalgebra $\hat{\obmf g}_i\subseteq\hat{\mf g}_L\oplus\hat{\mf g}_R$ and as such the gauge parameters should be identified\footnote{We allow the possibility that $v_{L,R}$ can mix mode expansions as we saw in {\bf example 1} of Section \ref{sect:examples}.} 
 \begin{equation}
 \lambda^{(i)}_{L,n}=\left(v_L\cdot\bar\lambda^{(i)}\right)_{n}\qquad\qquad \lambda^{(i)}_{R,n}=\left(v_R\cdot \bar\lambda^{(i)}\right)_{-n}, \qquad\qquad \bar\lambda^{(i)}\in\hat{\omf g}.
 \end{equation}
 Under this identification we demand
 \begin{equation}
 \frac{1}{4\pi}\left(\left(v_L^t\cdot\kappa_L\cdot\hat{\mc J}^L\right)_m\otimes \hat{\mathbbm 1}_R+\hat{\mathbbm 1}_L\otimes \left(v_R^t\cdot\kappa_R\cdot\hat{\mc J}^R\right)_{-m}\right)|\tilde\psi\rangle=0
 \end{equation}
 for each component of the circular interface.  We are free to redefine $\left({v_L}^t\cdot\kappa_L\cdot\hat{\mc J}^{L}\right)_m:=\kappa_{eff}\cdot\hat{\bmc J}_m$ and $\left({v_R}^t\cdot\kappa_R\cdot\hat{\mc J}^{R}\right)_m:=\kappa_{eff}\cdot\hat{\tbmc J}_m$, where from the arguments of Section \ref{sect:BCs}, the currents $\hat{\bmc J}_m$ and $\hat{\tbmc J}_m$ each separately satisfy a K\v ac-Moody current algebra with level-Killing form $\kappa_{eff}$.  The resulting condition of gauge-invariance is then
 \begin{equation}\label{eq:effgapcond}
 \frac{\kappa_{eff}^{\bar a\bar b}}{4\pi}\left(\hat{\bmc J}_{\bar b,m}\otimes\hat{\mathbbm 1}_R+\hat{\mathbbm 1}_L\otimes\hat{\tbmc J}_{\bar b,-m}\right)|\tilde\psi\rangle=0.
 \end{equation}
 \myfig{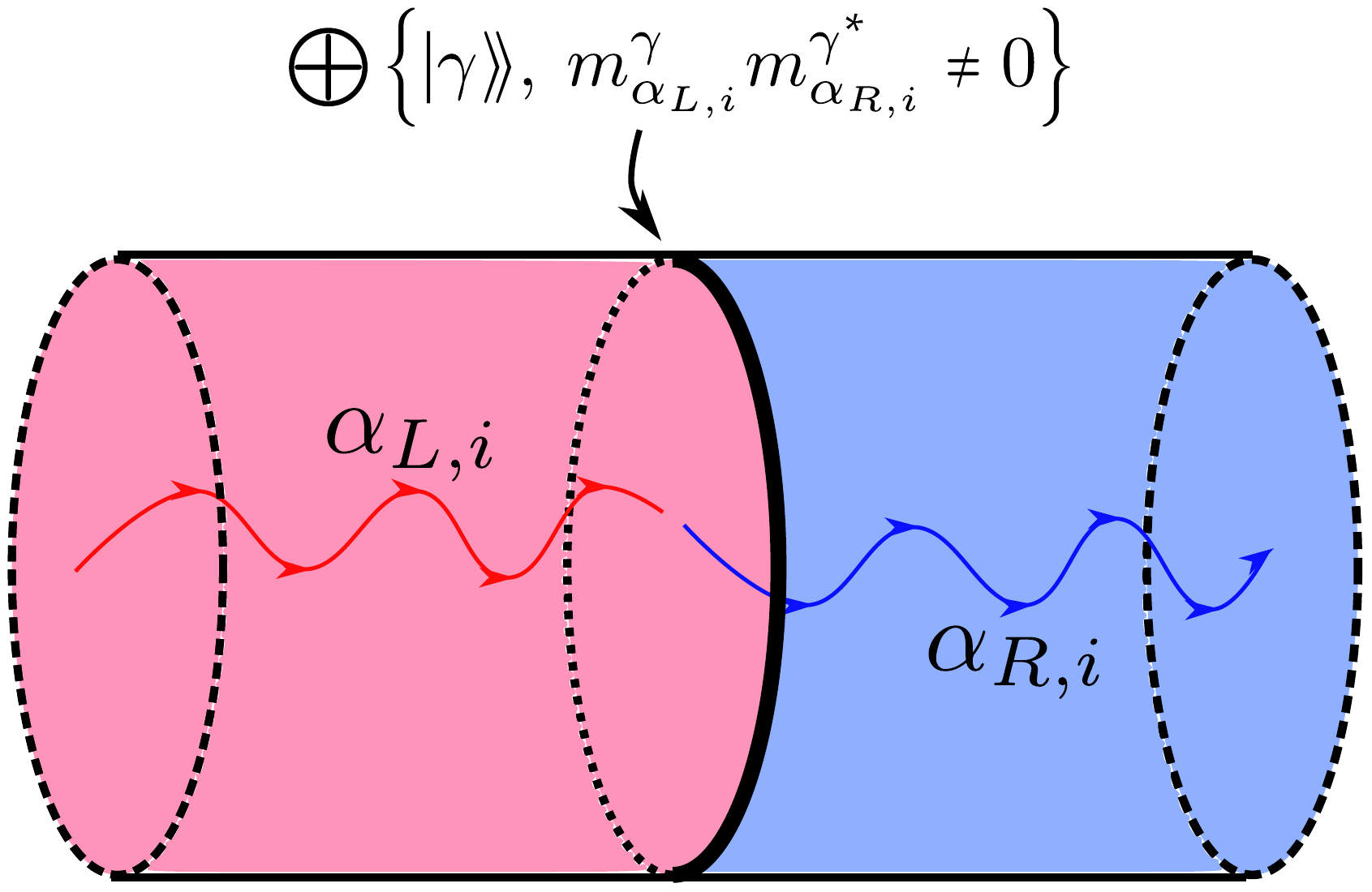}{7}{Anyon $\alpha_{L,i}$ branches into $\oplus_\gamma m_{\alpha_L}^\gamma\,\gamma$ when approaching the interface $\mc I_i$; a similar branching occurs for $\alpha_{R,i}$.  If they share a channel $m_{\alpha_{L,i}}^\gamma\times m_{\alpha_{R,i}^\ast}^{\gamma^\ast}\neq0$ for some $\gamma$ then this mediates the transmutation of $\alpha_{L,i}$ to $\alpha_{R,i}$ across the interface.  The effective set of Ishibashi states for this interface are spanned by these branching channels.}
It is clear that the solution to \eqref{eq:effgapcond} should be an Ishibashi state determined by the algebra $\hat{\omf g}_i$ at level $\kappa^{(i)}_{eff}$ but we also need to specify \emph{which} Ishibashi state it is, i.e., we need to specify a conformal primary.  In fact, unlike the above situation where the Chern-Simons theory is homogeneous (and all possible cuts are trivial interfaces), describing the Wilson line configuration on the interior of $\Sigma_g$ is not enough to specify a unique state.  This is a consequence of branching when the Wilson lines cross an interface.  Indeed, let us suppose that approaching our interface of interest, $\mc I_i$, from the left side is a Wilson line carrying a representation $\mc R_{\alpha_{L,i}}$, while on the right side, the Wilson line carrying representation $\mc R_{\alpha_{R,i}}$ emerges from the interface (see Figure \ref{fig:LRcircinterface_anyon}).  As we argued above in Section \ref{sect:BCs}, in order for this configuration to make sense a common representation must appear in the branchings of $\mc R_{\alpha_{L,R,i}}$ upon restriction to $\hat{\omf g}_i$.  In fact there might be several such channels for this branching and for each one we can choose an Ishibashi state for the corresponding conformal primary:
 \begin{equation}
 \left.\mc H_{A}\otimes\mc H_{A^c}\right|_{\mc I_i}=\text{span}\left\{|\gamma_i\rrangle\;\big|\;m^{\gamma}_{\alpha_{i,L}}m^{\gamma^\ast}_{\alpha^\ast_{i,R}}\neq 0\right\}
 \end{equation}
 
\section{Entanglement across interfaces}\label{sect:ExamplesQG}
Let us now take this extended Hilbert space prescription adapted for systems with interfaces and compute the entanglement entropy of a variety of scenarios.

\subsection{$S^2$ with a single equatorial interface}\label{sect:S2singleint}
We now consider a state in the Hilbert space of the configuration depicted in Figure \ref{fig:LR3sphereanyons} where the entangling surface is a single component taken along the equatorial interface.  As elaborated upon above, this configuration does not specify a unique state; the Hilbert space is spanned by the mutual branching channels of $\alpha_L$ and $\alpha_R^\ast$.  Moving to the extended Hilbert space, a generic state in $\mc H_{S^2}[\alpha_L,\alpha_R]$ is mapped to
\begin{equation}\label{eq:genstate}
|\psi\rrangle=\sum_{\gamma}\psi_\gamma m^{\gamma}_{\alpha_L}m^{\gamma^\ast}_{\alpha_R^\ast}|\gamma\rrangle_{\omf g,\kappa_{eff}}.
\end{equation}
where we have explicitly noted that the $|\gamma\rrangle$ are conformal primary Ishibashi states of $\hat{\omf g}$ with level $\kappa_{eff}$.  Note that we have conveniently included the branching coefficients in the wavefunctions and so the sum can be taken indiscriminately over all highest weight representations of $\hat{\omf g}$.  From here the left-right entanglement of such a state can easily be computed using standard techniques (see for example \cite{Wen:2016snr}).  For the sake of a self-contained discussion, we will repeat this calculation in this section only.  The computations in the later sections are wholly similar.

As is familiar, the Ishibashi state itself is non-normalizable and so the solution to \eqref{eq:effgapcond} is only formal and requires regularization.  Although we defined $|\gamma\rrangle_{\omf g,\kappa_{eff}}$ purely from bulk considerations, it is a simple fact that it is in natural correspondence with a particular Virasoro module of the CFT with Sugawara generators, which given our definitions of
  the currents $\bmc J$ and $\tbmc J$, are
\beq
{\bs L_n}=\frac{c}{2\dim\omf g}\left(\kappa^{(eff)}\right)^{ab}:\bmc J_{a,m}\bmc J_{b,n-m}:\qquad\qquad {\tilde{\bs L}}_{n}=\frac{c}{2\dim\omf g}\left(\kappa^{(eff)}\right)^{ab}:\tbmc J_{a,m}\tbmc J_{b,n-m}:
\eeq
A natural way to regularize a given Ishibashi state is with the CFT Hamiltonian
\begin{equation}
|\gamma\rrangle\rightarrow|\gamma^{(\varepsilon)}\rrangle=e^{-\varepsilon \bs H_{eff}}|\gamma\rrangle\qquad\qquad \bs H_{eff}=\frac{2\pi}{\ell}\left(\bs L_0\otimes \mathbbm 1+\mathbbm 1\otimes\tilde{\bs L}_0-\frac{c_L+c_R}{24}\right)
\end{equation}
where $\ell$ is the circumference of the circular interface. The introduction of $\varepsilon$ can equivalently be thought of as defining the states, as usual, by moving into complex time. 
Although we constructed ${\bs H}_{eff}$ from $\hat{\omf g}$ generators (embedded into $\hat{\mf g}_{L,R}$), as opposed to say from $L_0^{(\hat{\mf g}_L)}\otimes \mathbbm 1+\mathbbm 1\otimes L_0^{(\hat{\mf g}_R)}$, there is no ambiguity here: the virtue of a conformal embedding is a matching of not only the central charges $c_L=c_R=c$, but also the Sugawara stress tensor.  

The norm of $|\gamma^{(\varepsilon)}\rrangle$ is simply the character associated to the primary $\gamma$ in the effective CFT:
\begin{equation}
n_\gamma^2=\chi^{eff}_{\gamma}\left(e^{\frac{-8\pi\varepsilon}{\ell}}\right).
\end{equation}
These regulated Ishibashi states, once divided by their norm,  define a basis of normalized states for the Hilbert space on $S^2$ pierced by $\alpha_{L,R}$ equipped with a finite norm:
\begin{equation}\label{eq:HS2finitenorm}
\mc H_{S^2[\alpha_L,\alpha_R]}^{(\varepsilon)}=\text{span}\left\{\frac{|\gamma^{(\varepsilon)}\rrangle}{n_\gamma}\;\Big|\; m_{\alpha_L}^\gamma m_{\alpha_R^\ast}^{\gamma^\ast}\neq 0\right\}
\end{equation}
From here on, when we write equations of the form \eqref{eq:genstate}, we will really have in mind their regulated versions defined with the same coeffcients but taken with the orthonormal basis \eqref{eq:HS2finitenorm} equipped with finite norm:
\begin{equation}
|\psi\rrangle\rightarrow |\psi^{(\varpepsilon)}\rrangle=\sum_{\gamma}\psi_\gamma(m_{\alpha_L}^\gamma m_{\alpha_R^\ast}^{\gamma^\ast})\frac{|\gamma^{(\varepsilon)}\rrangle}{n_\gamma}
\end{equation}
The reduced density matrix is obtained by tracing out $\mc H_{A^c}$:
\begin{equation}
\rho_{red}^{(\varepsilon)}=\sum_{\gamma}\frac{|\psi_\gamma|^2(m_{\alpha_L}^\gamma m_{\alpha_R^\ast}^{\gamma^\ast})^2}{n_\gamma^2}
\sum_{M}e^{-\frac{8\pi\varepsilon}{\ell}(h_{eff}(\gamma)+N_M-\frac{c}{24})}|\gamma;M\rangle\langle\gamma;M|
\end{equation}
where $h_{eff}$ is the conformal dimension of $|\gamma;0\rangle$ under $\bs L_0$ and $N_M$ is the grade of $M$\footnote{Recall that a weight space of an affine hight-weight representation is specified by $r+2$ integers: the $r$ Dynkin labels of its highest weight, its grade, and the level of of the algebra.  For an excellent and thorough review of affine algebras see \cite{DiFrancesco:1997nk}.}.  The resulting replica trace gives the character of the representation $\gamma$:
\begin{equation}\label{eq:TrRhon_1}
\mTr_{\mc H_A}\left({\rho_{red}^{(\varepsilon)}}^n\right)=\sum_{\gamma}\frac{|\psi_\gamma|^{2n}(m_{\alpha_L}^\gamma m_{\alpha_R^\ast}^{\gamma^\ast})^{2n}}{n_\gamma^{2n}}\chi_\gamma^{(eff)}\left(e^{-\frac{8\pi n\varepsilon}{\ell}}\right).
\end{equation}
It is clear that the role of the regulator is to move a small distance away from the singular point $\tau=0$ --- effectively the regulator has fattened the interface into a torus.
The characters appearing in the numerator and denominator (via $n_\gamma^2$) of \eqref{eq:TrRhon_1} can be evaluated in the limit of $\varepsilon/\ell\rightarrow 0$ by performing a modular transformation; in the limit, only the identity representation of $\hat{\omf g}$ contributes to the sum:
\begin{equation}\label{eq:TrRho_n_S2singleint}
\lim_{\varepsilon/\ell\rightarrow0}\mTr_{\mc H_A}\left({\rho_{red}^{(\varepsilon)}}^n\right)=\sum_{\gamma}\frac{|\psi_\gamma|^{2n}(m_{\alpha_L}^\gamma m_{\alpha_R^\ast}^{\gamma^\ast})^{2n}}{\left({\left(\mc S^{eff}\right)_\gamma}^0e^{\frac{\pi c\ell}{48 \varepsilon}}\right)^n}{\left(\mc S^{eff}\right)_\gamma}^0\,e^{\frac{\pi c\ell}{48 n\varepsilon}}
\end{equation}
It follows that the $n^{th}$ R\'enyi entropy is
\begin{equation}\label{eq:SR_S2singleint}
\small
S_n=\frac{1}{1-n}\log\frac{\mTr\left({\rho_{red}^{(\varepsilon)}}^n\right)}{\left(\mTr\rho_{red}^{(\varepsilon)}\right)^n}=\frac{1+n}{n}\frac{\pi c\ell}{48\varepsilon}+\frac{1}{1-n}\log\left(\sum_{\gamma}|\psi_\gamma|^{2n}(m_{\alpha_L}^\gamma m_{\alpha_R^\ast}^{\gamma^\ast})^{2n}\left({\left(\mc S^{eff}\right)_\gamma}^0\right)^{1-n}\right)-\frac{n}{n-1}\log\left(\sum_{\gamma}|\psi_\gamma|^2(m_{\alpha_L}^\gamma m_{\alpha_R^\ast}^{\gamma^\ast})\right)
\end{equation}
and the $n\rightarrow 1$ limit gives the corresponding von Neumann entropy:
\begin{equation}\label{eq:S_S2singleint}
S_{EE}=\frac{\pi c_L\ell}{24\varepsilon}+\sum_\gamma p_\gamma \log\left({\left(\mc S^{eff}\right)_\gamma}^0\right)-\sum_\gamma p_\gamma\,\log p_\gamma\qquad\qquad p_\gamma\equiv \frac{|\psi_\gamma|^2(m_{\alpha_L}^\gamma m_{\alpha_R^\ast}^{\gamma^\ast})^2}{\sum_{\gamma}|\psi_\gamma|^2(m_{\alpha_L}^\gamma m_{\alpha_R^\ast}^{\gamma^\ast})^2}.
\end{equation} 
Thus we see that the correction to the area law is given by the weighted sum of topological entanglement entropies of the effective anyons threading the entangling surface plus the Shannon entropy of the coefficients of the choice of state.  This feature is generic; in all examples that we examine below, we will see an analogous correction appears for each circular interface of the entangling surface.

\subsection{$T^2$ with a single interface}\label{sect:T2singleint}
Now we consider the scenario depicted in Figure \ref{fig:torus1int_anyon_v3_1}, where again the entanglement cut is taken at the interface surrounding the $R$ topological phase.
\myfig{torus1int_anyon_v3_1}{5}{The space of states on $T^2$ with a single patch (with connected boundary) of the $R$ topological phase.  A generic state can be heuristically thought of as being produced by a the interior path integral with an anyon $\alpha_L$ running around the non-contractible cycle of the torus.  If the identity irrep of the $R$ phase can branch to a non-trivial $\gamma$ at the interface then it may be possible for it to pair with a $\beta_L$ which can fuse with the interior anyon. 
}
Within the extended Hilbert space approach, $\mc H_A$ is the Hilbert space of a $T^2$ in the $L$ topological phase subtracted a disc, while $\mc H_{A^c}$ is the Hilbert space of the remaining disc (now in the $R$ topological phase):
\begin{equation}
\mc H_{A}=\bigoplus_{\alpha_L\beta_L}{\mc N^L_{\alpha_L\alpha_L^\ast}}^{\beta_L^\ast}\mc H_{S^1}[\beta_L]
\qquad\qquad \mc H_{A^c}=\mc H_{S^1}[0]
\end{equation}
where $\mc N^L$ are the fusion coefficients of the $L$ topological phase\footnote{This result can be understood by thinking of $T^2\setminus D^2$ as the gluing of an annulus to two of the holes of the ``pair of pants": $S^2\setminus(3D^2)$.  The Hilbert space of the latter is a direct sum with coefficients given by $\mc N^L$, and the gluing to the annulus forces two of the indices to be conjugates of each other.} and $\mc H_{S^1}$ is the Hilbert space defined on the right-hand side of \eqref{eq:HSsurfacewithholes}.  The space of gauge invariant states inside the extended Hilbert space $\mc H_A\otimes\mc H_{A^c}$ is   
\begin{equation}\label{eq:embHST2oneint}
\tmc H=\bigoplus_{\alpha_L,\beta_L}{\mc N^L_{\alpha_L\alpha_L^\ast}}^{\beta_L}\,m_{\beta_L}^\gamma\,m_{0_R}^{\gamma^\ast}\text{span}\left\{|\gamma\rrangle^{\omf g,\kappa_{eff}}\right\}
\end{equation}
Note that the identity irrep, $0_R$, appearing in $\mc H_{A^c}$ can possibly branch into more than the identity representation of $\hat{\omf g}$.\footnote{For instance in {\bf example 2} of Section \ref{sect:examples} we saw that the identity of $\mf{su}(3)_1$ can branch into both $(2j=0)$ and $(2j=4)$ of $\mf{su}(2)_4.$}  
A generic state is given by 
\begin{equation}
|\psi\rrangle={\sum_{\alpha_L,\beta_L}}{\sum_{\gamma}}\psi_{\alpha_L,\beta_L}^\gamma |\gamma\rrangle_{\alpha_L,\beta_L}\qquad\qquad {\sum_{\alpha_L,\beta_L}}{\sum_{\gamma}}|\psi_{\alpha_L,\beta_L}^\gamma|^2=1
\end{equation}
(recall that we mean this in the regulated sense by the replacement $|\gamma\rrangle\rightarrow |\gamma^{(\varepsilon)}\rrangle/n_\gamma$ explained in Section \ref{sect:S2singleint}).  We leave implicit that the coefficients $\psi_{\alpha_L,\beta_L}^\gamma$ are nonzero for representations satisfying
\beq
{\mc N_{\alpha_L\alpha_L^\ast}^L}^{\beta_L},\;m_{\beta_L}^{\gamma},\;m_{0_R}^{\gamma^\ast}\neq0
\eeq
Additionally note that there might be multiple choices of $\alpha_L$ and $\beta_L$ giving rise to the same irrep $\gamma$ of $\hat{\omf g}$ and so we label the Ishibashi state by its corresponding sector.

The entanglement entropy is given by
\beq
S_{EE}=\frac{\pi c \ell}{24 \varepsilon}+{\sum_\gamma} p_\gamma\log\left({(\mc S^{eff})_\gamma}^0\right)-{\sum_{\gamma}} p_\gamma\,\log p_\gamma\qquad\qquad p_\gamma={\sum_{\alpha_L,\beta_L}}|\psi_{\alpha_L,\beta_L}^\gamma|^2.
\eeq
Let us briefly discuss the physics of this result: the identity of the $R$ topological phase constrains the possible anyons appearing in the effective theory at the interface.  This leads to a weighted sum of topological corrections to the area law.  
Although there might be multiple branching channels realizing the same effective anyon, the topological correction only detects the effective anyon threading the interface: the choice of branching channels only contributes to the Shannon term.  Here we see the interface adds something novel: for single entanglement cuts on $T^2$ in homogenous theories it was found that every state in $\mc H_{T^2}$ gives the same topological correction (again up to a Shannon term) \cite{Dong:2008ft, Wen:2016snr}.

\subsection{$T^2$ with two interfaces}
Now we consider states on $T^2$ with two interfaces.  There are two separate set-ups to consider; one with $A^c$  disconnected (and each interface is contractible on the surface of $T^2$) and one where $A^c$ is connected (and each interface runs around the meridian of $T^2$).  We consider these cases separately.

\subsubsection{$A^c$ disconnected}\label{sect:torus2int_cont}
\myfig{torus2intcont_v3_1}{7}{States on the torus with two contractible interfaces, separating the $L$ topological phase from two ``islands" in phases $R_1$ and $R_2$.}
We start with the case where the two interfaces separate disconnected islands ($R_1$ and $R_2$) within the $L$ topological phase.  This is depicted in Figure \ref{fig:torus2intcont_v3_1}.  We are allowed to choose possibly distinct TBCs at each of these two interfaces and furthermore $R_{1}$ and $R_2$ can also host distinct topological phases.  We now construct an extended Hilbert space for this set-up.

The Hilbert space factors are\footnote{Again, the factor $\mc H_A$ follows from thinking of $T^2\setminus (2D^2)$ as gluing two ``pairs of pants," ($S^2\setminus(3D^2)$) along their legs.  Each comes with their own fusion coefficient, and we must correctly identify the anyons running through the legs.  
}
\begin{equation}\label{eq:splitHStorus2int_cont}
\mc H_A=\bigoplus_{\alpha_L,\beta_L,\gamma_L,\delta_L}\left({\mc N^L_{\alpha_L\beta^\ast_L}}^{\gamma_L}{\mc N^L_{\alpha^\ast_L\beta_L}}^{\delta_L}\right)\mc H_{S^1}[\gamma_L]\otimes\mc H_{S^1}[\delta_L]\qquad\qquad \mc H_{A^c}=\mc H_{S^1}[0]_1\otimes\mc H_{S^1}[0]_2
\end{equation}
The imposition of gauge invariance is wholly similar to the previous example and we are led to an extended Hilbert space of the form
\begin{equation}
\tmc H=\bigoplus_{\alpha_L,\beta_L,\gamma_L,\delta_L,\sigma,\eta}{\mc N_{\alpha_L\beta_L^\ast}^L}^{\gamma_L}m_{\gamma_L}^{\sigma}m_{0_{R_1}}^{\sigma^\ast}{\mc N_{\alpha_L^\ast\beta_L}^L}^{\delta_L}m_{\delta_L}^{\eta}m_{0_{R_2}}^{\eta^\ast}\text{span}\left\{|\sigma\rrangle^{\omf g_1,\kappa_{eff,1}}\otimes|\eta\rrangle^{\omf g_2,\kappa_{eff,2}}\right\}
\end{equation}
A generic state in this Hilbert space is
\beq
|\psi\rrangle = {\sum_{\alpha_L,\beta_L,\gamma_L,\delta_L}}{\sum_{\sigma}}{\sum_{\eta}}\psi_{\alpha_L,\beta_L,\gamma_L,\delta_L}^{\sigma,\eta}\left(|\sigma\rrangle^{\omf g_1,\kappa_{eff,1}}_{\alpha_L,\beta_L,\gamma_L}\otimes|\eta\rrangle^{\omf g_2,\kappa_{eff,2}}_{\alpha_L,\beta_L,\delta_L}\right)
\eeq
where, as before, the coefficients, $\psi$, are non-zero only when fusion/branching is possible and chosen such that $|\psi\rrangle$ is normalized.  We also denote with subscripts (as we did before) the possibility of $|\sigma\rrangle$ or $|\eta\rrangle$ appearing in multiple sectors.  We again find that the topological correction from the area law comes from a weighted sum over the possible ``effective anyons" threading the interfaces plus a Shannon term coming from a choice of the state in the fusion/branching channels:
\begin{equation}
S_{EE}=\frac{\pi c \ell}{12\varepsilon}+{\sum_{\sigma}}{\sum_{\eta}}p(\sigma,\eta)\left(\log\left({(\mc S^{eff,1})_\sigma}^0\right)+\log\left({(\mc S^{eff,2})_\eta}^0\right)\right)-\sum_{\sigma}\sum_{\eta}p(\sigma,\eta)\log p(\sigma,\eta)
\end{equation}
with $p(\sigma,\eta)=\sum_{\alpha_L,\beta_L,\gamma_L,\delta_L}|\psi_{\alpha_L,\beta_L,\gamma_L,\delta_L}^{\sigma,\eta}|^2$.

\subsubsection{$A^c$ connected}\label{sect:torus2int_noncont}
\myfig{torus2intnoncont_1}{7}{The states on torus with two noncontractible interfaces can be generated by the path integral with Wilson line insertions $\alpha_L$ and $\alpha_R$ as long as they have mutual branching channels $\bar\beta_1$ and $\bar\beta_2$.}
As a final example, we consider the space of states on $T^2$ with two interfaces taken along the meridian (as depicted in Figure \ref{fig:torus2intnoncont_1}).  Since $A^c$ is now connected, there is only one $L$ phase and one $R$ phase. However, we can still allow for possibly distinct TBCs at the separate interfaces.  The extended Hilbert space is given by factors
\begin{equation}
\mc H_A=\bigoplus_{\alpha_L}\mc H_{S^1}[\alpha_L]\otimes\mc H_{S^1}[\alpha_L^\ast]\qquad\qquad \mc H_{A^c}=\bigoplus_{\alpha_R}\mc H_{S^1}[\alpha_R]\otimes\mc H_{S^1}[\alpha_R^\ast]
\end{equation}
and the quantum gluing determines the embedded Hilbert space to be
\begin{equation}
\tmc H=\bigoplus_{\alpha_L,\alpha_R,\bar \beta_1,\bar\beta_2}(m_{\alpha_L}^{\bar\beta_1}m_{\alpha_R}^{\bar\beta_1^\ast})(m_{\alpha_L^\ast}^{\bar\beta_2}m_{\alpha_R^\ast}^{\bar\beta_2^\ast})\text{span}\left\{|\bar\beta_1\rrangle^{\omf g_1,\kappa_{eff,1}}\otimes|\bar\beta_2\rrangle^{\omf g_2,\kappa_{eff,2}}\right\}
\end{equation}
A generic state in this Hilbert space will be of the form
\begin{equation}
|\psi\rrangle=\sum_{\alpha_L,\alpha_R} \psi_{\alpha_L,\alpha_R}^{\bar\beta_1,\bar\beta_2}\left(|\bar\beta_1\rrangle^{\omf g_1,\kappa_{eff,1}}_{\alpha_L,\alpha_R}\otimes|\bar\beta_2\rrangle^{\omf g_2,\kappa_{eff,2}}_{\alpha_L,\alpha_R}\right)
\end{equation}
and its entanglement entropy is
\beq
S_{EE}=\frac{\pi c\ell}{12\varepsilon}+\sum_{\bar\beta_1,\bar\beta_2}p(\bar\beta_1,\bar\beta_2)\left(\log\left({(\mc S^{eff,1})_{\bar\beta_1}}^0\right)+\log\left({(\mc S^{eff,2})_{\bar\beta_2}}^0\right)\right)-\sum_{\bar\beta_1,\bar\beta_2}p(\bar\beta_1,\bar\beta_2)\log p(\bar\beta_1,\bar\beta_2).
\eeq
with $p(\bar\beta_1,\bar\beta_2)=\sum_{\alpha_L,\alpha_R}|\psi_{\alpha_L,\alpha_R}^{\bar\beta_1,\bar\beta_2}|^2$.  

\section{Surgery}\label{sect:surgery}
In this  section, we describe how these results can be realized from surgery techniques.  Surgery provides an independent method for evaluating the entanglement entropy by realizing the R\'enyi entropies as path integrals on replicated geometries.  Here the full power of Chern-Simons theory as a topological field theory can be brought to bear: 
in \cite{Dong:2008ft}, these techniques allowed the authors to extend the Kitaev-Preskill / Levin-Wen results \cite{Kitaev:2005dm,Levin:2006zz} to a wide variety of manifolds with a wide variety of entanglement cuts.  This section is constructed in a similar spirit to that paper, although now with the inclusion of interfaces.

Before we describe this in detail, we will make a couple of remarks.  Firstly, 
surgery, as a tool, typically uses a formal description of TQFT, and as such, the results of \cite{Dong:2008ft} naturally exclude the area term and calculate the TEE correction exactly.  However since these corrections are inherently negative,\footnote{All of the statements in this section apply for compact groups, where $\mc S$ is a unitary matrix and as such $|{\mc S_0}^\alpha|<1$.  Interestingly enough, this does not have to be true for non-compact groups.  For instance, \cite{McGough:2013gka} continues the $S_{topo}=\log{\mc S_0}^\alpha$ result to the non-compact $SL(2,\mathbb R)$ Chern-Simons theory and finds a positive value for this correction matching the BTZ Bekenstein-Hawking entropy.  
} the interpretation of these replica path integrals as an entropy is dubious; it is the area law that guarantees positivity, even if it is non-universal.  Secondly, the issue of the non-factorizability of the Hilbert space is completely ignored in these replica path integrals.  When including interfaces, we will soon see that it will be necessary to regulate this formal description of surgery.  While this regulator is introduced to mitigate interface intersections, it provides a natural UV regulator and we will see that through our ``regulated surgery" an area law appears.  This method, however, still does not address the second remark. At the end of Section \ref{sect:surgeryS2interface}, we will draw a connection to the results of Sections \ref{sect:QG} and \ref{sect:ExamplesQG} which should be viewed as a more fine-grained, fundamental description of the entanglement across the interface.  That said, surgery remains a powerful tool in the interface theory and complements the fine-grained calculations with geometric intuition; we will see it easily verifies the above results and we will give an example in the discussion (Section \ref{sect:Discussion}) where surgery can evaluate $S_{topo}$ when the corresponding extended Hilbert space description would be complicated.

Let us recall that the standard description of surgery is predicated by the fact that the dimension of the Chern-Simons Hilbert space on the two-sphere (possibly with two anyon punctures in conjugate representations) is one-dimensional:
\begin{equation}
\dim\mc H_{S^2}=1.
\end{equation}
With this fact we can write the overlap of any two-states $|\psi\rangle,|\phi\rangle\in\mc H_{S^2}$ in terms of their overlap with any choice of fiducial state $|\chi\rangle\in\mc H_{S^2}$:
\begin{equation}\label{eq:regularsurgery}
\langle \phi|\psi\rangle=\frac{\langle\phi|\chi\rangle\langle\chi|\psi\rangle}{\langle\chi|\chi\rangle}
\end{equation}
When $|\psi\rangle$ and $|\phi\rangle$ are produced by the path integral on manifolds $M_\psi$ and $M_\phi$ (with boundaries $\pa M_\psi=\pa M_\phi=S^2$) then the left-hand side of \eqref{eq:regularsurgery} is the path integral on the manifold $M^p_\phi\#_{S^2}M_\psi$ (here the ``$p$" on $M^p_\phi$ indicates a flip in orientation).  The right-hand side however is the multiplication of the path integrals on the $M_\phi^p\#_{S^2}M_\chi$ and $M_\chi^p\#_{S^2} M_\psi$ divided by $M_\chi^p\#_{S^2}M_\chi$.  This is particularly nice when $|\chi\rangle$ is produced by the path integral on the interior of a three-ball (with possibly a pair of Wilson lines intersecting the boundary $S^2$).  In this case we simply ``cap off" the geometries $M_\phi$ and $M_\psi$ (which we will denote $\bar M_\phi$ and $\bar M_\psi$) and divide by the expectation value of Wilson lines in $S^3$.  Using this iteratively, the path integral on any compact three-manifold can eventually be evaluated as a rational expression of path integrals on a handful of simple ``ingredient" geometries (for instance $S^2\times S^1$ and $S^3$).

Now we repeat this procedure for a three-manifold with interfaces.  A natural guess for what should be done is to ``cut" and ``sew" along an interface, taking $|\chi\rangle$ to be produced on the 3-ball bisected by an interface.  It turns out that this is not a very helpful approach, since when anyon punctures are included
\begin{equation}
\dim\mc H_{S^2[\alpha_L,\alpha_R]}={\mc W_{\alpha_L}}^{\alpha_R}
\end{equation}
which isn't necessarily one if the interface is non-trivial.  The resolution to this is to always perform surgery \emph{along trivial interfaces} (i.e., within a single topological phase) in such a way as to isolate the non-trivial interfaces within a simple manifold (say $S^3$).  The price to pay with this is we now regard this $S^3$ as an input ingredient to the surgery and so need to evaluate it independently.  Once that has been done, we can use this procedure to ``chop down" any complicated three manifold with (isolated) interfaces into easily evaluated objects.  Figure \ref{fig:surgery} is provided as an illustration of this procedure.

\myfig{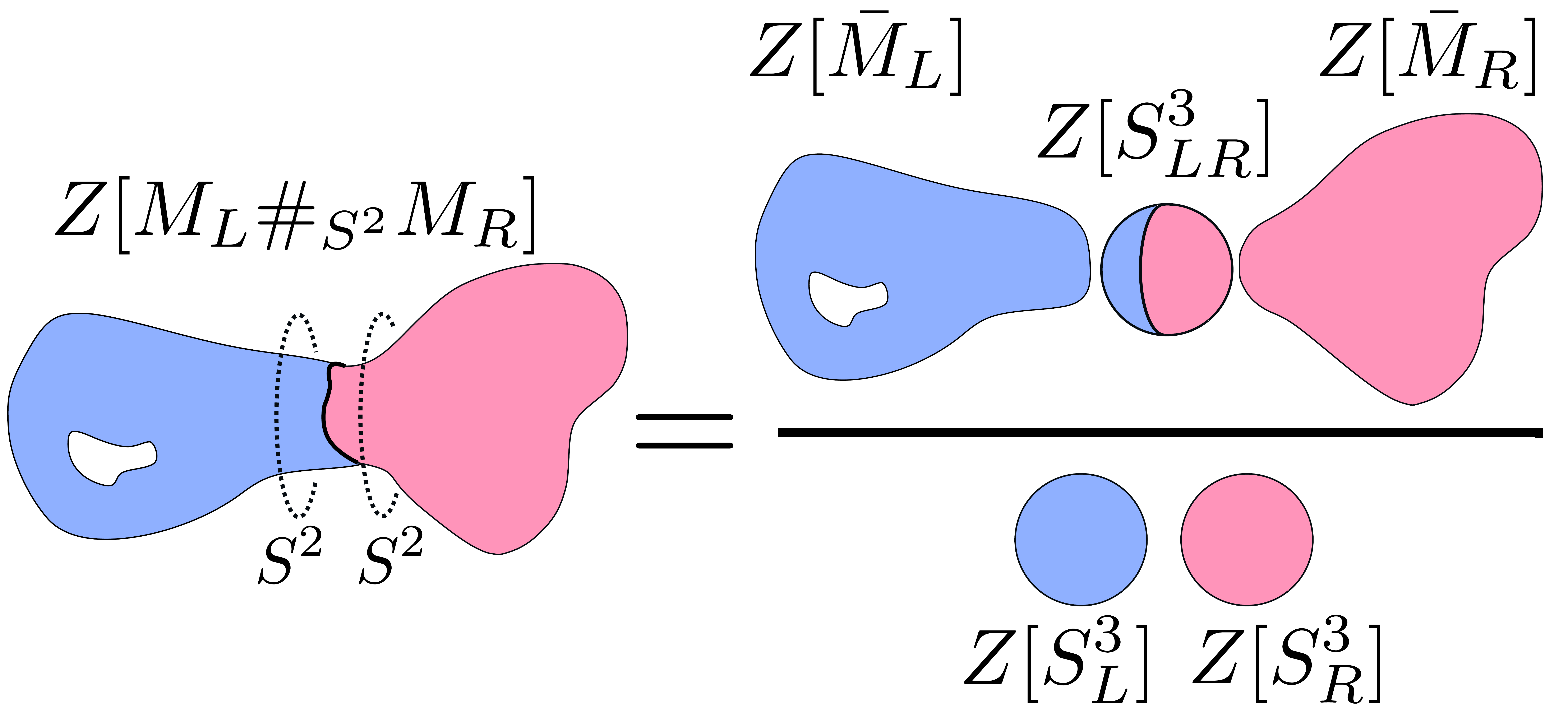}{10}{Manifolds $M_L$ and $M_R$ joined along an $S^2$ interface.  We can take surgery cuts just to the left and right of the interface; the price we pay is that we have to independently evaluate the path integral on $S^3$ with an $S^2$ interface (here denoted $Z[S^3_{LR}]$).}

\subsection{$S^2$: entanglement cut along the interface}\label{sect:surgeryS2interface}
\begin{figure}[h!]
\centering
 \begin{tabular}{ c c c c c }
    \includegraphics[align=c,height=4cm]{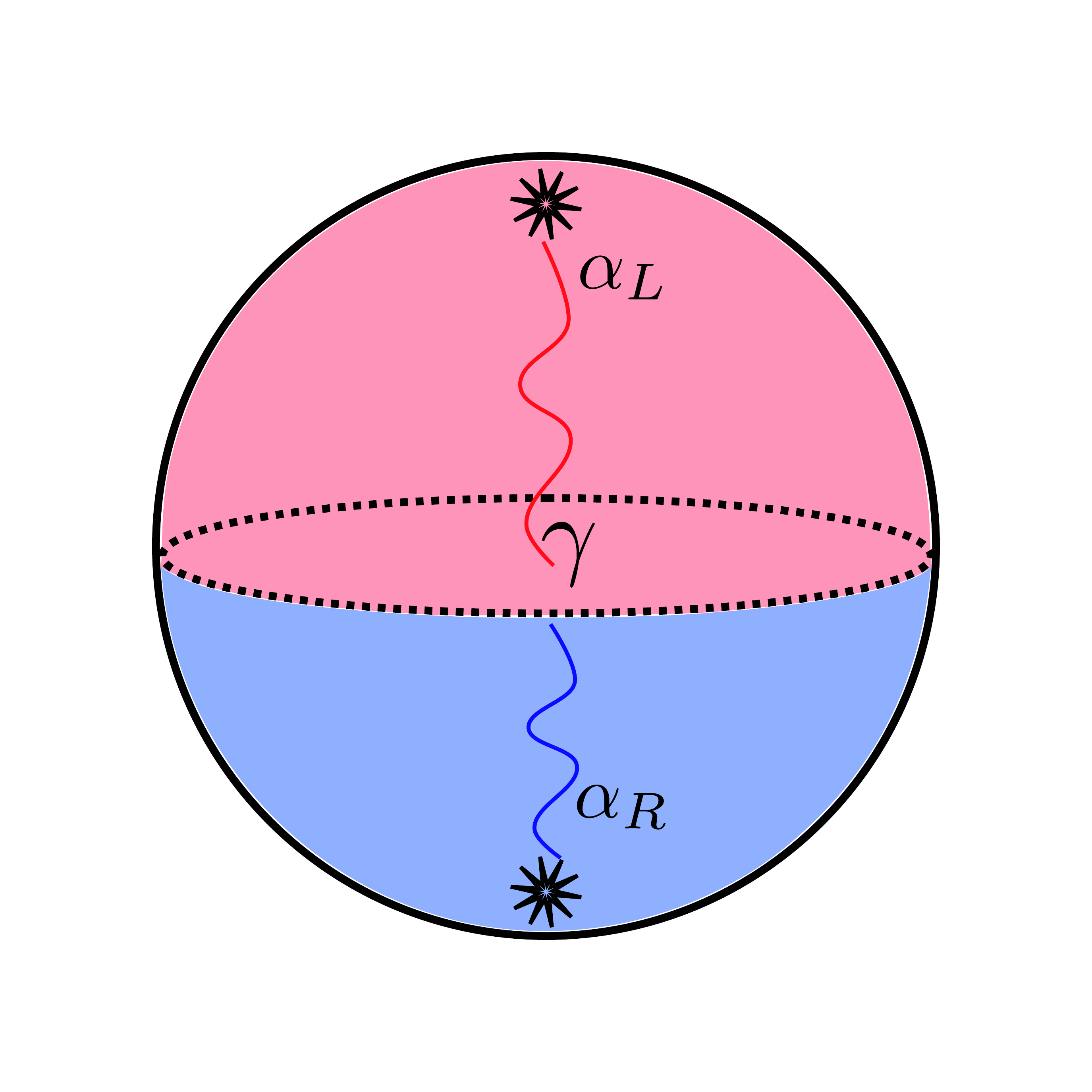} & \hspace{.5 cm} & \includegraphics[align=c,height=4cm]{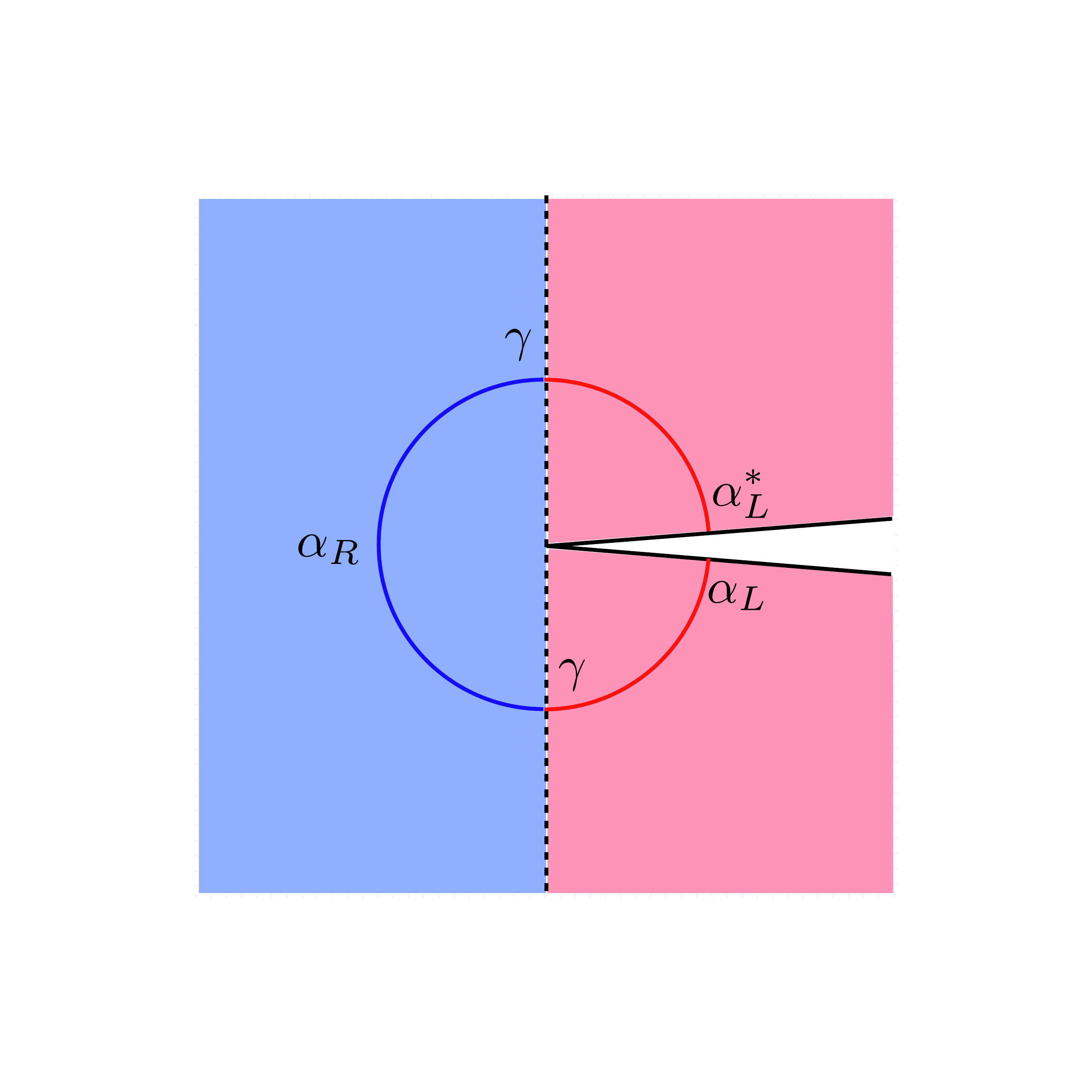} & \hspace{.5 cm} & \includegraphics[align=c,height=4cm]{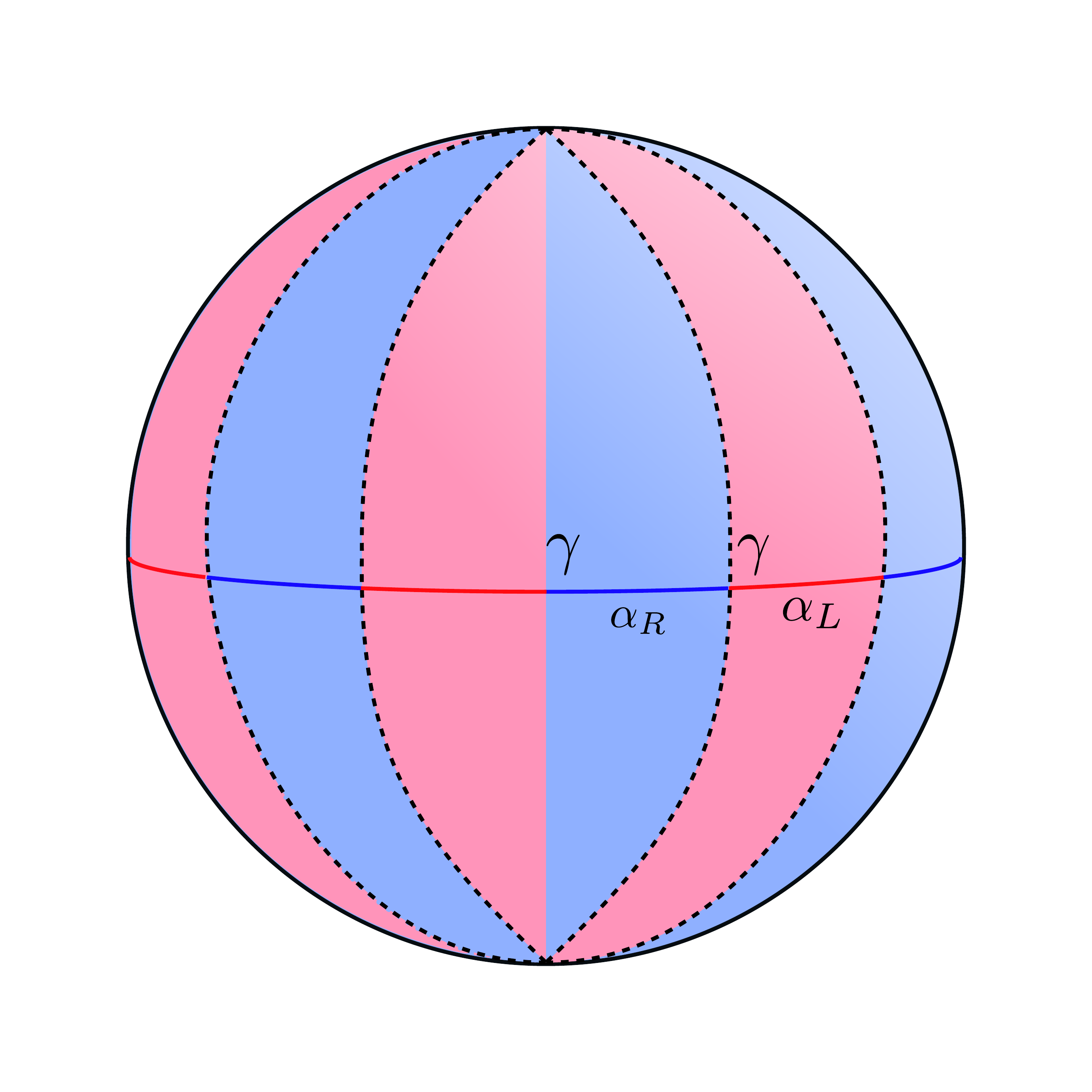}
    \end{tabular}\caption{\small{\textsf{(Left) The state prepared on the two-sphere with anyon punctures.  The middle point of the Wilson lines tunnel the interface through the effective theory in a {\it fixed} $\mc R_\gamma$ channel.  (Middle) The reduced density matrix (obtained by suppressing the azymuthal and conformally mapping to the Euclidean plane).  (Right) The replicated manifold representing $\mTr\rho_{red}^n$.  We are suppressing a dimension and so the intersection of the interfaces is inaccurately represented as two isolated points when they are actually a line fixed under the $\mathbb Z_n$ replica symmetry. 
    }}}\label{fig:ZS3LRn}
\end{figure}
Let us consider first the situation of a spatial two-sphere with an interface separating two anyon insertions $\alpha_L$ and $\alpha_R$.  As we stated above, there may be more than one state with this configuration depending on the branchings of $\alpha_L$ and $\alpha_R$ at the interface.  We pick a particular state formed by the path integral inside the three ball with Wilson line insertions restricted to the representation $\mc R_\gamma$ of $\hat{\omf g}$ at the interface for some $\gamma$ such that $m_{\alpha_L}^\gamma m_{\alpha_R^\ast}^{\gamma^\ast}\neq0$.  The replica path integral is $S^3$ with alternating wedges of $L$ and $R$ topological phases and a Wilson loop that tunnels through the interfaces in either the $\mc R_\gamma$ or the $\mc R_{\gamma^\ast}$ representations (depending on orientation).  This is depicted as a cartoon in Figure \ref{fig:ZS3LRn}.  We will call this object $Z_n\left[S^3_{LR}(\alpha_L,\alpha_R;\gamma)\right]$ or $Z_n\left[S^3_{LR}\right]$ for short.

\begin{figure}[h!]
\centering
 \begin{tabular}{ c c c c c }
    \includegraphics[align=c,height=4cm]{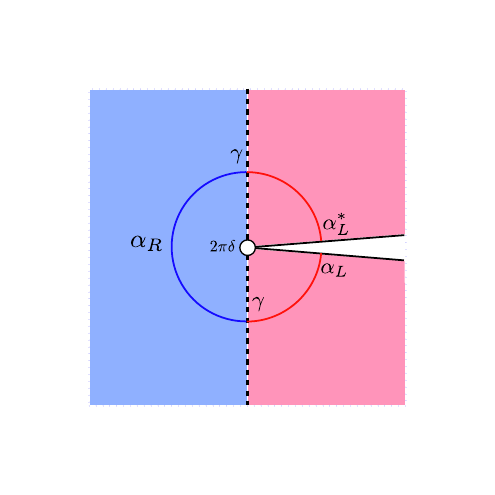} & \hspace{.5 cm} & \includegraphics[align=c,height=4cm]{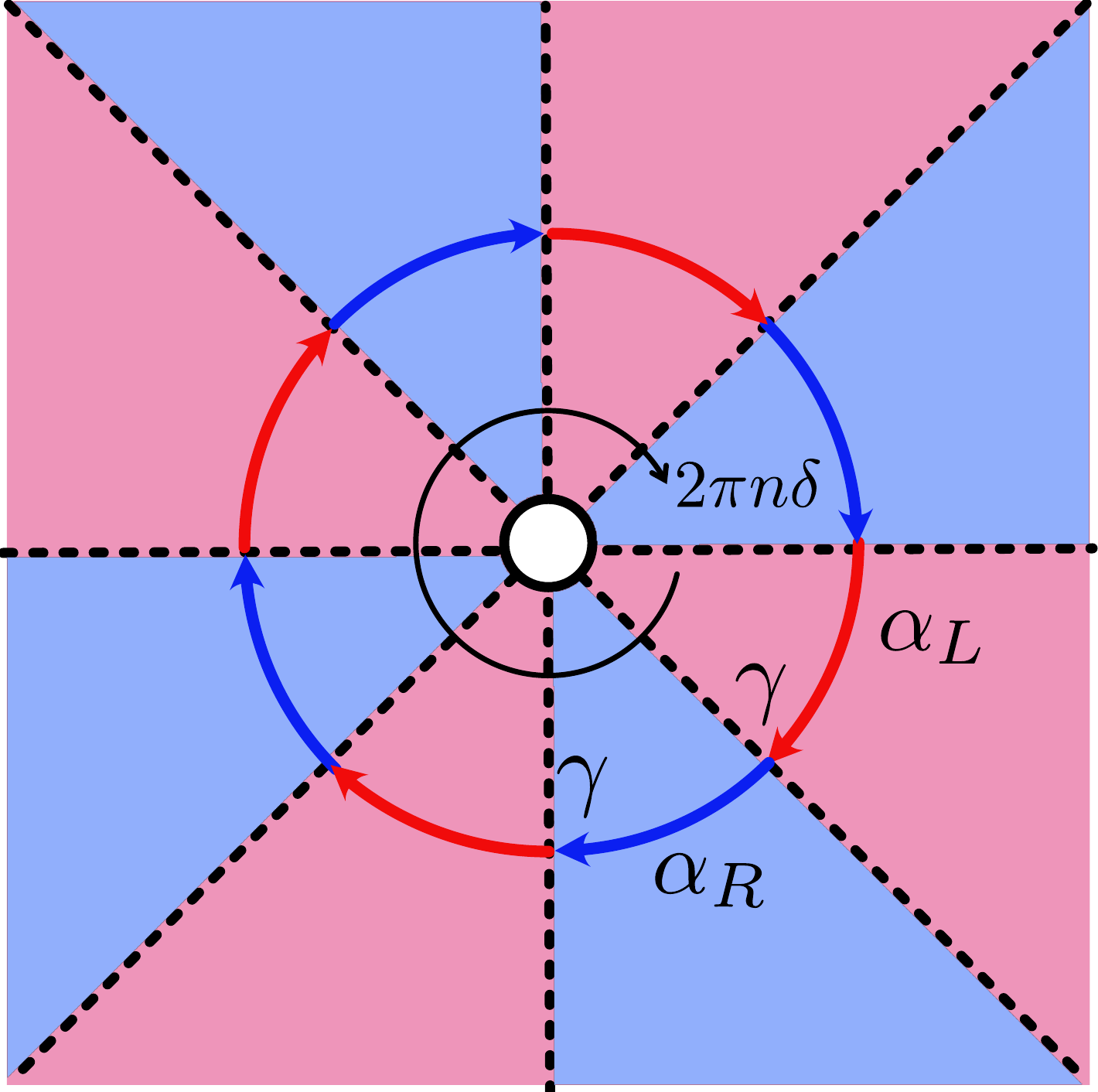} & \hspace{.5 cm} & \includegraphics[align=c,height=4cm]{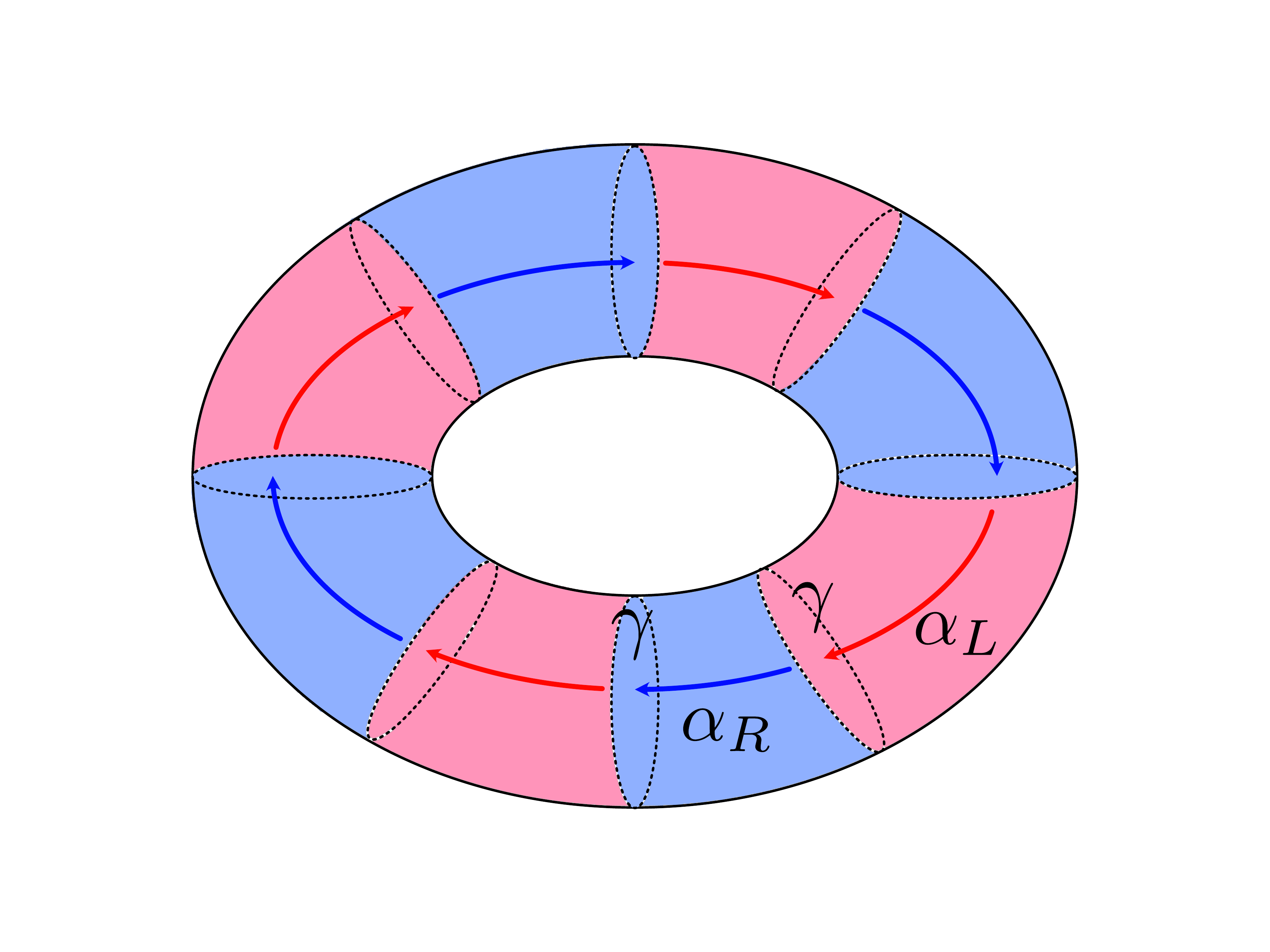}
    \end{tabular}\caption{\small{\textsf{(Left) A cartoon of reduced density matrix with the tubular neighborhood removed.  (Middle) The replica geometry; the mutual intersection of the interfaces has been regulated.  (Right) The geometry after mapping to $S^1\times D^2$.  The Wilson threads this solid torus, intersecting the interfaces transversely.
    }}}\label{fig:ZS3LRnkeyhole}
\end{figure}

It is clear from the geometry, that there is no meaningful way to surgically isolate the interfaces in the above path integral.  This is because they intersect at the line fixed by replica symmetry.  Because of this, $Z_n\left[S^3_{LR}\right]$ is the new independent ingredient we must supplement to the surgery method.  As we will see in later sections, once we evaluate $Z_n\left[S^3_{LR}\right]$, we can use it to surgically evaluate a large number of replica geometries.  To deal with the fixed line where the interfaces intersect, we will return to our reduced density matrix and excise a small tubular neighborhood of radius $\delta$ about this fixed line.  The replica path integral is then over the manifold, $S^3_{LR}\setminus\mc N_{n\delta}$.  We will conformally map this to $D^2\times S^1$ where $D^2$ is the Poincar\'e disc and the $S^1$ has geodesic length $2\pi n\delta$.\footnote{Although the theory is topological, this statement is easily seen with a cylindrically symmetric metric: $ds^2=dz^2+d\rho^2+\rho^2d\theta^2=\rho^2\left(d\theta^2+\frac{d\rho^2+dz^2}{\rho^2}\right)\sim ds^2_{S^1}+ds^2_{\mathbb H^2}$.  Note that the $\rho=\delta$ boundary is mapped to the boundary of $\mathbb H^2$.  The hyperbolic half-plane $\mathbb H^2$ can then be mapped to Poincar\'e disc.}.  Without loss of generality, we will take the Wilson lines to thread through the origin of the Poincar\'e disc.  This is illustrated in Figure \ref{fig:ZS3LRnkeyhole}.  To implement $\delta$ as a regulator, it is necessary to introduce a length scale, $\ell$, as the perimeter of the $D^2$.  We will be interested in taking $\delta/\ell\rightarrow0$ at the end of the computation.

Our expression for $Z_n[S^3_{LR}]$ is then
\beq\label{eq:ZnS3LRreg}
Z_n[S^3_{LR}]=\lim_{\delta/\ell\rightarrow 0}\left\langle \prod_{q=1}^n\mathbb P^{(\hat{\omf g})}_\gamma\hat W_{\alpha_L}[\theta_q,\theta_{q+1/2}]\mathbb P^{(\hat{\omf g})}_\gamma\hat W_{\alpha_R}[\theta_{q+1/2},\theta_{q+1}]\mathbb P^{(\hat{\omf g})}_\gamma\right\rangle_{D^2\times S^1}
\eeq
where we write $\big\langle\cdot\big\rangle_{D^2\times S^1}$ as short-hand for the insertion of the above into the path-integral on $D^2\times S^1$ and {\it not} the expectation value in a fixed state; in particular, operator insertions can be permuted cyclicly around the $S^1$.  For the Wilson line insertions, the subscripts denote their associated representation and the endpoints are denoted in the brackets (it is implicit that $\theta_{n+1}\equiv \theta_1$).  $\mathbb P^{(\hat{\omf g})}_{\gamma}$ is a projector onto the $\mc R_\gamma$ representation of the subalgebra $\hat{\omf g}$.  Since we are dealing with Wilson {\it line} operators, we pause to discuss possible issues of gauge invariance.  Even under a proper gauge transformation (that is, one with support localized in the bulk of the $D^2\times S^1$)
\beq
A^L\rightarrow g_L^{-1}\left(d+A^L\right)g_L
\eeq
the Wilson line operator is not invariant: it responds by conjugation
\beq
\hat W_{\alpha_L}[\theta_q,\theta_{q+1/2}]\rightarrow \pi^{(\hat{\mf g}_L)}_{\alpha_L}[g_L^{-1}(0,\theta_q)]\hat W_{\alpha_L}[\theta_q,\theta_{q+1/2}]\pi^{(\hat{\mf g}_L)}_{\alpha_L}[g_L(0,\theta_{q+1/2})]
\eeq
where $\pi^{(\hat{\mf g}_L)}_{\alpha_L}[g_L]$ is the representation of $g_L$ acting on $\mc R_{\alpha_L}$.  Wilson lines in the $R$ phase respond to gauge transformations of $A^R$ similarly.  The TBCs, however, break the bulk gauge invariance at the interfaces between these phases by identifying $A^L$ with $A^R$ through the isotropic subalgebra $\omf g$.  The unbroken gauge invariance preserves the TBCs: these are group elements limiting to the appropriate subgroup as they approach an interface.  For instance, approaching the interface positioned at $\theta_q$:
\beq
\lim_{\theta_L\rightarrow \theta_q}g_L(\theta_L)=\iota_L\circ\bar g(\theta_q)\qquad\qquad\lim_{\theta_R\rightarrow\theta_q}g_R(\theta_R)=\iota_R\circ\bar g(\theta_q)\qquad\qquad \bar g\in\exp\omf g.
\eeq
Upon this restriction $\pi^{(\hat{\mf g}_L)}_{\alpha_L}$ breaks up into irreps of this subgroup:
\beq
\pi^{(\hat{\mf g}_L)}_{\alpha_L}[\iota_L\circ \bar g]=\bigoplus_{\delta}\pi^{(\hat{\omf g})}_{\delta}[\bar g]
\eeq
of which $\mc R_\gamma$ appears in a particular block that is singled out by acting on $\mathbb P^{(\hat{\omf g})}_\gamma$.  A similar statement holds for $\pi_{\alpha_R}^{(\hat{\mf g}_R)}$.  Thus under generic gauge transformations preserving the TBCs, our expectation value \eqref{eq:ZnS3LRreg} remains invariant:
\begin{align}
Z_n[S_{LR}^3]\rightarrow&\lim_{\delta/\ell\rightarrow 0}\left\langle \prod_{q=1}^n\mathbb P^{(\hat{\omf g})}_{\gamma}\pi^{(\hat{\mf g}_L)}_{\alpha_L}[g_L^{-1}(\theta_q)]\hat W_{\alpha_L}\pi^{(\hat{\mf g}_L)}_{\alpha_L}[g_L(\theta_{q+1/2})]\mathbb P_{\gamma}^{(\hat{\omf g})}\pi_{\alpha_R}^{(\hat{\mf g}_R)}[g^{-1}_R(\theta_{q+1/2})]\hat W_{\alpha_R}\pi_{\alpha_R}^{(\hat{\mf g}_R)}[g_R(\theta_{q+1})]\mathbb P_{\gamma}^{(\hat{\omf g})}\right\rangle_{D^2\times S^1}\nonumber\\
=&\lim_{\delta/\ell\rightarrow 0}\left\langle \prod_{q=1}^n\mathbb P^{(\hat{\omf g})}_{\gamma}\pi^{(\hat{\omf g})}_{\gamma}[\bar g^{-1}(\theta_q)]\hat W_{\alpha_L}\pi^{(\hat{\omf g})}_{\gamma}[\bar g(\theta_{q+1/2})]\mathbb P_{\gamma}^{(\hat{\omf g})}\pi_{\gamma}^{(\hat{\omf g})}[\bar g^{-1}(\theta_{q+1/2})]\hat W_{\alpha_R}\pi_{\gamma}^{(\hat{\omf g})}[\bar g(\theta_{q+1})]\mathbb P_{\gamma}^{(\hat{\omf g})}\right\rangle_{D^2\times S^1}\nonumber\\
=&\lim_{\delta/\ell\rightarrow 0}\left\langle \prod_{q=1}^n\mathbb P^{(\hat{\omf g})}_\gamma\hat W_{\alpha_L}\mathbb P^{(\hat{\omf g})}_\gamma\hat W_{\alpha_R}\mathbb P^{(\hat{\omf g})}_\gamma\right\rangle_{D^2\times S^1}
\end{align}
since $\pi^{(\hat{\omf g})}_{\gamma}[\bar g]\mathbb P^{(\hat{\omf g})}_\gamma\pi^{(\hat{\omf g})}_\gamma[\bar g^{-1}]=\mathbb P^{(\hat{\omf g})}_\gamma$.  

We will now evaluate \eqref{eq:ZnS3LRreg}.  First we write out the projectors concretely,
\beq\label{eq:proj}
\mathbb P^{(\hat{\omf g})}_\gamma=\sum_{\bar m=0}^\infty\sum_{\bar i}|\gamma,\bar m,\bar i\rangle\langle\gamma,\bar m,\bar i|.
\eeq
Here we are explicitly notating the {\it grade} of a state by $\bar m$; a basis of states at this grade are labelled by $\bar i$.  Sandwiching these projectors in between Wilson line operators, our task is to evaluate overlaps of the form
\beq\label{eq:WLsandwich}
\langle\gamma,\bar m_1,\bar i_1|\hat W_{\alpha_L}[\theta_q,\theta_{q+1/2}]|\gamma,\bar m_2,\bar i_2\rangle
\eeq
As noted in \cite{Elitzur:1989nr} the path integral of Chern-Simons theory along $D^2\times [\theta_q,\theta_{q+1/2}]$, where the disc is punctured by a Wilson line in representation $\mc R_{\alpha_L}$ is equivalent to a chiral WZW path-integral on $S^1\times [\theta_q,\theta_{q+1/2}]$; this path integral is over group elements fixed in the conjugacy glass of the group element dual to integral weight $\alpha_L$.  States of this theory furnish the representation $\mc R_\alpha$.  In this vein, we treat $\hat W_{\alpha}[x_q,y_q]$ as an unrestricted WZW transition amplitude in the $\alpha_L$ sector.  As written, this boundary theory has no Hamiltonian and as such the expectation value \eqref{eq:ZnS3LRreg} diverges.  We will supplement this transition amplitude with the CFT Hamiltonian\footnote{This corresponds to the addition of the boundary term $S_{bndy}=\frac{1}{2\pi}\int_{\pa D_2\times I}\kappa_L\left(A_L,_{\wedge\ast}A_L\right)$ to the fictitious cutoff surface. Note that this boundary term introduces a metric on the cutoff surface and a corresponding geodesic length, $\ell$, around the circle; we shall see that this term is responsible for the area law.}, $\bs H^{\hat{\mf g}_L}=\frac{2\pi}{\ell}\left(\bs L^{\hat{\mf g}_L}_0-\frac{c_L}{24}\right)$.  Thus we arrive at
\begin{align}
\sum_{n_1,n_2}\sum_{j_1,j_2}\langle \gamma,\bar m_1,\bar i_1|\alpha_L,n_1,j_1\rangle&\langle\alpha_L,n_1,j_1|e^{-\delta \bs H^{\hat{\mf g}_L}}|\alpha_L,n_2,j_2\rangle\langle\alpha_L,n_2,j_2|\gamma,\bar m_2,\bar i_2\rangle\nonumber\\
&=\sum_{n_1}\sum_{j_1}\langle \gamma,\bar m_1,\bar i_1|\alpha_L,n_1,j_1\rangle e^{-\frac{2\pi\,\delta}{\ell}\left(h_{\alpha_L}+n_1-\frac{c_L}{24}\right)}\langle\alpha_L,n_1,j_1|\gamma,\bar m_2,\bar i_2\rangle
\end{align}
Now let us look at the overlap $\langle \gamma,\bar m_1,\bar i_1|\alpha_L,n_1,j_1\rangle$.  The representation $\mc R_\gamma$ appears in the restriction of $\mc R_{\alpha_L}$ and so the state $|\gamma,\bar m_1,\bar i_1\rangle$ appears at a particular grade of $\mc R_{\alpha_L}$.  Let us call this grade $n_{\bar m}$.  The nature of the conformal embedding is such that $\bs L^{\hat{\mf g}_L}_0$ acts on states of the $\hat{\omf g}$ CFT as $\bs L^{\hat{\omf g}}_0$ and so
\beq
h_{\alpha_L}+n_{\bar m}=h_{\gamma}+\bar m.
\eeq
Within the grades $n_{\bar m}$ and $\bar m$ there will be some rectangular change of basis matrix $\mathbb M^{(n_{\bar m})}$:
\beq
\langle \gamma,\bar m_1,\bar i_1|\alpha_L,n_1,j_1\rangle=\delta_{n_1,n_{\bar m_1}}{\mathbb M_{\bar i_1}}^{j_1}
\eeq
We will require that states of $\mc R_{\gamma}$ sit in the representation $\mc R_{\alpha_L}$ {\it isometrically} such that
\begin{align}
\delta_{\bar m_1,\bar m_2}\delta_{\bar i_1,\bar i_2}=&\langle \gamma,\bar m_1,\bar i_1|\gamma,\bar m_2,\bar i_2\rangle=\sum_{n}\sum_{j}\langle \gamma,\bar m_1,\bar i_1|\alpha_L,n,j\rangle\langle\alpha_L,n,j|\gamma,\bar m_2,\bar i_2\rangle\nonumber\\
=&\sum_{n,j}\delta_{n,n_{\bar m_1}}\delta_{n,n_{\bar m_2}}{\mathbb M_{\bar i_1}}^{j}{\mathbb M_{\bar i_2}^\ast}^j
\end{align}
Putting these facts together we have the following expression for \eqref{eq:WLsandwich}:
\beq
\sum_{n}\sum_j\delta_{n,n_{\bar m_1}}\delta_{n,n_{\bar m_2}}{\mathbb M_{\bar i_1}}^{j}{\mathbb M_{\bar i_2}^\ast}^je^{-\frac{2\pi\delta}{\ell}(h_\gamma+\bar m_1-\frac{\bar c}{24})}=\delta_{\bar m_1,\bar m_2}\delta_{\bar i_1,\bar i_2}e^{-\frac{2\pi\delta}{\ell}(h_\gamma+\bar m_1-\frac{\bar c}{24})}
\eeq
This is precisely the expression we would have arrived at evaluating the transition amplitude of a Wilson line in representation $\mc R_\gamma$ within a $\omf g$ Chern-Simons theory from the beginning.  By similar arguments we have for the $R$-phase Wilson lines
\beq
\langle\gamma,\bar m_2,\bar i_2|\hat W_{\alpha_R}[\theta_{q+1/2},\theta_{q+1}]|\gamma,\bar m_3,\bar i_3\rangle=\delta_{\bar m_2,\bar m_3}\delta_{\bar i_2,\bar i_3}e^{-\frac{2\pi\delta}{\ell}\left(h_\gamma+\bar m_2-\frac{\bar c}{24}\right)}
\eeq
Again this expression follows precisely because $\hat{\omf g}$ is conformally embedded into $\hat{\mf g}_R$ and $\mc R_\gamma$ appears in the restriction of $\mc R_{\alpha_R}$.  Putting this all together we arrive at
\beq
Z_n[S^3_{LR}]=\lim_{\delta/\ell\rightarrow0}\prod_{q=1}^n\sum_{\bar m_q}\sum_{\bar i_q}\delta_{\bar m_q,\bar m_{q+1}}\delta_{\bar i_q,\bar i_{q+1}}e^{-\frac{4\pi\delta}{\ell}\left(h_{\gamma}+\bar m_q-\frac{\bar c}{24}\right)}
\equiv \lim_{\delta/\ell\rightarrow0}\chi_\gamma^{(\hat{\omf g})}\left(e^{\frac{-8\pi\,n\delta}{\ell}}\right)
\eeq
Thus our path-integral in question is a character of $\mc R_\gamma$ in the effective subalgebra.  Taking $\delta/\ell\rightarrow 0$, we recover the expected answer
\beq
Z_n[S^3[\alpha_L,\alpha_R;\gamma]]={\left(S^{eff}\right)_\gamma}^0e^{\frac{\pi \bar c\ell}{48 n\delta}}
\eeq
That we arrived at this answer is probably no surprise: one might notice that if one views the projector \eqref{eq:proj} as a maximally mixed (unnormalized) density matrix over the $\mc R_\gamma$ representation, then its purification is {\it precisely the Ishibashi state} $|\gamma\rrangle$. Thus while in this computation, $Z_n$ is akin to a thermal partition function, the computations in Section \ref{sect:S2singleint} are the analogous computation using a pure-state purification.
\subsection{$T^2$ with a single interface}
\begin{figure}[h!]
\centering
\begin{tabular}{l l l}
\includegraphics[align=c,width=.3\textwidth]{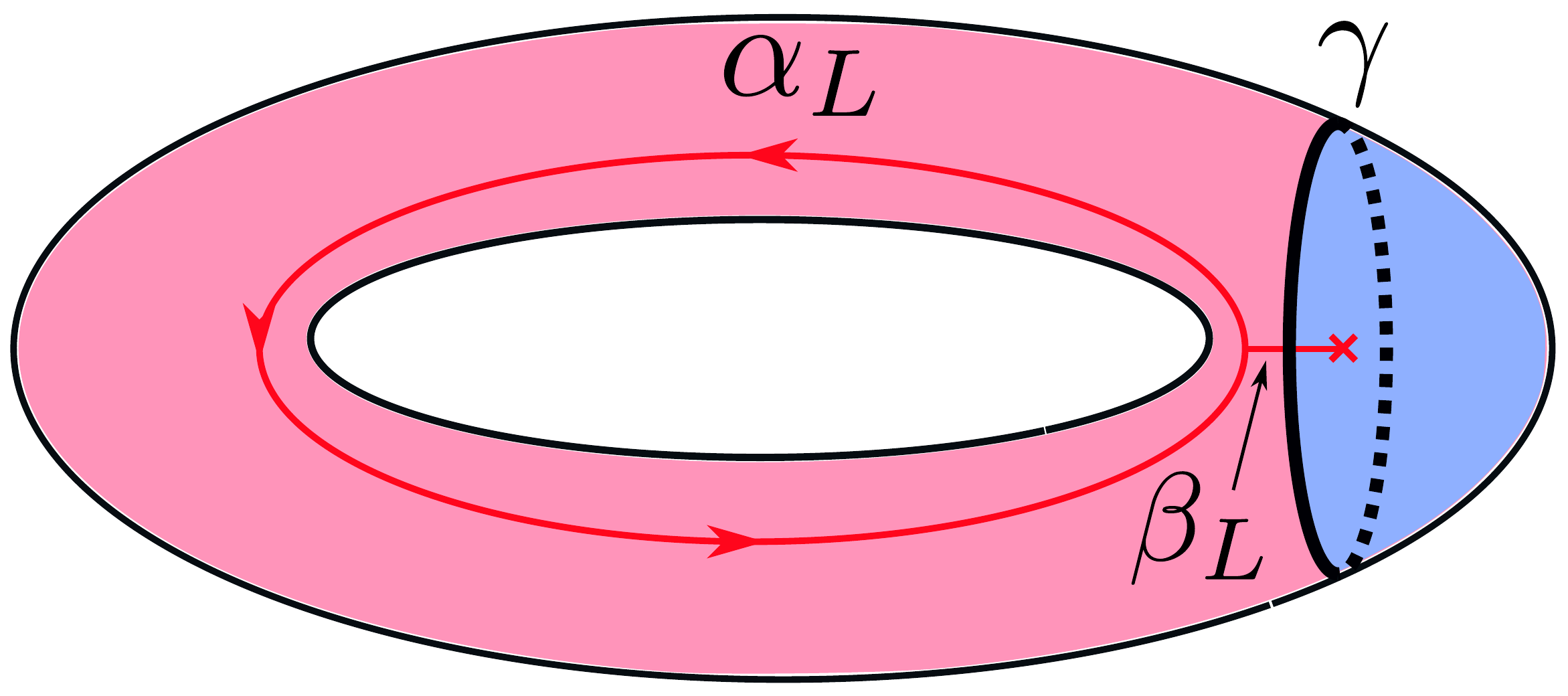}&
\includegraphics[align=c,width=.25\textwidth]{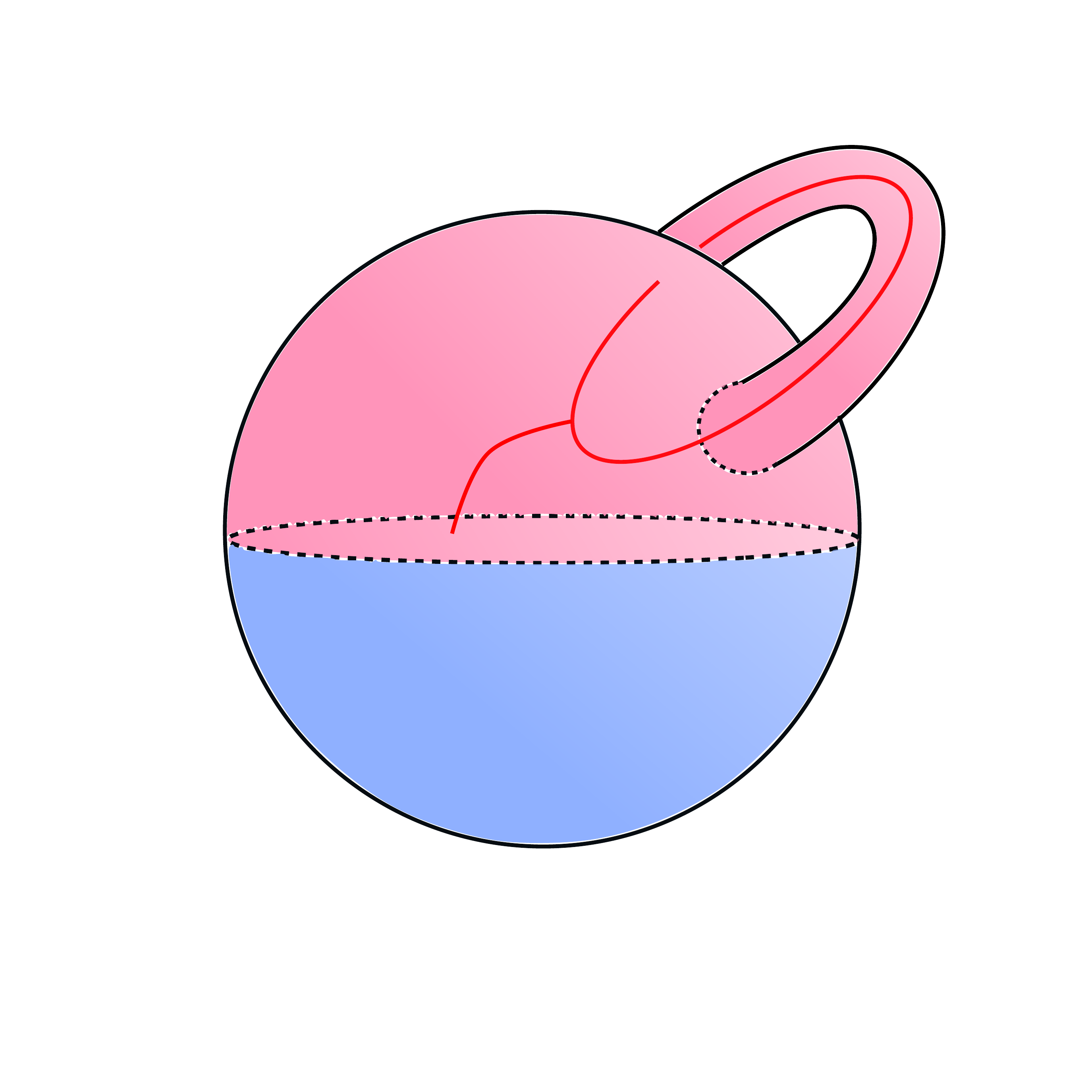}&
\includegraphics[align=c,width=.3\textwidth]{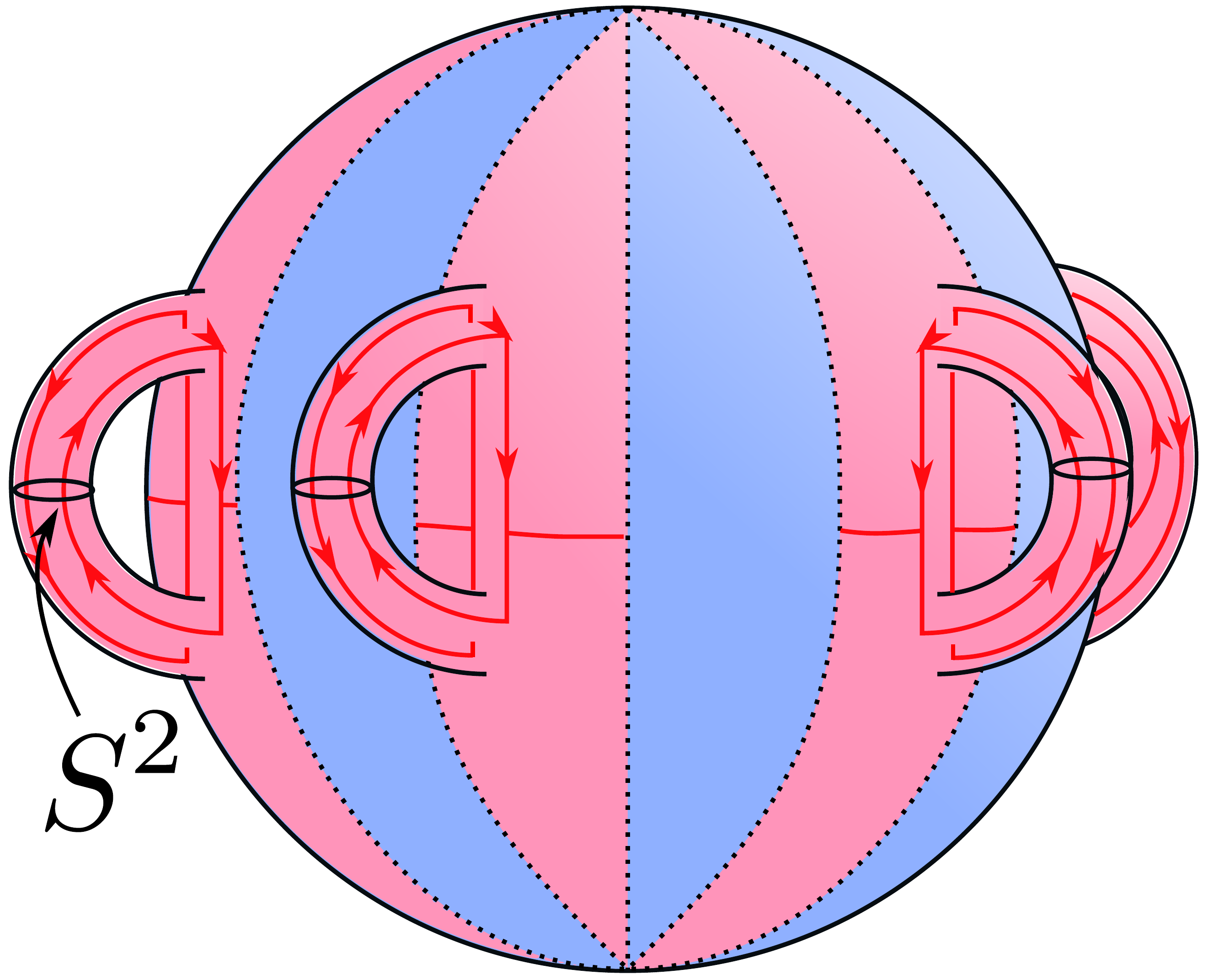}
\end{tabular}
\caption{\small{\textsf{(Left) The state on the torus and (middle) the same state redrawn suggestively.  (Right) The three-manifold of the replica path integral.}}}\label{fig:torus1int_surgery}
\end{figure}

Now we repeat the calculation of Section \ref{sect:T2singleint} from the viewpoint of surgery.  For convenience, in Figure \ref{fig:torus1int_surgery}, we recall the path integral picture for the state, as well as what the replicated geometry schematically looks like.  Unlike Section \ref{sect:T2singleint}, we will choose a state fixed by a {\it specific choice} of $\alpha_L$, $\beta_L$ and $\gamma$ illustrated on the left of Figure \ref{fig:torus1int_surgery}. 

\begin{figure}[h!]
\centering
\begin{tabular}{l l l}
\includegraphics[align=c,width=.4\textwidth]{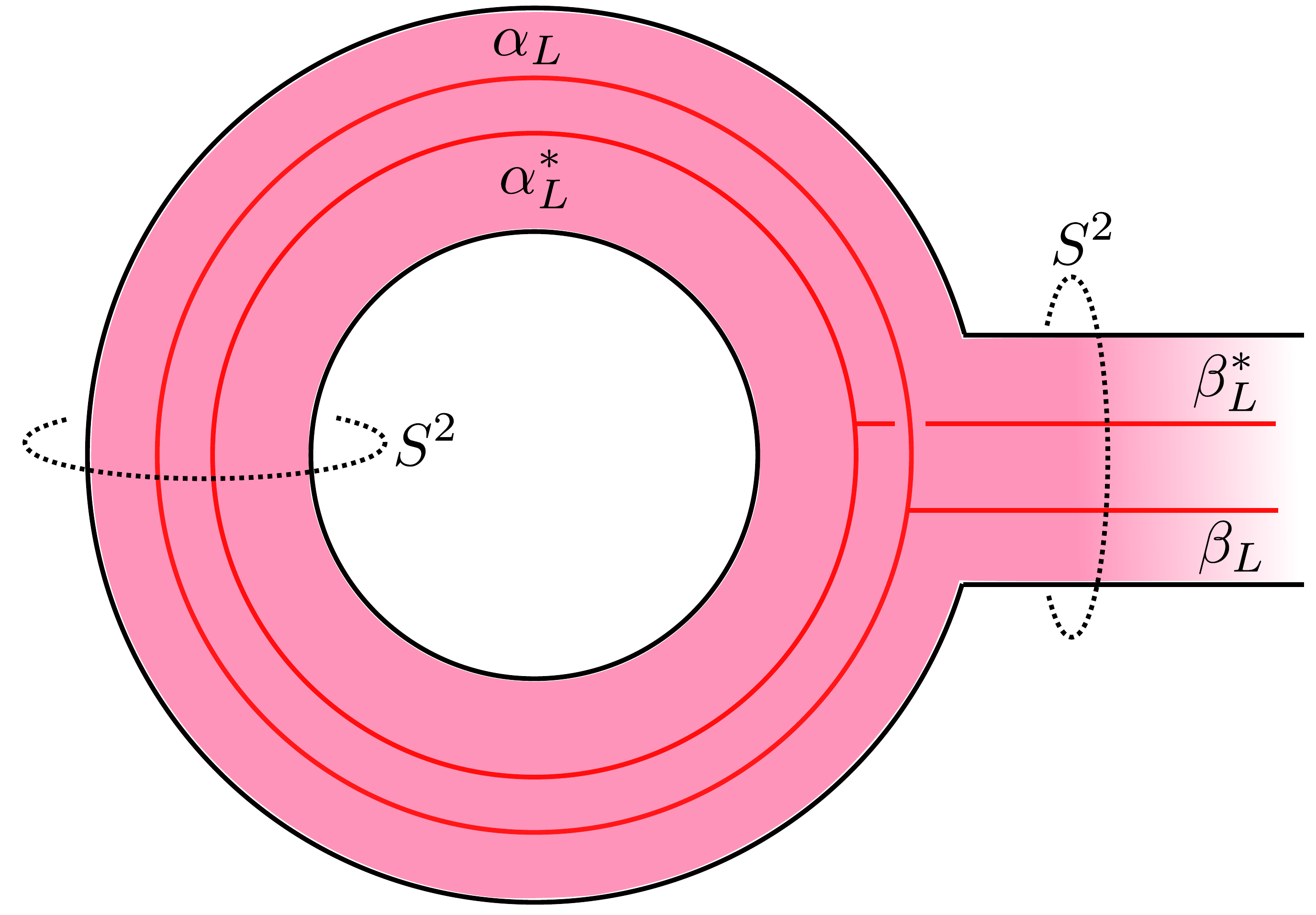}&&
\includegraphics[align=c,width=.3\textwidth]{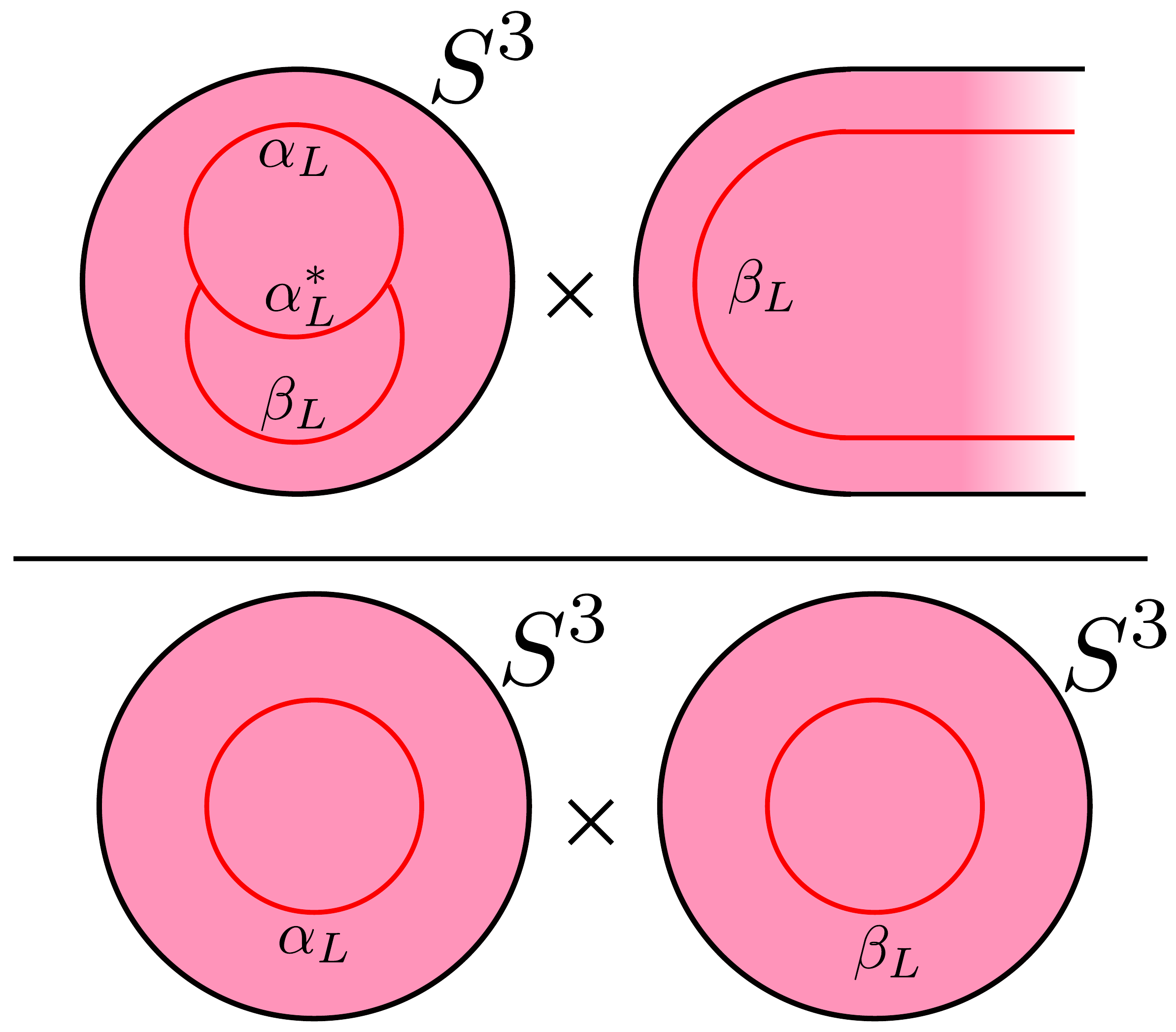}
\end{tabular}
\caption{\small{\textsf{(Left) One of the handles from the right subfigure of Figure \ref{fig:torus1int_surgery}, with $S^2$ surgery performed at the indicated locations.  (Right)  The equivalent path-integral expression.}}}
\label{fig:handlesurgery}
\end{figure}

Let us see how to deal with the handles in the right subfigure of Figure \ref{fig:torus1int_surgery}.  We first ``cut" at the midway point of the handle by inserting the one dimensional projector on $\mc H_{S^2[\alpha_L,\alpha_L^\ast]}$ (i.e., the two-sphere punctured by $\alpha_L$ and $\alpha_L^\ast$).  By pulling the handle away from the ``beachball", we ``cut" through the $\beta_L$ lines in the same way.  This is depicted in greater detail in Figure \ref{fig:handlesurgery}.  This isolates the $\alpha_L$ and $\beta_L$ fusion vertex within a single $S^3$.  Let us comment on this fusion vertex.  We will write this as an overlap of states on the two-sphere with a triple puncture (by supposition of ${\mc N^L_{\alpha_L\alpha_L^\ast}}^{\beta_L}\neq 0$, this Hilbert space is non-empty)
\beq
\langle \alpha_L,\alpha_L^\ast,\beta_L|_{S^2}|\alpha_L,\alpha_L^\ast,\beta_L\rangle_{S^2}\equiv |\Psi_{\alpha_L,\alpha_L^\ast,\beta_L}|^2.
\eeq
If ${\mc N^L_{\alpha_L\alpha_L^\ast}}^{\beta_L}>1$, then this overlap is ambiguous: we need to specify the fusion channel.  This is in fact data we need to supplement to completely specify our original state depicted in the left panel of Figure \ref{fig:torus1int_surgery}.  We will suppose this data has been fixed.  Regardless, as we shall soon see, this ambiguity does not enter into the R\'enyi entropy.

Doing this for each handle, we find the following expression for the R\'enyi path-integral
\begin{equation}\label{eq:torus1int_surgery}
Z_n=\frac{Z_n[S^3_{LR}[\beta_L,0;\gamma]]\;|\Psi_{\alpha_L,\alpha_L^\ast,\beta_L}|^{2n}}{\left(Z[S_L^3[\alpha_L]]\right)^{n}\left(Z[S_L^3[\beta_L]]\right)^n}
\end{equation}
where $Z[S^3_L[\alpha_L]]={\mc S^L_{\alpha_L}}^0$ is the path-integral on the three-sphere with an unknotted Wilson loop in the $\alpha_L$ representation.  As promised, the overlap $|\Psi_{\alpha_L,\alpha_L^\ast,\beta_L}|^2$ does not affect the R\'enyi entropies and we only pick up the contribution from the effective anyon, $\gamma$, at the interface:
\begin{align}
S_n=&\frac{1}{1-n}\log\frac{Z_n}{Z_1^n}=\frac{1}{1-n}\left(\log Z_n[S^3_{LR}[\beta_L,0;\gamma]]-n\log Z_1[S^3_{LR}[\beta_L,0;\gamma]]\right)=\frac{1+n}{n}\frac{\pi \bar c\ell}{48\delta}+\log{(\mc S^{eff})_\gamma}^0
\end{align}
This is consistent with the results from Section \ref{sect:T2singleint}: regardless of the branching channel from which it appears, the topological correction only knows about which effective anyon, $\gamma$, threads the interface.
\subsection{$T^2$ with two interfaces}
Now we consider states on the torus with two interfaces, evaluating the surgical expressions for each of the two possible configurations separately.
\subsubsection{$A^c$ disconnected}
\begin{figure}[h!]
\centering
\begin{tabular}{l l l}
\includegraphics[align=c,width=.3\textwidth]{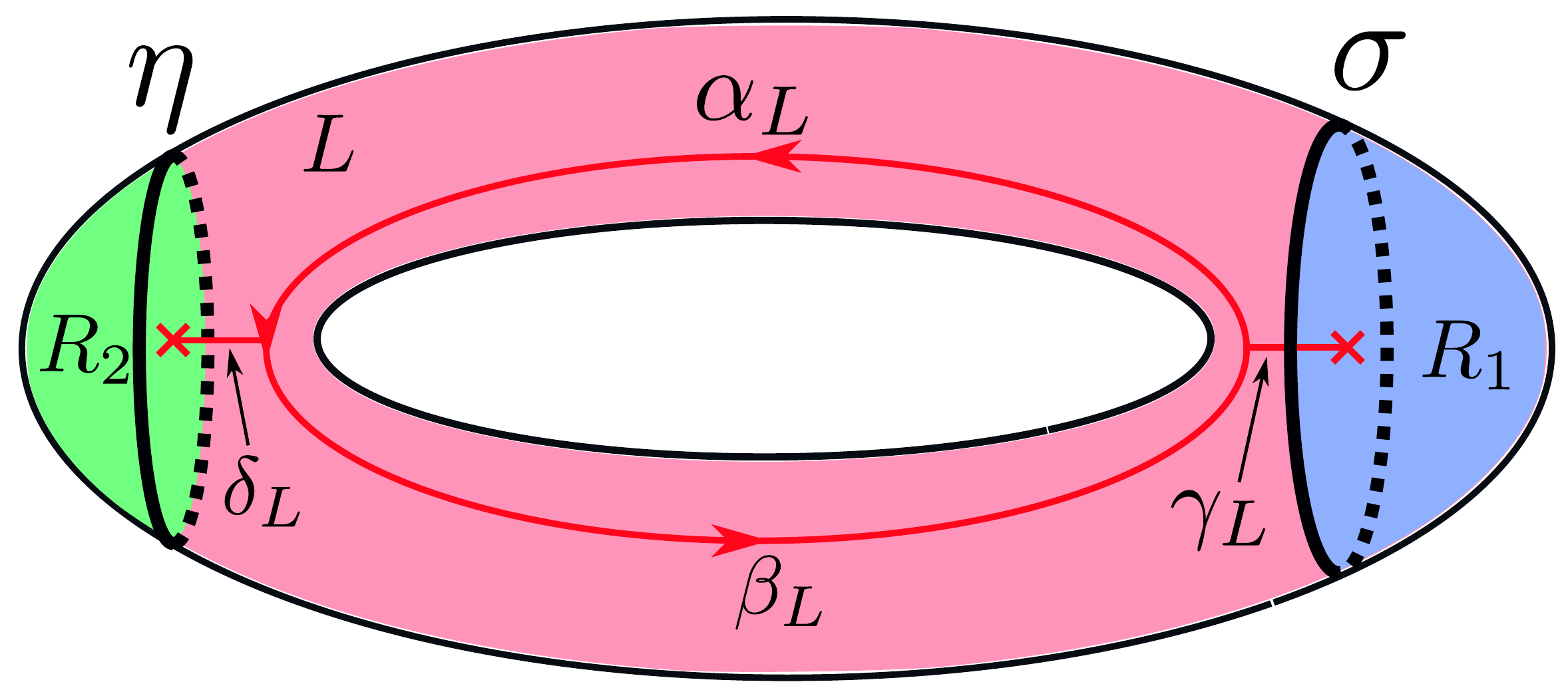}&
\includegraphics[align=c,width=.3\textwidth]{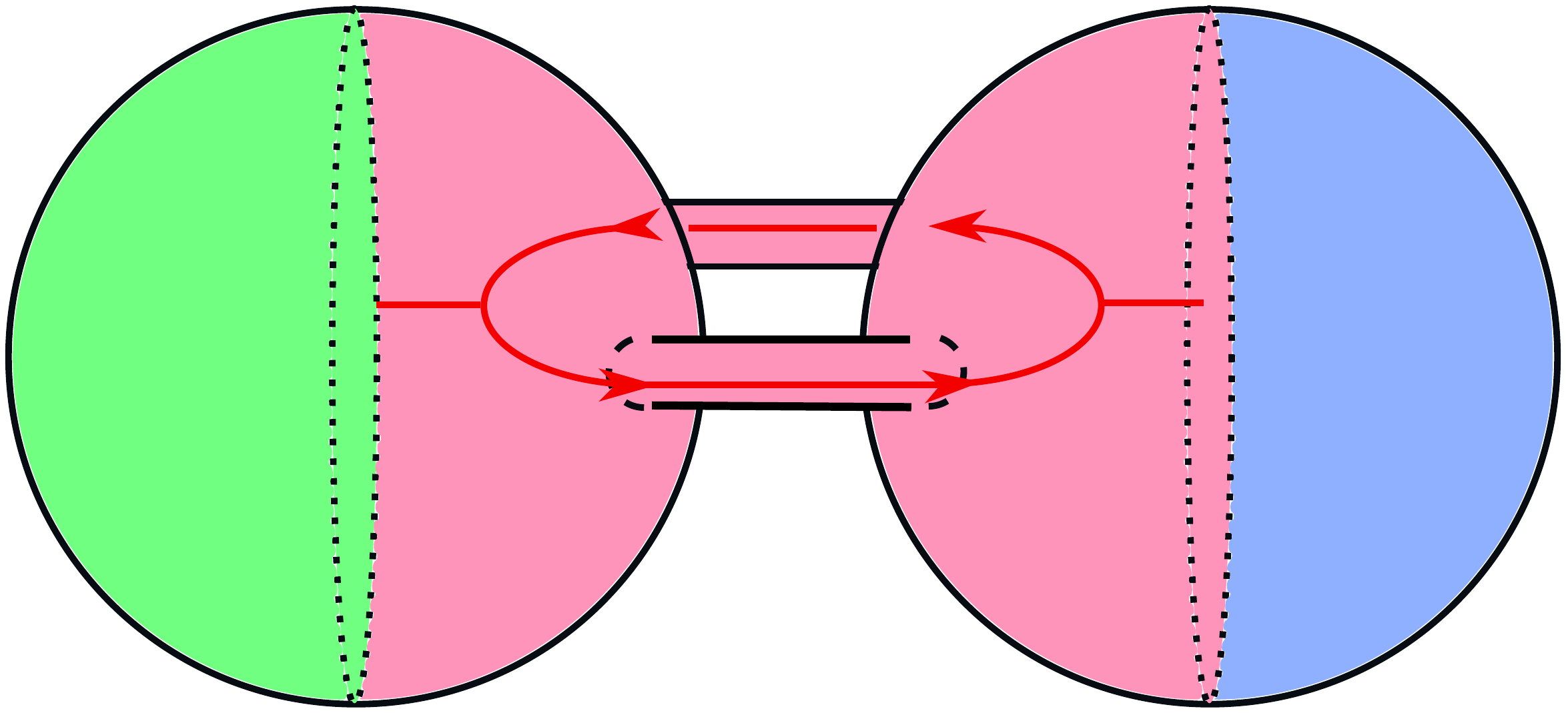}&
\includegraphics[align=c,width=.3\textwidth]{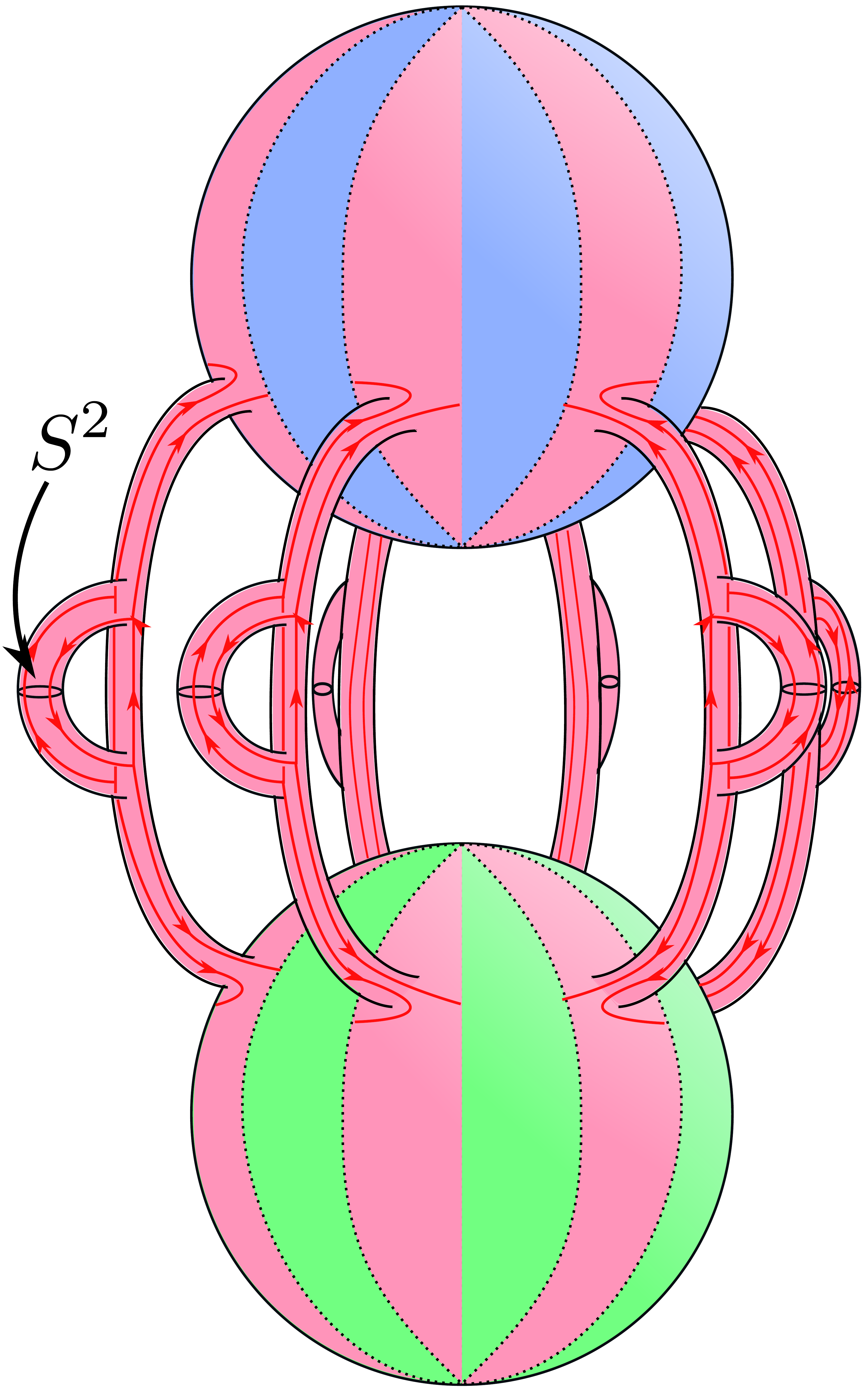}
\end{tabular}
\caption{\small{\textsf{(Left) The state on the torus and (middle) the same state redrawn suggestively.  (Right) The three-manifold of the replica path integral.}}}\label{fig:torus2intcont_surgery}
\end{figure}

First, we revisit the setup of Section \ref{sect:torus2int_cont}.  On the left of subfigure of \ref{fig:torus2intcont_surgery} we depict a path-integral preparation of a given state.  Again we are fixing a {\it particular set} $\{\alpha_L,\beta_L,\gamma_L,\delta_L,\sigma,\eta\}$ appearing in the fusion and branching channels.  On the right of the same figure we give a schematic visualization of the replica path integral.

The procedure for simplifying this is similar to the last section and we find the following expression for the replica path integral:
\begin{equation}\label{eq:torus2intcont_surgery}
Z_n=\frac{Z_n[S^3_{LR_1}[\gamma_L,0;\sigma]]Z_n[S^3_{LR_2}[\delta_L,0;\eta]]\,|\Psi_{\alpha_L,\beta_L^\ast,\delta_L}|^{2n}|\Psi_{\alpha_L^\ast,\beta_L,\gamma_L}|^{2n}}{\left(Z[S_L^3[\alpha_L]]\right)^n\left(Z[S_L^3[\beta_L]\right)^n\left(Z[S_L^3[\gamma_L]]\right)^n\left(Z[S^3_L[\delta_L]]\right)^n}
\end{equation}
Again we see that the overlaps $|\Psi_{\ldots}|^2$ do not contribute to the R\'enyi entropies which are completely controlled by the effective anyons at each interface:
\begin{align}
S_n=&\frac{1}{1-n}\left(\log Z_n[S^3_{LR_1}[\gamma_L,0;\sigma]-n\log Z_1[S^3_{LR_1}[\gamma_L,0;\sigma]+\log Z_n[S^3_{LR_2}[\delta_L,0;\eta]]-n\log Z_1[S^3_{LR_2}[\delta_L,0;\eta]]\right)\nonumber\\
=&\frac{1+n}{n}\frac{\pi \bar c\ell}{24\delta}+\log{(\mc S^{eff,1})_\sigma}^0+\log{(\mc S^{eff,2})_\eta}^0
\end{align}
as was expected from Section \ref{sect:torus2int_cont}.

\subsubsection{$A^c$ connected}
\begin{figure}[h!]
\centering
\begin{tabular}{l l l}
\includegraphics[align=c,width=.3\textwidth]{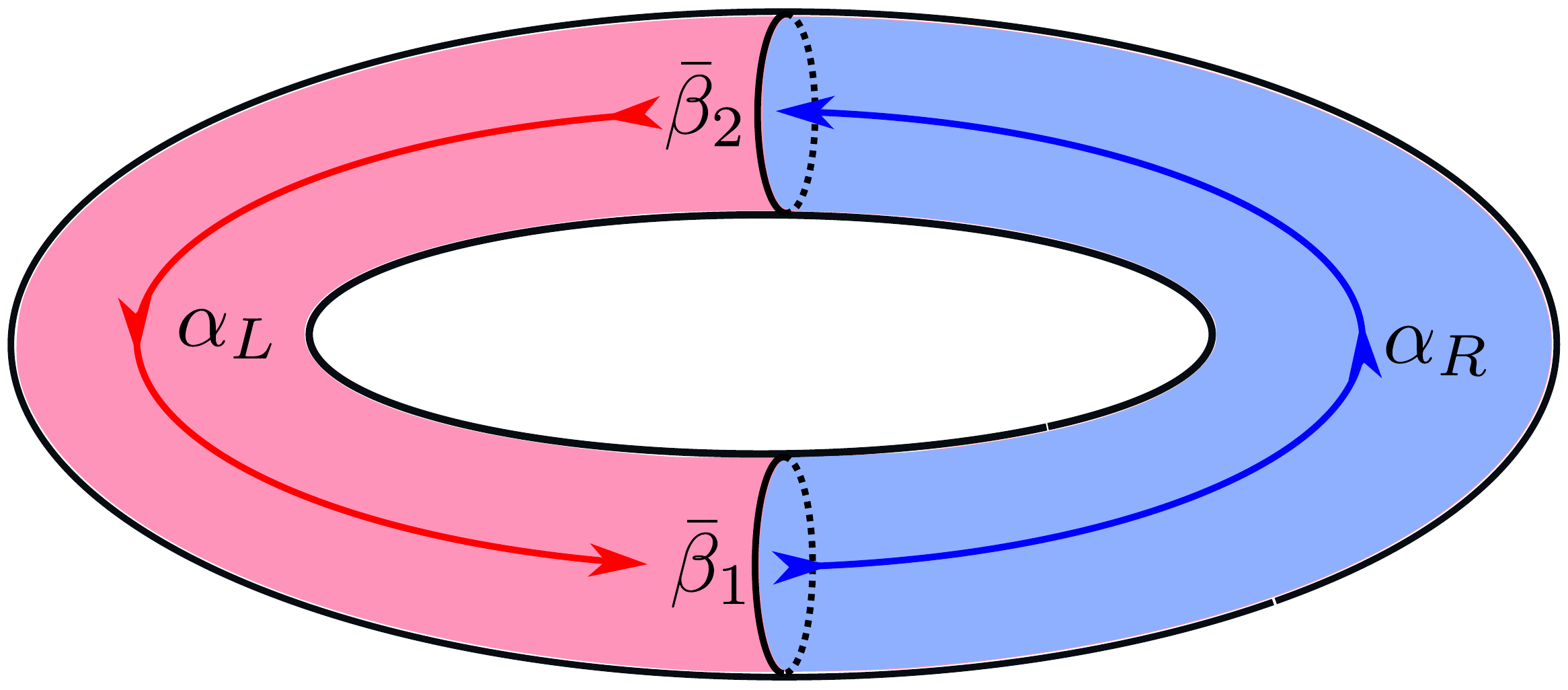}&
\includegraphics[align=c,width=.3\textwidth]{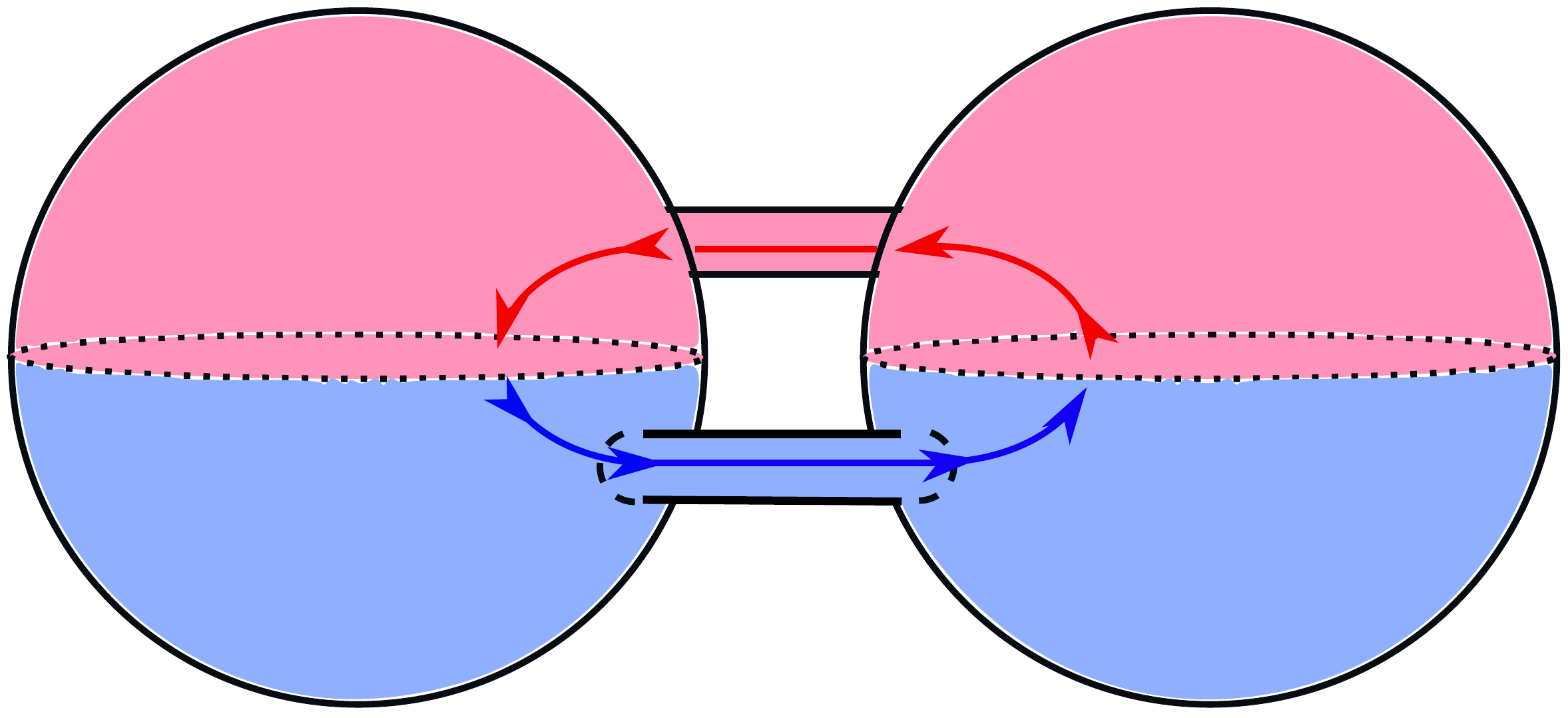}&
\includegraphics[align=c,width=.25\textwidth]{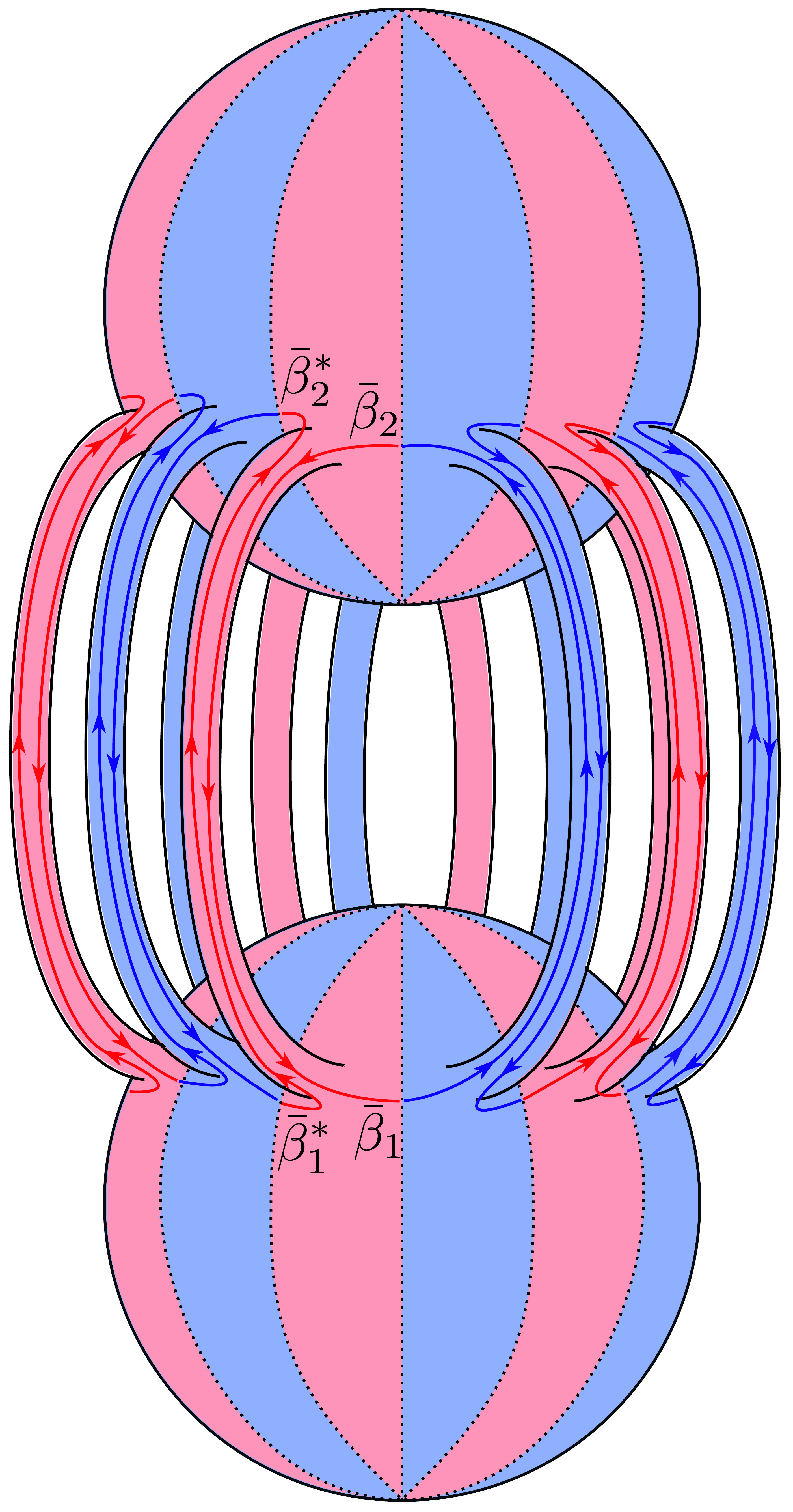}
\end{tabular}
\caption{\small{\textsf{(Left) The state on the torus and (middle) the same state redrawn suggestively.  (Right) The three-manifold of the replica path integral.}}}\label{fig:torus2int_noncont_surgery}
\end{figure}

Finally we revisit the states described in Section \ref{sect:torus2int_noncont}.  The path integral picture of the state is recalled on the left of Figure \ref{fig:torus2int_noncont_surgery}, while the replica path integral is schematically drawn on the right of the same figure.  As before, we are fixing the anyon content to a particular set appearing in the fusion and branching channels.

Surgically removing the $S^2$ handles from the geometry we have the following expression for the replica path integral:
\begin{equation}
Z_n=\frac{Z_n[S^3_{LR}[\alpha_L,\alpha_R;\bar\beta_1]]\,Z_n[S^3_{LR}[\alpha_L,\alpha_R;\bar\beta_2]]}{\left(Z[S^3_L[\alpha_L]]\right)^{n}\left(Z[S^3_R[\alpha_R]]\right)^{n}}
\end{equation}
and consequently the expression for the R\'enyi entropies,
\begin{align}
S_n=&\frac{1}{1-n}\left(\log \frac{Z_n[S^3_{LR}[\alpha_L,\alpha_R;\bar\beta_1]]}{\left(Z_1[S^3_{LR}[\alpha_L,\alpha_R;\bar\beta_1]]\right)^n}+\log \frac{Z_n[S^3_{LR}[\alpha_L,\alpha_R;\bar\beta_2]]}{\left(Z_1[S^3_{LR}[\alpha_L,\alpha_R;\bar\beta_2]]\right)^n}\right)\nonumber\\
=&\frac{1+n}{n}\frac{\pi \bar c\ell}{24\delta}+\log{(\mc S^{eff,1})_0}^{\bar\beta_1}+\log{(\mc S^{eff,2})_0}^{\bar\beta_2}
\end{align}
which matches our results from Section \ref{sect:torus2int_noncont}.

\section{Discussion}\label{sect:Discussion}
In this paper, we have extended the story of gapped interfaces in Chern-Simons theory found in \cite{Fliss:2017wop} in several different directions.  First, in considering non-Abelian theories we showed that the topological boundary conditions of the Abelian theories carry over in a natural way and allow us to define the notion of isotropic interfaces.  We addressed how these boundary conditions affect the anyon excitations of the theory as  the interface is approached; this is consistent with other descriptions in the literature \cite{Lou:2019heg,Shen:2019rck}.  We have also explained how these boundary conditions fit into a resolution to the obstruction of Hilbert space factorization in gauge theories by providing a definition of an extended Hilbert space.  From this we were able to compute the entanglement entropy across an interface, agreeing with the results of \cite{Lou:2019heg}.  We have  also extended our analysis away from states defined on the two-sphere.  Although our examples stopped at interfaces on a torus, in principle the construction of extended Hilbert spaces that we provide in appendix \ref{app:B} gives a clear road map for similar calculations on any Riemann surface.  We then outlined a notion of surgery for interface theories and used this alternative perspective to verify our previous results.  There are several natural open avenues for research that we describe below.

{\bf Further utilizing surgery}

Although the above section on surgery is presented as a novel verification of the exact results from the extended Hilbert space approach, we remark that the full ``power of topology" found in surgery methods was not leveraged in this paper.  Indeed one can envision configurations of interfaces and entangling surfaces, while easy to manipulate as R\'enyi path integrals, whose states are difficult to describe analytically inside an extended Hilbert space.  One such scenario is an entangling surface passing through, say, $p$ interfaces \emph{transversely} (while possibly leaving $q$ interfaces as spectators).  One such configuration on the plane is pictured in Figure \ref{fig:LpqRreplica.pdf}.  For this simple case we can easy use surgery to evaluate the R\'enyi path integral as
\begin{equation}
Z_n=\frac{Z[S^3_L]\prod_{i=1}^pZ[S^3_{LR_i}[\mc I_i]]\prod_{j=1}^{q}Z[S^3_{LR_j}[\mc I_j]]^n}{Z[S^3_L]^{p+nq}}\qquad\qquad S_n=(1-p)\log\left({{\mc S^{(\kappa_L)}}_0}^0\right)+\sum_{i=1}^p\log\left({{\mc S^{(\kappa_{eff}^{(\mc I_i)})}}_0}^0\right).
\end{equation}
\myfig{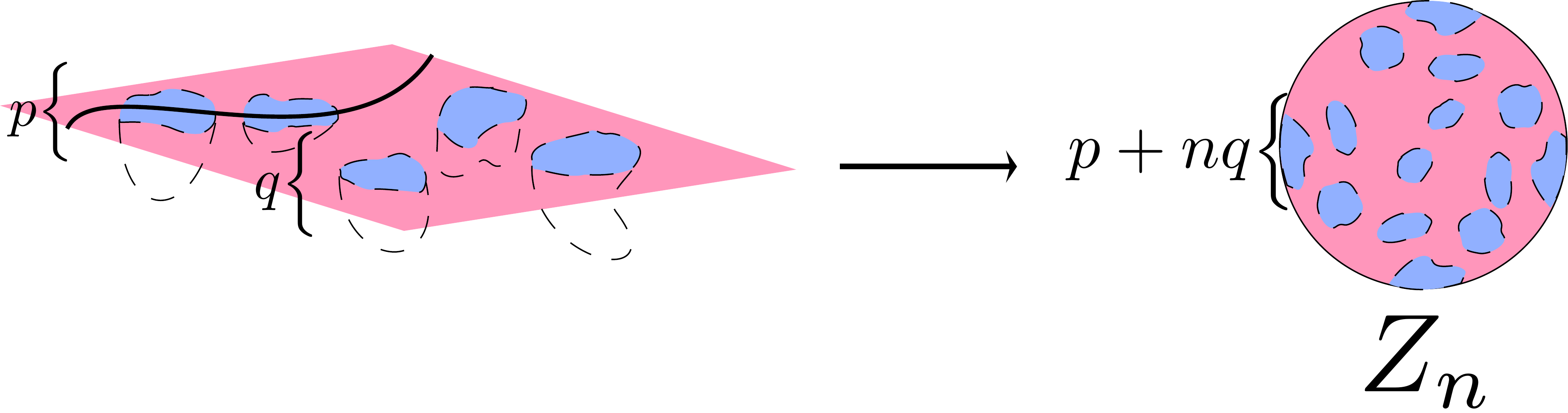}{12}{(Left) The state on the plane has $p+q$ interfaces, and we take the entangling surface to pass through $p$ of them transversely.  The replica manifold (Right) is easy to describe as a three-sphere with $p+nq$ islands.}
We see that the $q$ spectator interfaces add no contribution to the R\'enyi entropy as expected, but we also get an interesting dependence on the $p$ transversely intersected interfaces as well as the background $\kappa_L$ phase.  It will be interesting to include anyon punctures into this story and see if we can understand and extend the results of \cite{Shen:2019rck} from the effective field theory perspective.  We leave this to future work.

{\bf Algebraic entropy}

We have heavily employed the notion of an extended Hilbert space to construct the Hilbert spaces of these interface theories and to define and compute their entanglement entropy.  Let us make some comments as to whether these results can be similarly interpreted in terms of the complementary \emph{algebraic entropy} program \cite{Casini:2013rba}.

Recall that instead of positing a Hilbert space sub-region factorization, this program instructs one to define a local algebra of operators, $\mc A_A$, associated to a sub-region, $A$.  A direct consequence of non-factorization is that such local algebras generically have a non-trivial intersection with their commutant.  Note that there may be more than one choice of algebra associated to a region and hence more than one choice for the resulting center.  Instead of embedding the state in an extended space, one performs the following.  Given a state $\rho$, there exists a unique ``reduced state" $\rho_A\in\mc A_A$ reproducing all expectation values in $A$:
\beq
\mTr\left(\rho_A\mc O_A\right)=\mTr\left(\rho\mc O_A\right)\qquad\qquad\forall \mc O_A\in\mc A_A.
\eeq
$\rho_A$ can be block-diagonalized with respect to the center:
\begin{equation}
\rho_A=\left(\begin{array}{ccc}\lambda_1\rho_1&0&\ldots\\ 0&\lambda_2\rho_2&\ldots\\\vdots&\vdots&\ddots\end{array}\right)\qquad\qquad\mTr_{\mc H_i}\rho_i=1\qquad\sum_i\lambda_i=1.
\end{equation}
with $\mc H_i$ a particular eigenspace of the center.  The von Neumann entropy of $\rho_A$ then has a natural split into the weighted sum of the von Neumann entropies associated to each $\rho_i$ and a Shannon entropy arising from the classical distribution, $\{\lambda_i\}$, with respect to the central eigenspaces.  The relation between the extended Hilbert space, possible definitions for local algebras, and replica path integrals (which seems to ignore both subtleties) has begun to be explored \cite{Lin:2018bud,Delcamp:2016eya}. 
One might guess that the finite dimensionality of Chern-Simons' Hilbert spaces makes this an ideal arena in which to explore these relations.  
\myfig{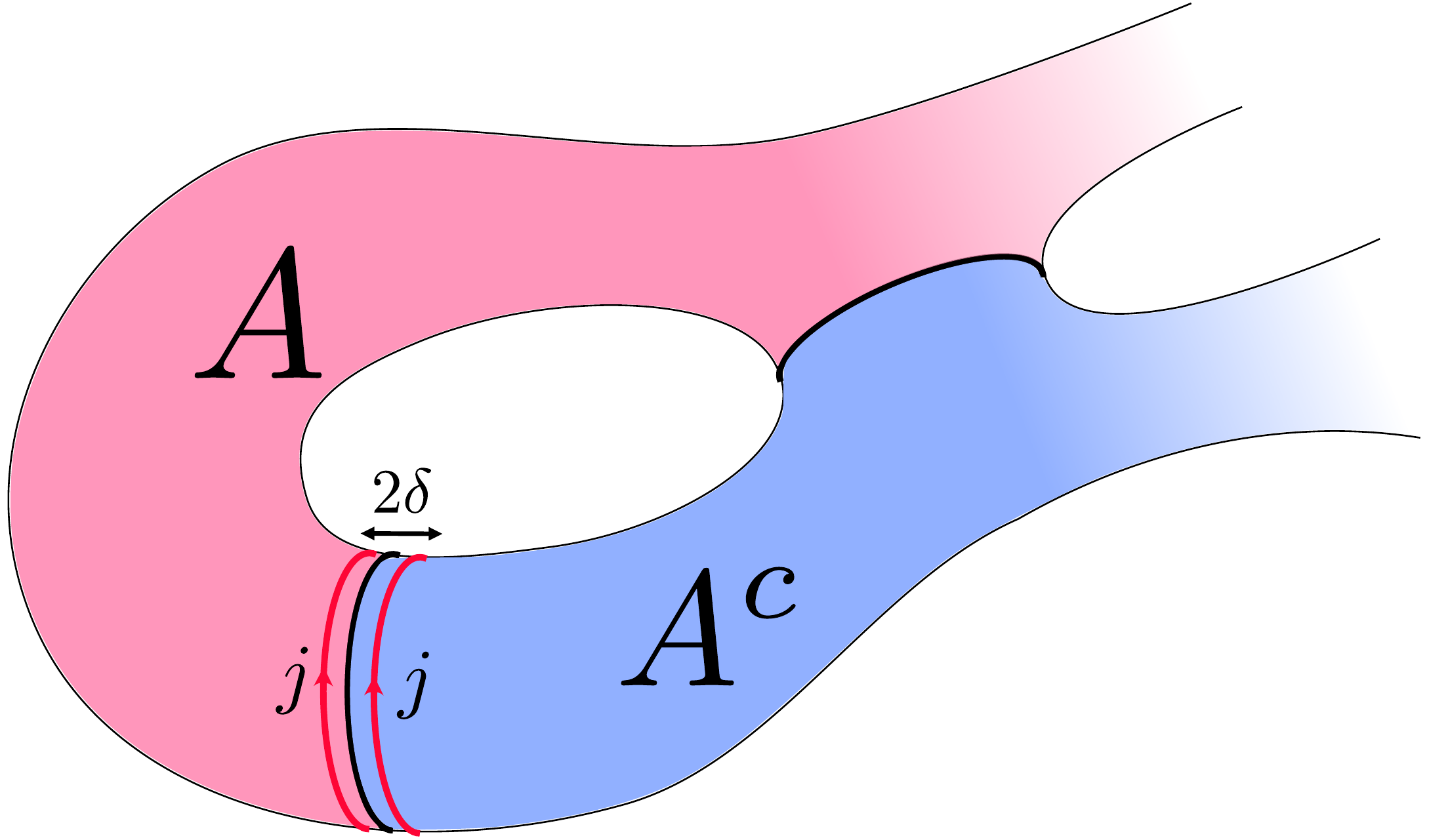}{7}{
In a homogeneous $\mf{su}(2)_k$ theory, $j$ and $j$ uniquely fuse to the identity operator as $\delta\rightarrow0$.  Thus Wilson loop operators acting the edge of $A$ can be represented as an operator acting on the edge of $A^c$.}

Indeed, intuition suggests a natural center generated by Wilson line operators parallel to the entangling surface, however the dimensionality of the Hilbert spaces on Riemann surfaces seems to be too restrictive to make this idea fruitful.  As a trivial example, on the two-sphere $\dim\mc H_{S^2}=1$, and the \emph{entire} operator algebra is proportional to the identity operator.  Thus, for instance, if we declare $\mc A_A$ to be Wilson loop operators acting exclusively on the northern hemisphere, the center is both trivial and everything.

For states on the torus, the Hilbert space is more non-trivial and labelled by integrable representations, $\mc H_{T^2}=\left\{|\alpha\rangle\right\}$.  The operator algebra acting on $\mc H_{T^2}$ is the universal algebra generated by Wilson loops acting on the meridian and on the longitude of the surface of the torus:
\beq
\mc A=\mc U\left[\hat W^{(m)}_\beta\,,\hat W^{(\ell)}_{\gamma}\right]
\eeq
Now consider the bipartition of $T^2$ with two non-contractible entangling surfaces similar to Section \ref{sect:torus2int_noncont} (for simplicity in the homogenous theory).  A natural declaration of the algebra for the $A$ region is the one generated by meridian Wilson operators acting on that region
\beq\label{eq:alg=merWLA}
\mc A_A=\mc U\left[\left.\hat W^{(m)}_\beta\right|_{m\in A}\right]
\eeq
as depicted in Figure \ref{fig:algtorus.pdf}.
\myfig{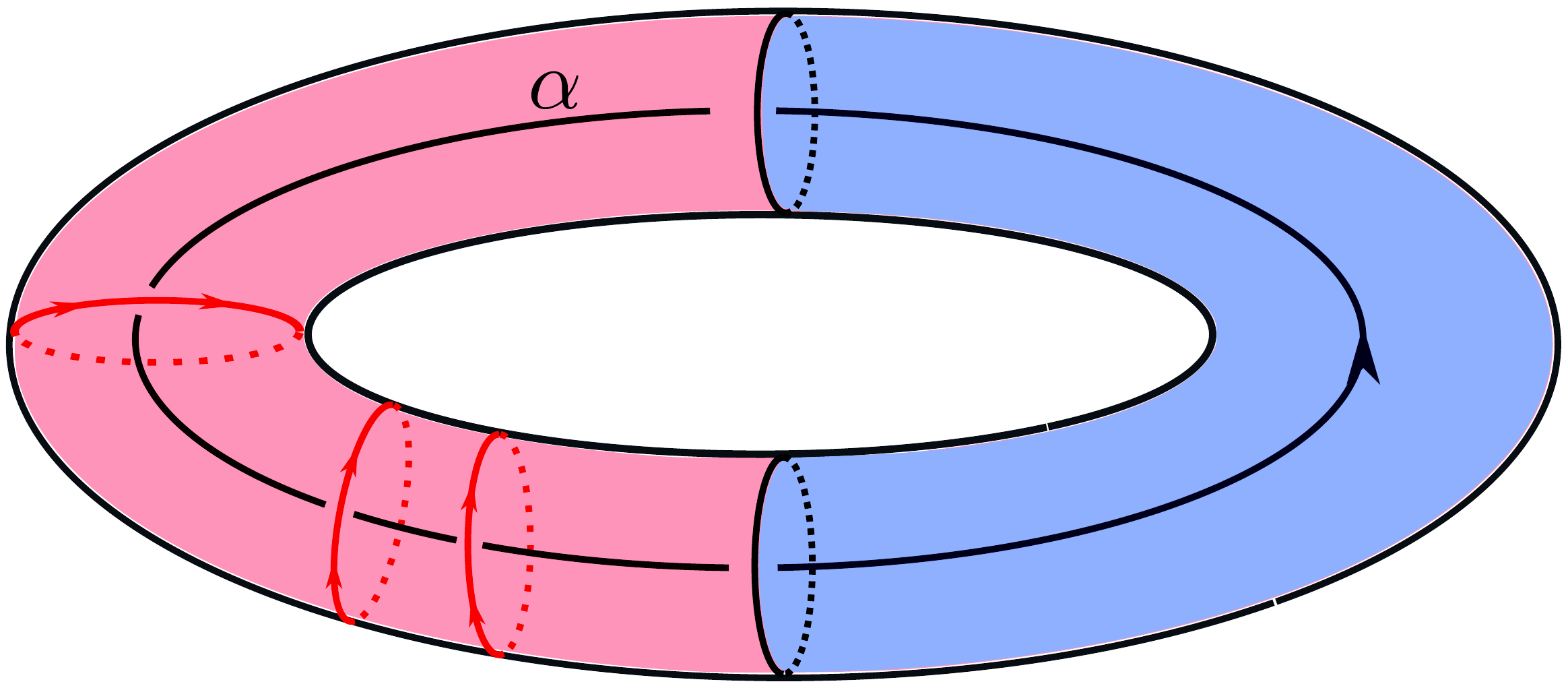}{7}{States on the torus can be prepared via the interior path integral with a longitudinal Wilson line insertion ($\alpha$ in this figure).  For this particular bipartition, a natural candidate for $\mc A_A$ is generated by meridian Wilson loop operators acting in the $A$ region (depicted here in red).}

The problem is that these operators already act diagonally on the standard basis of states \cite{Isidro:1992fz}:
\beq
\hat W_{\beta}^{(m)}|\alpha\rangle=\frac{{\mc S_\beta}^{\alpha}}{{\mc S_0}^{\alpha}}|\alpha\rangle
\eeq
so while it is possible to construct a ``reduced" matrix strictly from Wilson loop operators acting in the $A$ region that reproduces all the expectation values of a state $|\alpha\rangle$:
\beq
\rho^{(\alpha)}_A=\frac{1}{2}\sum_{\gamma}\left({\mc S_0}^\alpha\,{\mc S^\dagger_\gamma}^\alpha\,\hat W_\gamma^{(m)}+\text{h.c.}\right)\in\mc A_A
\eeq
it is easy to see that this reduced density matrix is in fact the full density matrix and its von Neumann entropy vanishes.  Of course the origin of this annoyance is the fact that the operators appearing in \eqref{eq:alg=merWLA} are identical to meridian operators acting {\it in any region}.  This is because the theory is topological.

It is curious that in this case, where the extended Hilbert space description works out so well, the algebraic approach seems to yield no leverage.  The center that we described above is akin to the ``magnetic center" described in \cite{Casini:2013rba}; in the topological $\mathbb Z_2$ lattice gauge theory, the authors also found zero algebraic entropy for this choice of center.  In that paper, the authors also constructed an ``electric center" correctly reproducing both the area law and the TEE.  However this center seems to only be available in the microscopic description (and not, say, from the effective $K=2\sigma_x$ Abelian Chern-Simons theory \cite{Kou:2008rn}).  In general, without embedding the Chern-Simons theory into some larger Hilbert space, we expect that the algebraic entropy corresponding to a bipartition of a closed surface cannot reproduce the TEE.  We can understand this, heuristically, from the following reasoning: the TEE is an intrinsically {\bf negative} contribution and can only appear in an entropic quantity because it is subleading to a {\bf divergent} area law.  It is then unclear how this can appear from the algebraic entropy which is (i) {\bf positive} by definition, and (ii) bounded above by $\log\text{dim}\mc H$ and therefore {\bf finite}.  Setups where the algrebraic approach could possibly yield interesting results will be a subject of future research.  This includes states on surfaces with boundary (such that $\dim\mc H=\infty$) and Chern-Simons theories with a non-compact gauge group.

{\bf AdS$_3$/CFT$_2$}

These  questions lead to another natural area of inquiry: the AdS$_3$/CFT$_2$ correspondence.  Investigations in this direction were initiated in \cite{Gutperle:2018fkz} for Abelian Chern-Simons theories propagating on AdS.  Our construction of isotropic interfaces for non-Abelian theories allows for deeper questioning.  Indeed, the bulk matterless theory itself has a natural description in terms of two $SL(2,\mathbb R)$ Chern-Simons connections.  This fact makes three-dimensional holography an ideal testing ground for exploring questions of bulk entanglement and bulk factorization (or more precisely, the lack thereof).  Interesting work has already appeared in this direction \cite{Belin:2019mlt}.  
It would be interesting if our construction (with suitable generalization), can provide precise realizations of entanglement wedge reconstruction, quantum error correction, and the ``area operator" in 3d holography.  Additionally we hope inquiries along these lines can shed light on recent appearances of Ishibashi(-like) states in AdS$_3$/CFT$_2$ and what role they play in both entanglement and Wilson line expectation values \cite{Castro:2018srf,Nakayama:2015mva,Verlinde:2015qfa,Verlinde:2020upt}.  A challenge to these follows-up is the extension of this work to non-compact gauge groups.  This is a non-trivial and interesting subject in its own right.

After the present work was completed, Ref. \cite{Sohal:2020ojl} appeared, which explores an interesting example of non-Abelian interfaces between distinct Moore-Read states. 

\vskip .5cm
{\bf Acknowledgements:} 
We would like to thank Taylor Hughes and Onkar Parrikar for useful discussions at the onset of this work.  JRF and RGL are supported by the United States Department of Energy QuantISED program, under contract DE-SC0019517.  JRF is also supported by the ERC Starting Grant GenGeoHol. 

\appendix\numberwithin{equation}{section}

\section{Appendix: Hilbert space construction and ground state degeneracy}\label{app:B}

In this appendix we explore the role of our extended Hilbert space construction as a regulated form of the ``gluing" procedure familiar in axiomatic TQFT.  That is to say, it provides an identification of the constituent Hilbert spaces (as an extended, tensor product space) and the final product manifold Hilbert space (as an embedded Hilbert space).  This prescription is both precise and effective: below we show that it can reproduce the known ground state degeneracy (GSD) of the field theory on a Riemann surface and then also extend it to Riemann surfaces with isolated interfaces.  In fact, for this latter case, the extended Hilbert space {\it provides a  principled definition} of the Hilbert space of these theories, reproducing the ground state degeneracy (GSD) counting in \cite{Lan:2014uaa}.

\textbf{Homogenous theories}

Let us begin with a homogeneous theory on a Riemann surface, $\Sigma_g$, of genus $g$.  As a brief description, we want to decompose this Riemann surface into a collection of simpler surfaces with circular boundaries.  Associated to each circular boundary is a WZW Hilbert space in a fixed conformal module (intuitively the primary associated with an anyon threading the circle) and the Hilbert space on the surface is given by the fusion space of these conformal primaries.  The full Hilbert space $\mc H_{\Sigma_g}$ will be realized as an embedded subspace of the tensor product of the constituent Hilbert spaces; this subspace is isolated by the kernel of appropriate gapping operators.  To be specific, let's try to realize $\Sigma_g$ as a two-sphere with $2g$ circular boundaries glued to $g$ annuli.  The extended Hilbert space prescription tells us to realize $\mc H_{\Sigma_g}$ as
\begin{equation}
\mc H_{\Sigma_g}\hookrightarrow \tmc H_{\Sigma_g}\subset \mc H_{S^2\setminus (D^2)^{2g}}\otimes \left(\bigotimes_{i=1}^{g}\mc H_{S^2_i\setminus (D^2)^2}\right)
\end{equation}
where each factor is given by
\begin{align}\label{eq:HS2holes}
\mc H_{S^2\setminus (D^2)^n}=&\bigoplus_{\alpha_1\ldots\alpha_n}\left(\mc V_{\alpha_1,\ldots,\alpha_n}\mc H^{(WZW)_{\mf g_k}}_{S^1}[\alpha_1]\otimes \ldots\otimes \mc H^{(WZW)_{\mf g_k}}_{S^1}[\alpha_n]\right)
\end{align}
where $\mc H_{S^1}^{(WZW)_{\mf g_k}}[\alpha]$ is an infinite dimensional module of the $\hat{\mf g}_k$ K\v ac-Moody with affine weight $\alpha$ and $\mc V_{\alpha_1,\ldots,\alpha_n}$ is the fusion space of the conformal primaries labelled $\left\{\alpha_1,\ldots,\alpha_n\right\}$ (or equivalently the Hilbert space dimension of the $S^2$ punctured by sources $\left\{\alpha_1,\ldots,\alpha_n\right\}$); it is a finite number.  The above embedding is uniquely specified by restoring bulk gauge invariance of the CS theory on $\Sigma_g$; this is enforced at each circular boundary.  Without loss of generality, let us pick an ordering of the holes on the sphere into pairs (labelled by an index $i$) that will be matched with a particular constituent annulus (also labelled by $i$).  Then to be precise
\begin{equation}
\tmc H_{\Sigma_g}=\text{ker}\left\{\mathbb Q_{i}^{1,2}\right\}_{i=1,\ldots, g}
\end{equation}
where $\mathbb Q_{i_{1,2}}$ is shorthand for a collection of operators
\begin{align}\label{eq:mathbbQdef}
\mathbb Q_{i}^{1}\equiv& (\overbrace{\mathbbm 1\otimes\ldots\otimes\underbrace{\mc J_{a,n}\otimes \mathbbm 1}_{i}\otimes\ldots}^{\mc H_{S^2\setminus (D^2)^{2g}}}\overbrace{\mathbbm 1\otimes\mathbbm 1\otimes\ldots}^{\bigotimes_i\mc H_{S^2_i\setminus (D^2)^2}})+(\overbrace{\mathbbm 1\otimes\ldots}^{\mc H_{S^2\setminus (D^2)^{2g}}}\otimes \overbrace{\mathbbm 1\otimes \mathbbm 1}^{\mc H_{S^2_1\setminus(D^2)^2}}\otimes\ldots\otimes\overbrace{\mc J_{a,-n}\otimes \mathbbm 1}^{\mc H_{S^2_i\setminus (D^2)^2}}\otimes\ldots)\nonumber\\
\mathbb Q_{i}^2\equiv& (\overbrace{\mathbbm 1\otimes\ldots\otimes\underbrace{\mathbbm 1\otimes\mc J_{a,n}}_{i}\otimes\ldots}^{\mc H_{S^2\setminus (D^2)^{2g}}}\overbrace{\mathbbm 1\otimes\mathbbm 1\otimes\ldots}^{\bigotimes_i\mc H_{S^2_i\setminus (D^2)^2}})+(\overbrace{\mathbbm 1\otimes\ldots}^{\mc H_{S^2\setminus (D^2)^{2g}}}\otimes \overbrace{\mathbbm 1\otimes \mathbbm 1}^{\mc H_{S^2_1\setminus(D^2)^2}}\otimes\ldots\otimes\overbrace{\mathbbm 1\otimes\mc J_{a,-n}}^{\mc H_{S^2_i\setminus (D^2)^2}}\otimes\ldots)\qquad n,m \in \mathbb Z
\end{align}
This kernel forces an identification of the K\v ac-Moody weights at the glued interfaces.  It is instructive to focus on the action of \eqref{eq:mathbbQdef} (ignoring the extraneous $\mathbbm 1$'s) on a particular block in the decomposition of \eqref{eq:HS2holes}
\begin{equation}
\left.\text{ker}\left\{\mc J_{a,n}\otimes\mathbbm 1+\mathbbm 1\otimes\mc J_{a,-n}\right\}\right|_{rest.}\subset\mc H_{S^1}^{(WZW)_{\hat{\mf g}_k}}[\alpha]\otimes\mc H_{S^1}^{(WZW)_{\hat{\mf g}_k}}[\beta]\equiv \tmc H_{S^2[\alpha,\beta^\ast]}
\end{equation}
and note that it is precisely how we described the embedded Hilbert space of a two-sphere decomposed into two hemipheres punctured by anyons $\alpha$ and $\beta^\ast$.  Hence the dimension of this kernel is $\dim\left(\text{ker}\left(\mc J_{a,n}\otimes\mc J_{a,-n}\right)\right)=\dim\tmc H_{S^2[\alpha,\beta^\ast]}=\delta_{\alpha,\beta}$.  Thus the gluing operators pick out a unique state (per primary module) at each circular interface.  This construction of $\tmc H_{\Sigma_g}$ is heuristically correct; indeed we can now count the GSD at genus $g$:
\begin{equation}
\dim\tmc H_{\Sigma_g}=\sum_{\alpha_1,\ldots,\alpha_g}\mc V_{\alpha_1,\alpha_2,\ldots,\alpha_g, \alpha_1^\ast,\alpha_2^\ast,\ldots,\alpha_g^\ast}.
\end{equation}
Let us massage the above fusion space by fusing $\left\{\alpha_1,\ldots,\alpha_g\right\}$ into the representation $\gamma$ and similarly $\left\{\alpha_1^\ast,\ldots,\alpha_g^\ast\right\}$ into $\gamma^\ast$:
\begin{align}
\dim\tmc H_{\Sigma_g}=&\sum_{\alpha_1\ldots\alpha_g}\sum_{\gamma}{\mc N^\gamma}_{\alpha_1\ldots\alpha_g}{\mc N^{\gamma^\ast}}_{\alpha_1^\ast\ldots\alpha_g^\ast}\nonumber\\
=&\sum_{\alpha_1\ldots\alpha_g}\sum_{\gamma}\sum_{\beta_1,\beta_2}\frac{1}{\left({\mc S_0}^{\beta_1}\right)^{g-1}}{\mc S_{\alpha_1}}^{\beta_1}\ldots{\mc S_{\alpha_g}}^{\beta_1}{{\mc S^\dagger}_{\beta_1}}^{\gamma}\frac{1}{\left({\mc S_0}^{\beta_2}\right)^{g-1}}{\mc S_{\alpha_1^\ast}}^{\beta_2}\ldots{\mc S_{\alpha_g^\ast}}^{\beta_2}{{\mc S^\dagger}_{\beta_2}}^{\gamma^\ast}\nonumber\\
=&\sum_{\beta}\frac{1}{|{\mc S_0}^{\beta}|^{2g-2}}
\end{align}
where the second line follows from the Verlinde formula \cite{Verlinde:1988sn} and the third from the unitarity of the modular $S$ matrix.

\textbf{Interface theories}

Now we move to theories defined on Hilbert spaces with isolated interfaces (that we will for simplicity, take to be circular) that we will index by the pair $A,B$: $\left\{\mc I_{AB}\right\}$.  Given the discussion of the paper the generalization is entirely clear: each interface $\mc I_{AB}$ is determined by a topological boundary condition that i) maps the current algebras on either side to a consistent diagonal subalgebra $\hat{\obmf g}$ and ii) determines the set of permeable anyons through their mutually nonzero branching channels upon restriction to representations of $\omf g$; the latter is enumerated by branching coefficients and the sum of these channels are the tunneling matrices $\mc W^{(\mc I_{AB})}$.

Now let us imagine a Riemann surface, $\Sigma_{total}$, constructed from a collection of compact two-manifolds $\Sigma_{A}$ residing in topological phases whose low-energy descriptions are $\mf g_{A}$ CS theories (with level-Killing forms $\kappa_{A_i}$) by gluing them along their circular interfaces (here labelled by the index $i$), $\{\mc I_{A_iB_i}\}$.  As we saw above, this itself is a non-trivial affair.  Every pair $(\Sigma_A,\Sigma_B)$ that are glued together along a boundary must be commensurate, that is they must support an isotropic subalgebra.  If there are several choices of subalgebras then one must be specified.  We will assume that these details have been sorted and describe the resulting Hilbert space.

The Hilbert space of each constituent $\Sigma_A$ is simple enough to describe:
\begin{equation}\label{eq:HSigmaA}
\mc H_{\Sigma_A}=\bigoplus_{\alpha_{A,1},\alpha_{A,2},\ldots}\mc V^{(\Sigma_A)}_{\alpha_{A,1},\alpha_{A,2},\ldots}\; \mc H_{S^1}^{(WZW)_{\hat{\mf g}_A,\kappa_{A}}}[\alpha_1]\otimes \mc H_{S^1}^{(WZW)_{\hat{\mf g}_A,\kappa_A}}[\alpha_{A,2}]\otimes\ldots
\end{equation}
where $\mc V^{(\Sigma_A)}_{\alpha_{A,1},\alpha_{A,2},\ldots}$ is the fusion space of the compact manifold formed from $\Sigma_A$ by shrinking its circular boundaries to anyon punctures with the respective representation.  Following the discussion in Section \ref{sect:QGint} and the preceding section, the Hilbert space on $\Sigma_{total}$ should be realized as the embedded space
\begin{equation}
\tmc H_{\Sigma_{total}}\subset \bigotimes_{\{A_i\}}\mc H_{\Sigma_{A_i}}\bigotimes_{\{B_i\}}\mc H_{\Sigma_{B_i}}
\end{equation}
defined by the gapping operators given in Section \ref{sect:QGint}:
\begin{equation}\label{eq:defHSigmatotal}
\tmc H_{\Sigma_{total}}=\text{ker}\left\{\mathbb Q_{A_i,B_i}\right\}\qquad\qquad \mathbb Q_{A_i,B_i}\sim\left(v_{A_i}^t\cdot\kappa_A\cdot \mc J^{A_i}\right)_m\otimes \mathbbm 1_{B_i}+\mathbbm 1_{A_i}\otimes\left(v_{B_i}^t\cdot \kappa_{B_i}\cdot \mc J^{B_i}\right)_{-m}
\end{equation}
where by ``$\sim$" we leave implicit all the $\mathbbm 1$'s acting on the additional tensor factors.  As we have seen the kernel of a particular $\mathbb Q_{A_i,B_i}$ is spanned by Ishibashi states of the effective K\v ac-Moody algebra $\omf g_{(A_i,B_i)}$ with level-Killing form $\kappa_{eff,(A_i,B_i)}=v_{A_i}^t\cdot\kappa_{A_i}\cdot v_{A_i}=v_{B_i}^t\cdot \kappa_{B_i}\cdot v_{B_i}$ where the Ishibashi states in question are primary states associated to mutually nonzero branching channels.  In principle, once the topological boundary conditions have been specified, this data uniquely determines this span of effective Ishibashi states and we can regard  \eqref{eq:defHSigmatotal} as \emph{definition} of the full Hilbert space.

The counting of the associated GSD is facilitated by again noting that $\mathbb Q_{A_i,B_i}$ restricted to a fixed block appearing in \eqref{eq:HSigmaA}:
\begin{equation}
\left.\text{ker}\left\{\left(v_{A_i}^t\cdot \kappa_{A_i}\cdot\mc J^{A_i}\right)_{n}\otimes\mathbbm 1+\mathbbm 1\otimes \left(v_{B_i}^t\cdot\kappa_{B_i}\cdot\mc J^{B_i}\right)_{-n}\right\}\right|_{\text{rest.}}\subset\mc H_{S^1}^{(WZW)_{\hat{\mf g}_{A_i},\kappa_{A_i}}}[\alpha_{A_i}]\otimes\mc H_{S^1}^{(WZW)_{\hat{\mf g}_{B_i},\kappa_{B_i}}}[\alpha_{B_i}]
\end{equation}
is the extended Hilbert space description of the two-sphere with an equatorial interface $\mc I_{A_iB_i}$ separating anyons $\alpha_{A_i}$ and $\alpha_{B_i}$.  The dimension of this space is no longer necessarily $\leq 1$ but instead given by the tunneling matrix elements, ${\left(\mc W^{(\mc I_{A_i,B_i})}\right)_{\alpha_{A_i}}}^{\alpha_{B_i}}$.  Although the precise GSD counting will depend on the specific configuration of interfaces, this gives an effective algorithm for computing it: each interface comes associated with $\mc W^{(\mc I_{AB})}$ that we must contract over the free indices of the fusion spaces $\mc V^{(\Sigma_A)}$ and $\mc V^{(\Sigma_B)}$.

To see how this works in practice, let us extend our homogeneous example and take a two-sphere (with $2g$ discs excised) in a phase labelled by algebra $\mf g_0$ and level-Killing form $\kappa_0$ and we attach $g$ annuli in phases described by $\{\mf g_i,\kappa_{i}\}_{i=1,\ldots, g}$.  Each annulus comes associated with two possible interfaces, $\mc I_i^1$ and $\mc I_i^2$, and corresponding tunneling matrices $\mc W^{(\mc I_i^1)}$, and $\mc W^{(\mc I_i^1)}$.  Turning the crank we find
\begin{align}
\dim\tmc H_{\Sigma_{total}}=&\sum_{\alpha_{1}\ldots\alpha_{g}}\sum_{\beta_1\ldots\beta_g}\sum_{\gamma_1\ldots\gamma_g}\sum_{\delta_1\delta_2}\sum_{\epsilon}\frac{1}{\left({{\mc S^{(\kappa_0)}}_0}^{\delta_1}\right)^{g-1}}{{\mc S^{(\kappa_0)}}_{\beta_1}}^{\delta_1}\ldots{{\mc S^{(\kappa_0)}}_{\beta_g}}^{\delta_1}{{{\mc S^{(\kappa_0)}}^\dagger}_{\delta_1}}^\epsilon\nonumber\\
&\qquad\qquad\qquad\times\frac{1}{\left({{\mc S^{(\kappa_0)}}_0}^{\delta_2}\right)^{g-1}}{{\mc S^{(\kappa_0)}}_{\gamma_1}}^{\delta_1}\ldots{{\mc S^{(\kappa_0)}}_{\gamma_g}}^{\delta_1}{{{\mc S^{(\kappa_0)}}^\dagger}_{\delta_2}}^{\epsilon^\ast}\prod_{i=1}^g{{\mc W^{(\mc I_i^1)}}_{\alpha_i}}^{\beta_i}{{\mc W^{(\mc I_i^2)}}_{\alpha_i^\ast}}^{\gamma_i}\nonumber\\
=&\sum_{\alpha_1\ldots\alpha_g}\sum_{\delta}\frac{1}{\left|{{\mc S^{(\kappa_0)}}_0}^\delta\right|^{2g-2}}\prod_{i=1}^g{\left(\mc W^{(\mc I_i^1)}\cdot\mc S^{(\kappa_0)}\right)_{\alpha_i}}^\delta{\left(\mc W^{(\mc I_i^2)}\cdot\mc S^{(\kappa_0)}\right)_{\alpha_i^\ast}}^\delta\nonumber\\
=&\sum_\delta\frac{1}{\left|{{\mc S^{(\kappa_0)}}_0}^\delta\right|^{2g-2}}\prod_{i=1}^g{\left({\mc W^{(\mc I_i^1)}}^t\cdot\mc W^{(\mc I_i^2)}\right)_{\delta}}^\delta
\end{align}
where we've used the fact that $\mc W$ intertwines modular $S$ matrices as well as denoted ${{\mc W}_\alpha}^\beta={{\mc W^t}_{\beta}}^{\alpha}$.  In the cases where the above example matches an example in \cite{Lan:2014uaa} (e.g. $g=1$: a torus with two non-contractible interfaces), the GSD matches.

\providecommand{\href}[2]{#2}\begingroup\raggedright\endgroup

\end{document}